\documentclass[apj]{emulateapj}

\usepackage{natbib}
\usepackage{color}
\usepackage{graphicx}
\usepackage{epstopdf}

\newcommand{\Mjup}{\ensuremath{\mathrm{M}_{\mathrm{Jup}}\!}}
\newcommand{\Msun}{\ensuremath{\mathrm{M}_{\odot}\!}}

\newcommand{\ipfilt}{\ensuremath{i^{\prime}}}

\newcommand{\Ifilt}{\ensuremath{I}}
\newcommand{\Kfilt}{\ensuremath{K}}
\newcommand{\Kpfilt}{\ensuremath{K^{\prime}}}
\newcommand{\Ksfilt}{\ensuremath{K_s}}
\newcommand{\Kcfilt}{\ensuremath{K_c}}

\newcommand{\Hfilt}{\ensuremath{H}}

\newcommand{\Jcfilt}{\ensuremath{J_c}}

\renewcommand{\deg}{\ensuremath{^{\circ}}}
\renewcommand{\arcsec}{\ensuremath{^{\prime\prime}}}
\renewcommand{\arcmin}{\ensuremath{^{\prime}}}
\newcommand{\ee}[1]{\ensuremath{\times10^{#1}}}
\newcommand{\unit}[1]{\ensuremath{\,\textrm{#1}}}

\slugcomment{AJ, 153, 242}

\shorttitle{Companions to RV detected giant planets}
\shortauthors{Ngo et al.}

\begin{document}


 \title{No difference in orbital parameters of RV-detected giant planets \\between 0.1 and 5 au in single vs multi-stellar systems}

\author{Henry Ngo\altaffilmark{1}, Heather A. Knutson\altaffilmark{1}, Marta L. Bryan\altaffilmark{2}, Sarah Blunt\altaffilmark{3,4,5}, Eric L. Nielsen\altaffilmark{3,5}, Konstantin Batygin\altaffilmark{1}, Brendan P. Bowler\altaffilmark{6,7}, Justin R. Crepp\altaffilmark{8}, Sasha Hinkley\altaffilmark{9}, Andrew W. Howard\altaffilmark{2}, and Dimitri Mawet\altaffilmark{2,10}}
\email{hngo@caltech.edu}

\altaffiltext{1}{Division of Geological and Planetary Sciences, California Institute of Technology, Pasadena, CA, USA}
\altaffiltext{2}{Department of Astronomy, California Institute of Technology, Pasadena, CA, USA}
\altaffiltext{3}{SETI Institute, Carl Sagan Center, Mountain View, CA, USA}
\altaffiltext{4}{Department of Physics, Brown University, Providence, RI, USA}
\altaffiltext{5}{Kavli Institute for Particle Astrophysics and Cosmology, Stanford University, Stanford, CA, USA}
\altaffiltext{6}{McDonald Observatory and the Department of Astronomy, The University of Texas at Austin, Austin, TX, USA}
\altaffiltext{7}{Hubble fellow}
\altaffiltext{8}{Department of Physics, University of Notre Dame, Notre Dame, IN, USA}
\altaffiltext{9}{Department of Physics and Astronomy, University of Exeter, Exeter, UK}
\altaffiltext{10}{Jet Propulsion Laboratory, California Institute of Technology, Pasadena, CA, USA}


\begin{abstract}
Our Keck/NIRC2 imaging survey searches for stellar companions around 144 systems with radial velocity (RV) detected giant planets to determine whether stellar binaries influence the planets' orbital parameters. This survey, the largest of its kind to date, finds eight confirmed binary systems and three confirmed triple systems. These include three new multi-stellar systems (HD 30856, HD 86081, and HD 207832) and three multi-stellar systems with newly confirmed common proper motion (HD 43691, HD 116029, and HD 164509). We combine these systems with seven RV planet-hosting multi-stellar systems from the literature in order to test for differences in the properties of planets with semimajor axes ranging between 0.1-5 au in single vs multi-stellar systems. We find no evidence that the presence or absence of stellar companions alters the distribution of planet properties in these systems. Although the observed stellar companions might influence the orbits of more distant planetary companions in these systems, our RV observations currently provide only weak constraints on the masses and orbital properties of planets beyond 5 au. In order to aid future efforts to characterize long period RV companions in these systems, we publish our contrast curves for all 144 targets. Using four years of astrometry for six hierarchical triple star systems hosting giant planets, we fit the orbits of the stellar companions in order to characterize the orbital architecture in these systems. We find that the orbital plane of the secondary and tertiary companions are inconsistent with an edge-on orbit in four out of six cases.
\end{abstract}




\section{Introduction}
Gas giant exoplanets have been found to orbit their host stars over a wide range of orbital separations, spanning more than four orders of magnitude from close-in ``hot Jupiters'' to distant directly imaged planetary mass companions~\citep{Fischer2014b,Bowler2016}. Conventional core accretion models~\citep[e.g.][]{Pollack1996} have argued that giant planet formation is most favorable just beyond the water ice line, where the increased density of solids allows for the rapid formation of cores large enough to accrete a significant gas envelope. If correct, this would suggest that most short period gas giant planets formed at intermediate separations and then migrated inwards to their present-day locations~\citep[e.g.][]{Lin1996}. However, new modeling work motivated by the numerous close-in super-Earth exoplanetary systems~\citep[e.g.][]{Fressin2013,Mulders2015} has suggested that it may be possible to form close-in gas giant planets {\it in situ}, providing an alternative to the migration-driven hypothesis~\citep{Bodenheimer2000,Boley2016,Batygin2016}. We note that the conglomeration of the rocky core itself is a separate process from the accretion of the gaseous envelope. In other words, local formation of the core, followed by extended gas accretion as well as long range migration of the core followed by rapid gas accretion at close-in separations both represent viable {\it in situ} formation scenarios. It is unclear what role, if any, stellar companions might play in these processes. However, the fact that approximately 44\% of field stars~\citep{Duquennoy1991,Raghavan2010} are found in multiple star systems makes this a crucial question for studies of giant planet formation and/or migration.

There had been many recent imaging surveys carried out to determine the frequency of outer stellar companions in systems with close-in ($a<0.1$\unit{au}) transiting giant planets~\citep{Wang2015b,Woellert2015a,Woellert2015b,Ngo2015,Ngo2016,Evans2016}. However, there are relatively few studies that have examined the architectures of systems with intermediate separation ($0.1-5$ au) planets. Giant planets at these intermediate separations have a different migration history than their their short-period counterparts. Some dynamical interactions depend strongly on orbital separations. For example, close-in planets are more tightly coupled to the host star and would therefore experience more rapid tidal circularization than planets on more distant orbits. In addition, the environment of the protoplanetary disk varies as a function of radial separation so these intermediate planets may be the product of different formation pathways. Therefore, it is important to study formation and migration processes on a wide range of planetary separations.

\citet{Eggenberger2007} carried out the most comprehensive survey thus far, searching around 56 known RV-planet host stars as well as a control group of 74 stars without a planetary signal. Both their planet-hosting and control samples were from a CORALIE RV planet survey~\citep{Udry2000}. Considering only companion candidates they have assessed as likely or truly bound, the planet sample had a companion rate of 6/56 while the control group had a larger companion rate of 13/74. Since this study, there have only been a few other surveys~\citep{Ginski2012,Mugrauer2015,Ginski2016} searching for companion stars to RV-detected planet hosts, all with similar sample sizes and target lists. In total, these surveys found 17 systems with RV-detected giant planets and stellar companions within 6\arcsec. 

In this work, we used the Keck Observatory to conduct the largest stellar companion search around RV-detected giant planet host stars to date. These stars host giant planets with orbital semimajor axes ranging from 0.01 to 5 AU, including hot Jupiters, warm Jupiters, and cool Jupiters. Because \citet{Eggenberger2007} and \citet{Mugrauer2015} conducted their diffraction-limited AO surveys with the VLT in the southern hemisphere, our sample of 144 targets contains 119 unique new targets without previous diffraction-limited imaging from observatories similar in size to Keck. \citet{Ginski2012,Ginski2016} carried out a ``lucky imaging'' survey with the 2.2m Calar Alto observatory in the northern hemisphere. Although lucky imaging surveys are less sensitive to close stellar companions, our sample contains 72 unique targets not present in either the VLT or Calar Alto surveys. In \citet{Bryan2016}, we searched for long term RV trends around the same stars to find planetary companions; however, we excluded 23 stars with fewer than 12 Keck RV measurements in order to ensure good constraints on detected RV trends. For the three triple star systems in our sample, we combine our new astrometric measurements with previous measurements in order to fit the orbits of the binary star companions around their center of mass. We also include additional observations of three triple systems with transiting planets which were detected in previous surveys. Unlike them relatively wide separation binaries in our sample, the secondary and tertiary companions in these hierarchical triple systems have a much shorter mutual orbital period, allowing us to detect orbital motion with a several year baseline. For transiting planet systems, we show that imaging of these triple systems can constrain the inclination of the stellar orbits relative to that of the planetary orbit.

In Section~\ref{sec:obs}, we describe our observational campaign. In Section~\ref{sec:analysis}, we describe our photometric and astrometric analysis of candidate stellar companions and provide detection limits for all observed stars. In Section~\ref{sec:individual_systems}, we discuss each detected multi-stellar system individually. In Section~\ref{sec:discuss}, we compare our results to other surveys, discuss the implications on giant planet formation and characterize the orbits of companion stars in our hierarchical triple systems. Finally, we present a summary in Section~\ref{sec:summary}.

\section{Observations}
\label{sec:obs}
We obtained infrared AO images of 144 stars with RV-detected giant planets in order to search for stellar companions. This sample includes the set of AO images used to constrain the masses and orbits of the RV detected companions described in \citet{Bryan2016}, except for two systems. We exclude HD 33636 and HD 190228 because subsequent studies revealed that their companions are actually stars on very close orbits. The companion to HD 33636 is a M-dwarf star on a 2117 day orbit~\citep{Bean2007} and the companion to HD 190228 is a brown dwarf on a 1146 day orbit~\citep{Sahlmann2011}.  All target stars are part of the California Planet Survey~\citep{Howard2010}. We conducted our survey with the NIRC2 instrument (instrument PI: Keith Matthews) on Keck II using Natural Guide Star AO~\citep{Wizinowich2013} from August 2013 to September 2016. The observations are listed in Table~\ref{tab:obs}. We follow the procedure in \citet{Ngo2015}, which we briefly describe here. We operated NIRC2 in natural guide star mode and used the narrow camera setting which has a pixel scale of 10\unit{mas}\unit{pixel}$^{-1}$. A majority of our targets were bright enough (K magnitudes from 1.8 to 8.1) to saturate the NIRC2 detector in \Ksfilt band, so we used the narrower $\Kcfilt$ bandpass ($2.2558-2.2854\,\mu$m) instead to search for companions. For systems where we detected a candidate companion we also obtained $\Jcfilt$ images to measure a $\Jcfilt-\Kcfilt$ color. We determine whether each candidate companion is physically bound using a second epoch of $\Kcfilt$ images taken 1-3 years later, which allows us to check for common proper motion. We flat-field and dark-subtract our data as well as apply a spatial filer to remove bad pixels, as described in~\cite{Ngo2015}. We made photometric and astrometric measurements using individually calibrated frames and computed contrast curves using a median stack of these individually calibrated frames.

In addition to the 144 RV-detected planet host stars in our main survey sample, we also obtained images of three additional transiting planet-host stars previously known to be in triple systems. Two of these triple systems (HAT-P-8, WASP-12) were previously discovered by imaging surveys~\citep{Bergfors2013,Ginski2013} and later characterized as part of our ``Friends of hot Jupiters'' program~\citep{Bechter2014,Ngo2015}. The other triple system (KELT-4A, also known as HIP-51260) was recently reported by \citet{Eastman2016}. Although we do not include these additional triple systems when determining the overall multiplicity rate for planet-hosting stars, we obtain and process the images of these additional systems in the same way as our survey targets. Table~\ref{tab:system_params} lists the properties of the stars from our survey with detected companions as well as the separate sample of previously published triple systems. 

\begin{figure*}
\epsscale{1.18}
\plotone{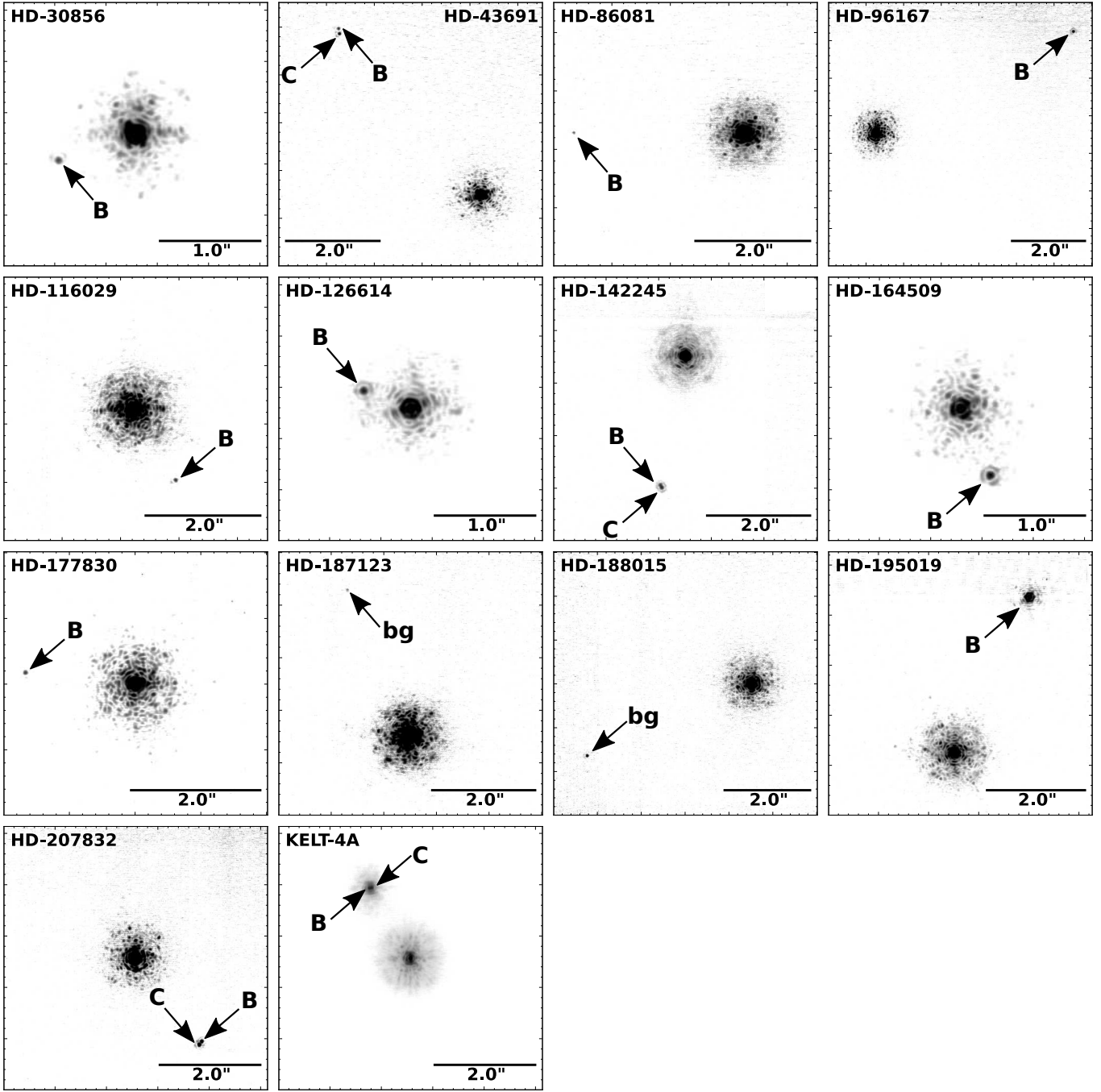}
\caption{Median-stacked K band image for each candidate mutli-stellar system. Each image is oriented such that North points up and East to the left. Confirmed comoving companions are indicated by capital letters while candidates determined to be background objects are labelled as ``bg''. The KELT-4A triple system was not part of our main survey (see Section~\ref{sec:obs}).
\label{fig:stamps}}
\end{figure*}

\begin{figure*}
\epsscale{1.18}
\plotone{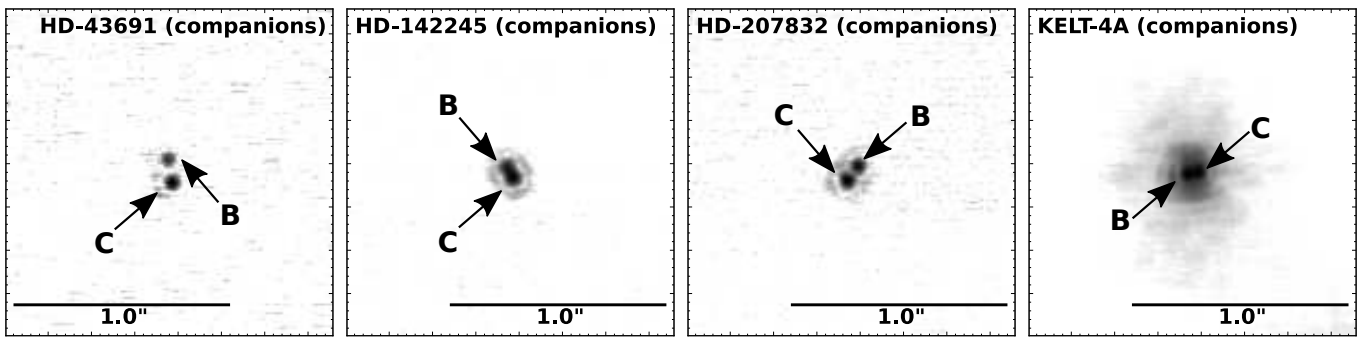}
\caption{Same as Figure~\ref{fig:stamps}, but only showing a close-up view of the secondary and tertiary components of the three triple systems from our survey and a newly reported triple system, KELT-4A.
\label{fig:stamps_triple}}
\end{figure*}

\clearpage
\section{Analysis}
\label{sec:analysis}
\subsection{Photometry and astrometry of candidate multi-stellar systems}
We detect candidate companions around 13 stars in our survey (see Figures~\ref{fig:stamps} and~\ref{fig:stamps_triple}). We summarize the stellar parameters for stars with detected companions as well as our determination of the companion's bound or background status in Table~\ref{tab:system_params}. As described in~\citet{Ngo2015}, we model the stellar point-spread function (PSF) as a combination of a Moffat and Gaussian function. We use a maximum likelihood estimation routine to find the best fit parameters of a multiple-source PSF for each candidate multi-stellar system and determine the flux ratio of the candidate companion to the primary star, as well as the on-sky separation. On 2015 April 13, the optics in the Keck II AO bench were realigned to improve performance. We account for the NIRC2 detector distortion and rotation using astrometric solutions from~\citet{Yelda2010} for data taken prior to this realignment work and from~\citet{Service2016} for data taken afterwards. To determine the stability of the \citet{Yelda2010} solution  (based on data from 2007 to 2009), \citet{Service2016} also computed a distortion solution for data taken just prior to the NIRC2 realignment. The \citet{Service2016} and \citet{Yelda2010} solutions are consistent within 0.5 milliarcseconds, demonstrating that the \citet{Yelda2010} solution is suitable for all of our NIRC2 data taken prior to 2015 April 13 UT. Our reported uncertainties include both measurement errors and the uncertainty contributed by the published astrometric solution. We report the fluxes and astrometry for each candidate companion in Tables~\ref{tab:comp_phot} and~\ref{tab:comp_astr}. 

\subsection{Common proper motion check}
We obtained a second epoch of $\Kcfilt$ images of all candidate companions to determine whether these companions are gravitationally bound to the primary star. As described in~\citet{Ngo2015}, we show the measured projected separation and position angle of each candidate companion as a function of time and compare it to the predicted tracks for a bound companion and an infinitely distant background object in Figures~\ref{fig:astrom_bound1} and~\ref{fig:astrom_bound2}. Predicted tracks are computed using stellar proper motions from \citet{vanLeeuwen2007} and start from the epoch with the smallest uncertainties. When available, we also include previously published astrometric measurements and their corresponding uncertainties in Figures~\ref{fig:astrom_bound1} and~\ref{fig:astrom_bound2} and Table~\ref{tab:comp_astr}. After reviewing the available astrometry, we conclude that 11 out of 13 candidate multi-stellar systems are gravitationally bound. We discuss the astrometric measurements for each individual system separately in Section~\ref{sec:individual_systems}.

\begin{figure*}
\epsscale{1.18}
\plotone{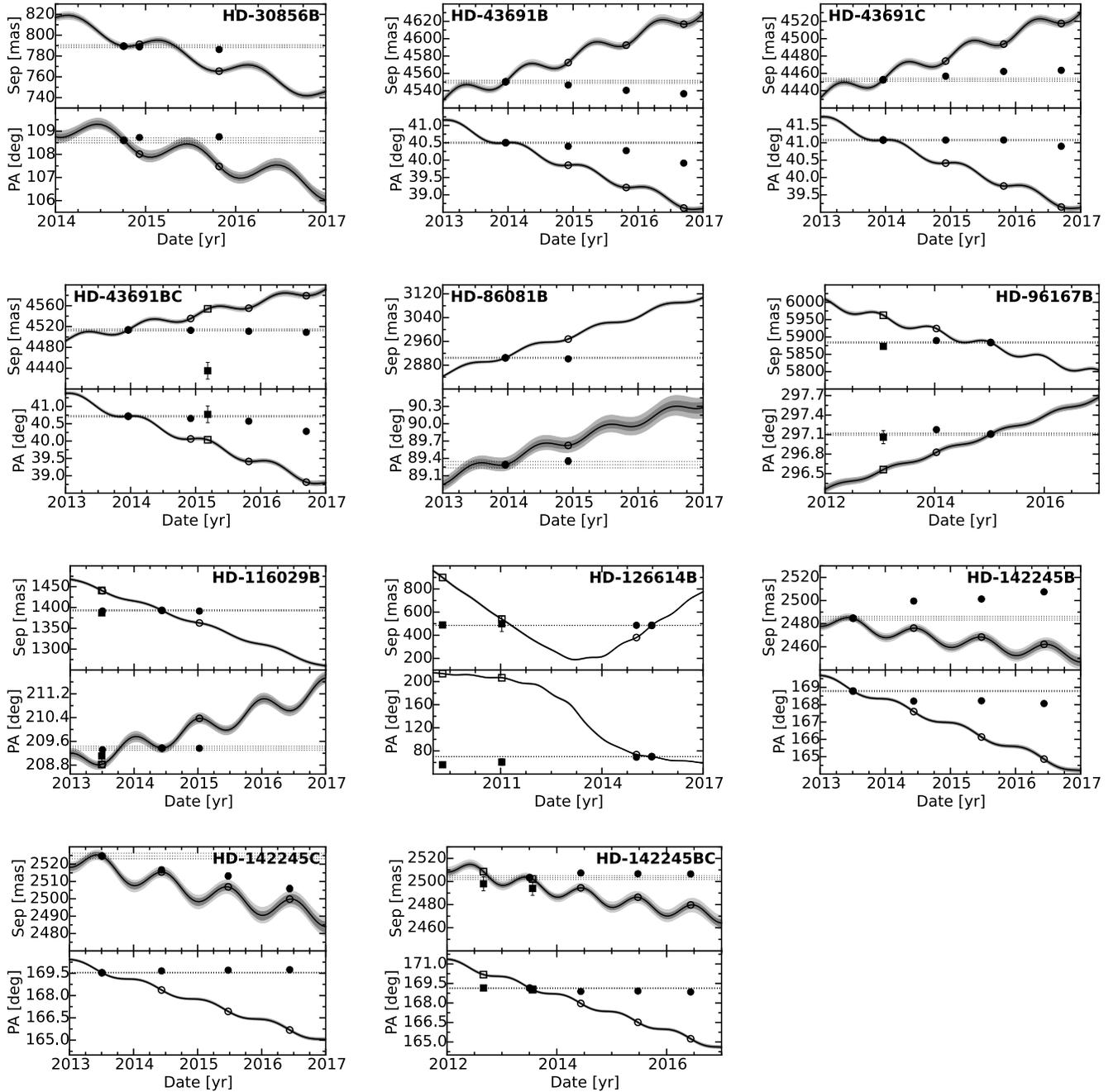}
\caption{Top and bottom panels show the projected separation and position angle of each companion star relative to the primary star. This figure and the following figure includes all confirmed common proper motion companions and background objects from our survey. The solid line shows the expected evolution of separation and position angle for an infinitely distant background object. The dark grey and light grey shaded regions represent the $68\%$ and $95\%$ confidence regions. We use a Monte Carlo routine accounting for uncertainties in our measurements, the primary star's celestial coordinates, proper motion, and parallax. The horizontal dashed lines represent a trajectory with no change in separation or position angle. Filled symbols show measured positions of companions while open symbols show the expected position of an object if it were a background source. Circles represent data from this work and squares represent data from the literature. The data used in this figure can be found in Table~\ref{tab:comp_astr}. Companion candidates that were determined to be physically bound are labelled as the ``B'' or ``C'' components with the center of mass of the two companions denoted as ``BC''.
\label{fig:astrom_bound1}}
\end{figure*}

\begin{figure*}
\epsscale{1.18}
\plotone{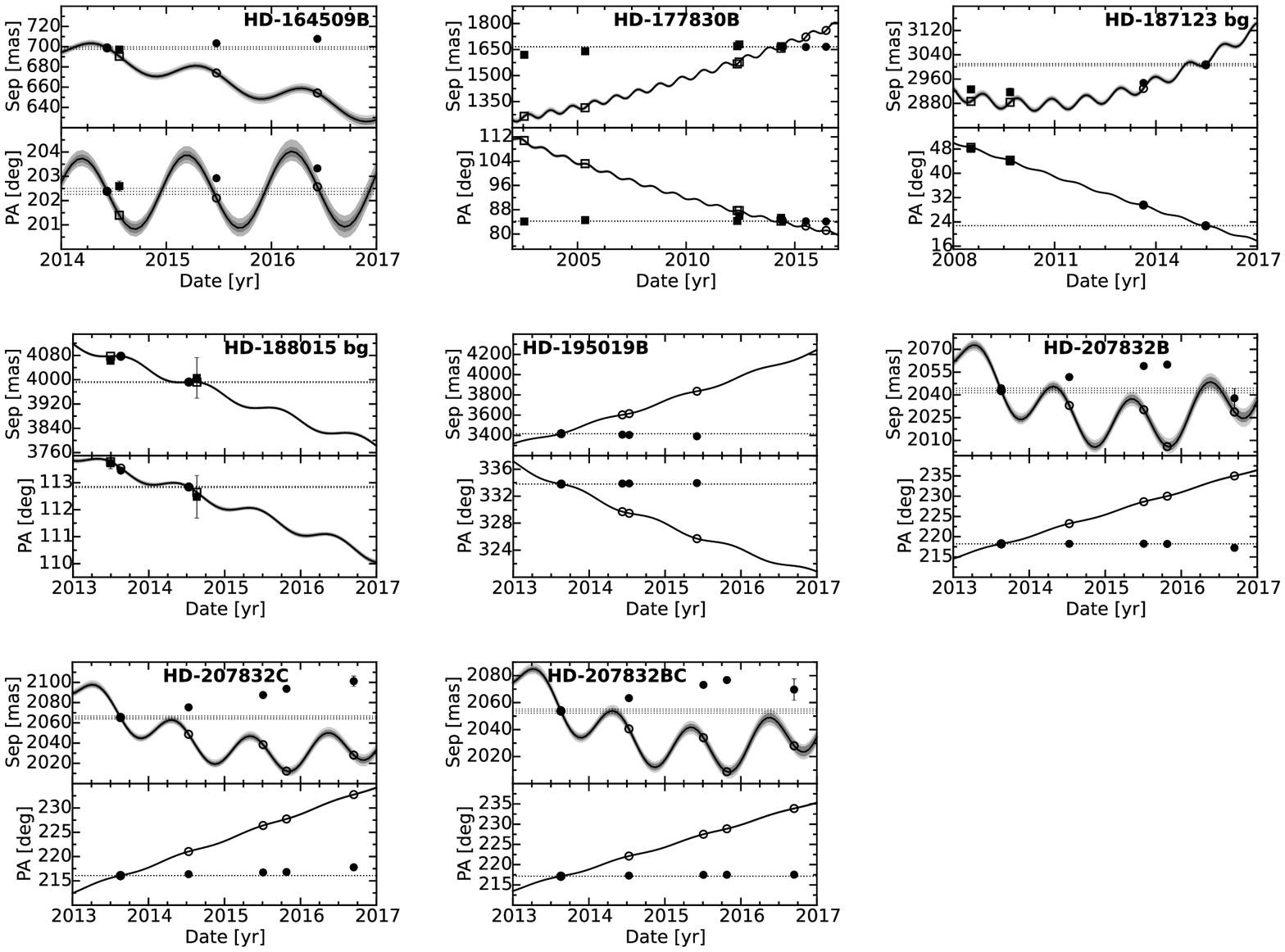}
\caption{Continued from Figure~\ref{fig:astrom_bound1}. These panels also include two background objects, which are labelled as ``bg''.
\label{fig:astrom_bound2}}
\end{figure*}

\subsection{Companion star masses and separations}
For the 11 confirmed multi-stellar systems, we follow the procedure described in~\citet{Ngo2015} to compute the companion star's physical parameters. Here, we will describe our method briefly. We model the stars with PHOENIX synthetic spectra~\citep{Husser2013} assuming solar compositions for both stars ([Fe/H]=0, [$\alpha$/H]=0) and calculate fluxes from each star by integrating the chosen spectra over the observed bandpass. For the primary star, we interpolate PHOENIX spectra to get a model with the corresponding stellar mass, radius, and effective temperature as reported in previous studies, summarized in Table~\ref{tab:system_params}. Using the published parallax and corresponding distance to each system and the flux of the primary star, we solve for the companion temperature that best fits our measured photometric flux ratio. With this best fitting effective temperature, we calculate the corresponding companion mass and radius using zero-age main sequence star models~\citep{Baraffe1998}. We report the properties of each companion star in Table~\ref{tab:interp_sec} with uncertainties calculated from the uncertainty in the measured flux ratio. These errors do not account for systematic uncertainties from our use of PHOENIX spectra or the zero-age main sequence model. We find that errors introduced by assuming solar metallicities and compositions are much smaller than the uncertainties on the measured contrast ratio. Similarly, some error may be introduced from using literature values for primary star mass and radius as these measurements may not have included the effects of the secondary star. Because the secondary stars are several magnitudes or more fainter, this effect is also smaller than the uncertainties. For planet population and orbit fit analyses presented in Section~\ref{sec:discuss}, we use the epoch with the smallest measurement error as the final measurement for each system.

\subsection{Contrast curves for all systems}
\begin{figure*}
\epsscale{1.0}
\plottwo{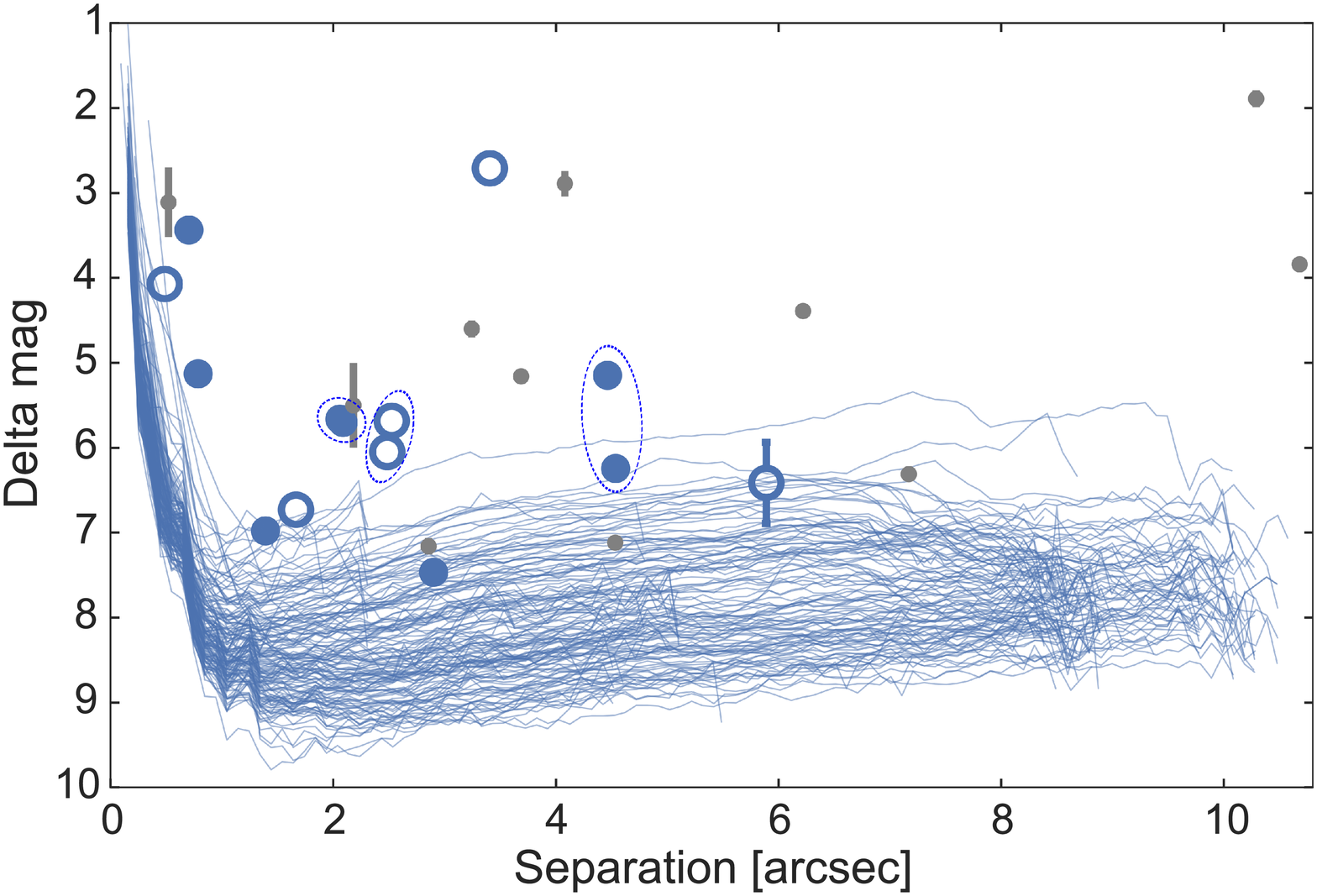}{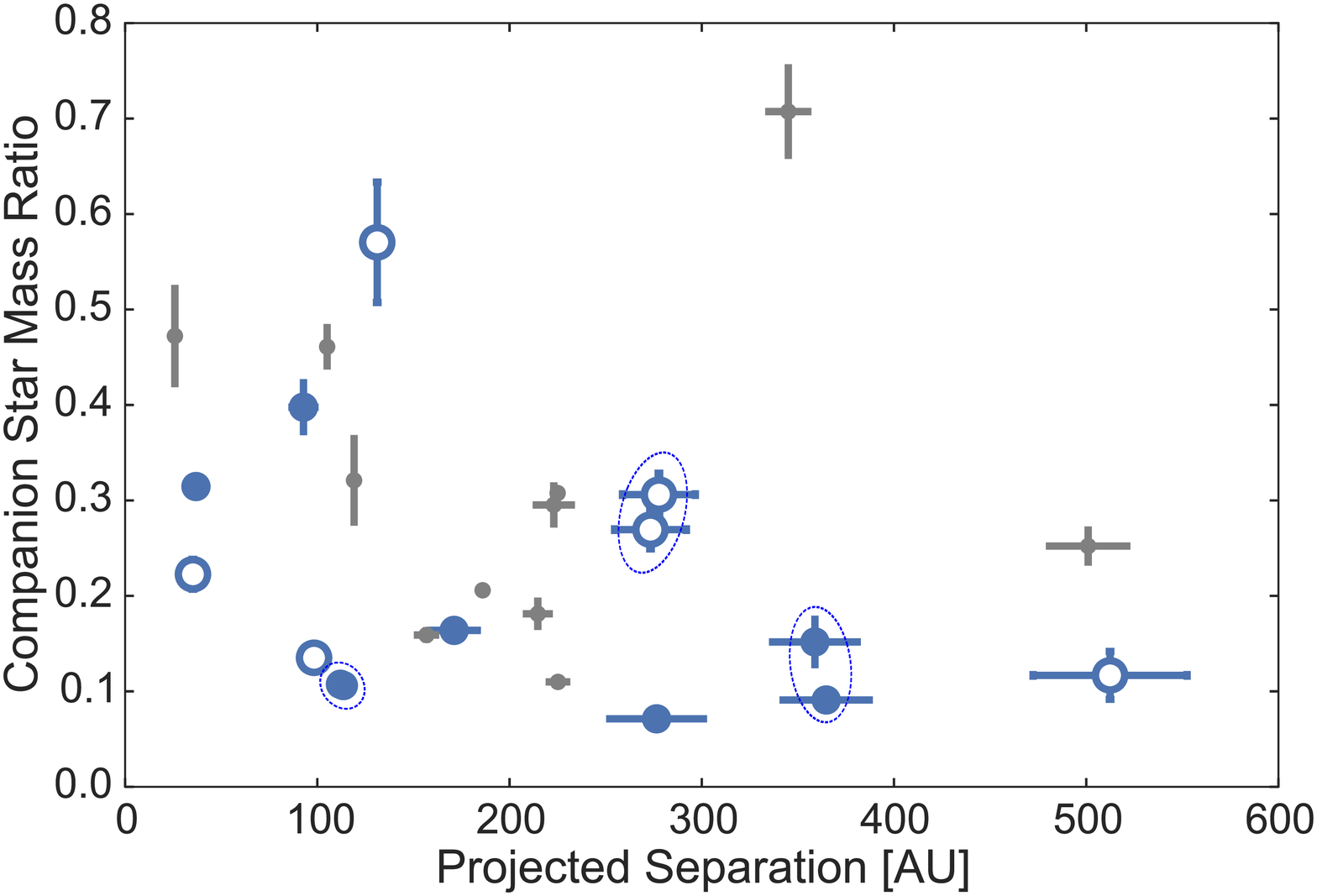}
\caption{In both plots, filled blue symbols are companions that are either new or newly confirmed to be comoving in this work. Open blue symbols are companions previously reported in other studies. Dashed ellipses encompass the two components of the three triple systems in our survey (the two companions to HD 207832 are similar enough that their points almost completely overlap). Small gray symbols show confirmed comoving companions found only in other surveys~\citep{Eggenberger2007,Ginski2012,Mugrauer2015,Ginski2016}. Two of these surveys, \citet{Ginski2012} and \citet{Ginski2016}, are conducted in $i$ band rather than $K$ band. Some of these studies do not report measurement uncertainties and in some cases, mass ratios are estimated based on the reported brightness difference and primary star spectral type. In the left plot, the blue lines show contrast curves for all 144 surveyed RV-host stars out to 10\arcsec. This figure excludes one companion detected by \citet{Ginski2012}. With $\Delta I = 7.5\pm0.5$ and a separation of $1\arcsec.139\pm0\arcsec.005$, this object would be the lowest flux ratio companion on this plot. However, the target star is HD 176051, which is a previously known binary hosting an astrometry detected planet, not a RV-detected planet.
\label{fig:comp_survey}}
\end{figure*}

We report the 5-sigma $\Kcfilt$-band detection limits for each star in our survey as a function of the projected separation. For systems with a companion, we mask out the companion before calculating this detection limit. Figure~\ref{fig:comp_survey} shows the contrast curves for the stars in this work. We compute the contrast curves from the standard deviation of pixel values in a series of annuli, following the procedure described in~\citet{Ngo2015}, and provide a complete list of these curves for each individual system in Table~\ref{tab:contrast}. For targets imaged on the full 1024x1024 array, we do not have coverage in all directions beyond 5\arcsec\ and drop to 90\% directional completeness at 6\arcsec. This limit is smaller for targets imaged on smaller subarrays. This lack of directional completeness results in fewer frames imaged in that region, which increases the standard deviation of stacked pixels and leads to lower contrast.

\section{Individual systems}
\label{sec:individual_systems}
Tables~\ref{tab:system_params}--\ref{tab:interp_sec} summarize our survey targets' properties, measured companion photometry, measured companion astrometry and calculated companion properties, respectively. Table~\ref{tab:comp_astr} also includes astrometric measurements from other studies, when available. The following paragraphs provide additional notes on each of the eleven confirmed multi-stellar systems from our survey as well as candidate companions that were found to be background objects. In total, we report the discovery of three new multi-stellar planet-hosting systems (HD 30856, HD 86081, and HD 207832) and the first confirmation of common proper motion for three additional systems (HD 43691, HD 116029, and HD 164509). 

{\bf HD 30856.} This binary system is reported for the first time in this work. Our images from 2014 and 2015 confirm this companion is comoving with its host star.

{\bf HD 43691.} Our images from 2013 through 2016 provide the first confirmation that this is a comoving hierarchical triple system. The secondary and tertiary components have a projected separation of 360\unit{au} from the primary star, and a mutual projected separation of 10\unit{au}. \citet{Ginski2016} reported a single companion to this system that is consistent with our detection. They did not resolve the individual secondary and tertiary components but they did note that the companion appeared to have an elongated PSF. We are unable to use their astrometric measurements in our analysis because they did not resolve the two companion stars. We label the primary star as ``A'', the northernmost companion as ``B'' and the other companion as ``C''.

{\bf HD 86081.} This binary system is presented for the first time in this work. In \citet{Bryan2016}, we report a long term RV trend of $-1.3\pm0.25\unit{m}\unit{s}^{-1}\unit{yr}^{-1}$ in this system corresponding to a companion with a minimum mass of 0.69 Jupiter masses at a separation of 4.6\unit{au}. Here, we report a companion with a mass of $88\pm2$ Jupiter masses ($0.0840\pm0.002$\Msun) and a projected separation of $280\pm30$\unit{au}. To determine whether or not our imaged companion could be responsible for the measured RV trend, we calculate the minimum companion star mass required to produce the observed RV trend at this projected separation using Equation 6 from \citet{Torres1999}. This minimum mass is $1.4^{+0.6}_{-0.5}\Msun$, indicating that the companion star is not responsible for the RV trend. The non-detection of an additional companion in the AO images set an upper limit on the RV trend companion to be 72 Jupiter masses at 124\unit{au}. Our images from 2013 and 2014 indicate that the stellar companion is comoving with its host star.

{\bf HD 96167.} This binary system was previously reported by \citet{Mugrauer2015} to be a comoving companion to its host star. Their astrometric measurements date back to 2013 and are consistent with our measurements in 2014 and 2015.

{\bf HD 116029.} \citet{Ginski2016} reported this as a candidate binary system based on their 2013 image but were unable to confirm that the companion was comoving because they only had one epoch of astrometry. Our measured separation and position angle from our 2013, 2014 and 2015 images are consistent with their measurement. We provide the first confirmation that this system is a comoving binary pair.

{\bf HD 126614.} The close companion star in this system was first detected in 2009 by \citet{Howard2010}. \citet{Ginski2012} also imaged this system in 2011 and concluded that the companion is comoving. Our images from 2014 and 2015 agree with this assessment. We also found a long term RV trend for this system in \citet{Bryan2016} that is consistent with the imaged stellar companion. Finally, this system also has an additional distant common proper motion companion, NLTT 37349, at 41\arcsec.8~\citep[3000 au;][]{Lodieu2014} that is outside of our survey's field of view.

{\bf HD 142245.} Our images from 2013 through 2016 provide the first images that resolve the individual stars in this hierarchical triple system. The two companion stars have a projected separation of 280\unit{au} from the primary star and a mutual projected separation of 6\unit{au}. \citet{Mugrauer2015} reported images from 2012 and 2013 showing a companion with an elongated PSF that is consistent with our measurements. They determined that this companion was comoving with the host star but were unable to resolve the individual components. We label the primary star as ``A'', the northernmost companion as ``B'' and the other companion as ``C''.

{\bf HD 164509.} Our images from 2014, 2015 and 2016 provide the first confirmation that this is a comoving binary system. \citet{Wittrock2016} also report a companion from their 2014 image that is consistent with our detection. With only one epoch, they were unable to confirm whether the companion is bound, but they noted that the color of the companion was consistent with a lower mass star at the same distance as the target star. We also found a long term RV trend for this system in \citet{Bryan2016} that is consistent with the imaged stellar companion.

{\bf HD 177830.} The companion to this star was first reported by \citet{Eggenberger2007}. They used images from 2004 (\Hfilt-band) and 2005 (\Kfilt-band) to determine that this is a comoving binary system. \citet{Roberts2011,Roberts2015} later combined their images of this binary system with additional observations dating back to 2002. Our images in 2014, 2015 and 2016 recover the same companion reported in these previous studies and support the conclusion that the companion is bound.

{\bf HD 195019.} \citet{Fischer1999} reported a companion around this star, but did not have precise measurements of its photometry or astrometry. \citet{Eggenberger2004} subsequently noted that both components are comoving based on archival data from \citet{Fischer1999}, \citet{Allen2000} and \citet{Patience2002}. \citet{Roberts2011} published additional images from 2002, but did not report uncertainties on their astrometry. Our images from 2013 through 2015 are consistent with all of the previous detections and also confirm that this is a comoving binary system.

{\bf HD 207832.} This hierarchical triple system is reported for the first time in this work. The two companions have a projected separation of 110\unit{au} from the primary star and a mutual projected separation of 4\unit{au}, making this system the most compact RV-planet hosting triple system. Our images from 2013 through 2016 confirm that both companion stars are comoving with their host star. This system also has an extremely wide stellar companion, at 38\arcmin.6~\citep{Lodieu2014}, which is far outside of our survey's field of view. With the fourth star at a projected separation of 126000\unit{au}, we only consider the inner hierarchical triple system for further analysis in this work. We label the primary star as ``A'', the northernmost companion as ``B'' and the other companion as ``C''.

{\bf Background objects.} Two candidate companions were determined to be background objects rather than comoving multi-stellar systems. HD 187123 has a background object approximately 3\arcsec to the north-east. \citet{Ginski2012} concluded that this companion was a background star based on their 2008 and 2009 images. Our new images from 2013 and 2015 independently confirm this conclusion. We found a source approximately 4\arcsec to the north-west in our 2013 and 2014 images of HD 188015 that we determined to be a background object. \citet{Ginski2016} found two candidate companions with similar projected separations, one of which was consistent with our detection, and determined both of them to be background sources. HD 188015 does have a distant comoving companion at 11\arcsec~\citep{Raghavan2006}, but this companion is outside our field of view and we therefore do not include it when calculating the frequency of stellar companions in our sample.

{\bf Companions beyond our survey's field of view.} Six systems in our survey host companions that were outside of our survey's field of view, but reported in the literature. \citet{Eggenberger2007} report companions with separations of 6\arcsec.2 and 10\arcsec.3 around HD 16141 and HD 46375, respectively. The Washington Double Star Catalog~\citep[WDS;][]{Mason2001} shows that HD 109749, HD 178911, HD 186427 (also known as 16 Cyg B) each have a companion at separations of 8\arcsec.3, 16\arcsec.1 and 39\arcsec.6, respectively. Finally, the WDS also reports three companions around GJ 667C at separations of 31\arcsec.2, 32\arcsec.5 and 36\arcsec.4.

\section{Discussion}
\label{sec:discuss}
\subsection{Stellar companion fraction}
We find a raw companion fraction of 11 multi-stellar systems out of 144 surveyed stars, corresponding to a multiplicity rate of $7.6\%\pm2.3\%$. For the typical target, we are sensitive to stellar-mass companions in all directions with projected separations between 0\arcsec.3 and 6\arcsec (at 90\% directional completeness), corresponding to projected separations of 15\unit{au} and 300\unit{au} for a star at 50\unit{pc}. We are sensitive to companions in limited directions up to 10\arcsec. The most distant companion detected in our survey was found at $512\pm43$\unit{au} around HD 96167. It was found at the outer edge of our survey limit, at a separation of $5\arcsec.9$. Our raw companion fraction is consistent, within $1\sigma$, with results from other direct imaging surveys for stellar companions around RV-detected planet host stars, as reported in Table~\ref{tab:RVsurveys}. This companion fraction is lower than the \citet{Eggenberger2007} control sample's companion fraction of $17.6\%\pm4.9\%$, at a significance of $1.9\sigma$. It is not certain whether the difference is by chance, is due to different companion vetting by different RV planet surveys, or if it suggests an anti-correlation between intermediate distance giant planet and a stellar companion. Figure~\ref{fig:comp_survey} compares the projected separations, flux ratios, and mass ratios for the companions in our survey to those reported in these other imaging surveys. 

Prior to this study there were sixteen confirmed wide-binary multiple star systems with separations less than 6\arcsec that were known to host RV planets. Our observations increase this number by six, for a total of 22 systems. This work also increases the number of confirmed multi-stellar systems with companions within 200\unit{au} by four; bringing the total number of such systems to twelve. As indicated by our contrast curves, our survey is more sensitive at small separations than these previous surveys. Figure~\ref{fig:comp_survey} shows that a majority of our new confirmed multi-stellar systems have relatively small flux ratios and projected separations as compared to the sample of previously published planet-hosting multiple star systems. All known multi-stellar RV-planet hosting systems with companions within 6\arcsec\ are listed in Table~\ref{tab:RV_pl}. For each system, we calculate $\nu/n$, the ratio of the planet's precession due to the companion star divided by the planet's mean motion as a proxy for the companion star's ability to dynamically influence the planet. We sort the multi-stellar systems by this metric in order to highlight the most interesting systems for future dynamical studies.

The stellar companion rate for our population of RV-detected giant planet host stars is much lower than the companion fraction of $47\%\pm7\%$ that we reported for transiting hot Jupiter systems~\citep{Ngo2016}. This is most likely due to the relatively severe biases against multiple star systems in the target selection process for RV surveys. Unfortunately, these biases are neither well characterized nor fully reported, and we are therefore unable to report a completeness corrected stellar multiplicity rate for the RV-detected planet population. Unlike transit surveys, RV surveys such as the California Planet Survey~\citep{Howard2010} and the HARPS survey~\citep{Lagrange2009}, vet potential targets for known companion stars that are close enough (generally within 2\arcsec) to fall within the spectrograph slit and are bright enough to have detectable spectral lines at the optical wavelengths where most RV surveys operate. Although it is possible to measure RV shifts for double-lined spectroscopic binaries, RV pipelines developed to search for planets are not typically designed to accommodate a second set of spectral lines and therefore avoid these kinds of systems. Nearby stars are generally identified via archival surveys such as the Washington Double Star catalog~\citep{Mason2001}. RV surveys also discard targets that show large RV variations, effectively eliminating close binaries from their samples. Although RV-planet survey target selection is performed with some quantitative and objective metrics, there are also any number of subjective choices made over the years which are difficult to quantify retroactively~\citep{Clanton2014b}. We therefore conclude that we cannot reliably compare stellar multiplicity rates for planet-hosting stars from RV surveys with similar results for samples of planets detected by transit surveys.

\begin{figure*}
\epsscale{1.0}
\plottwo{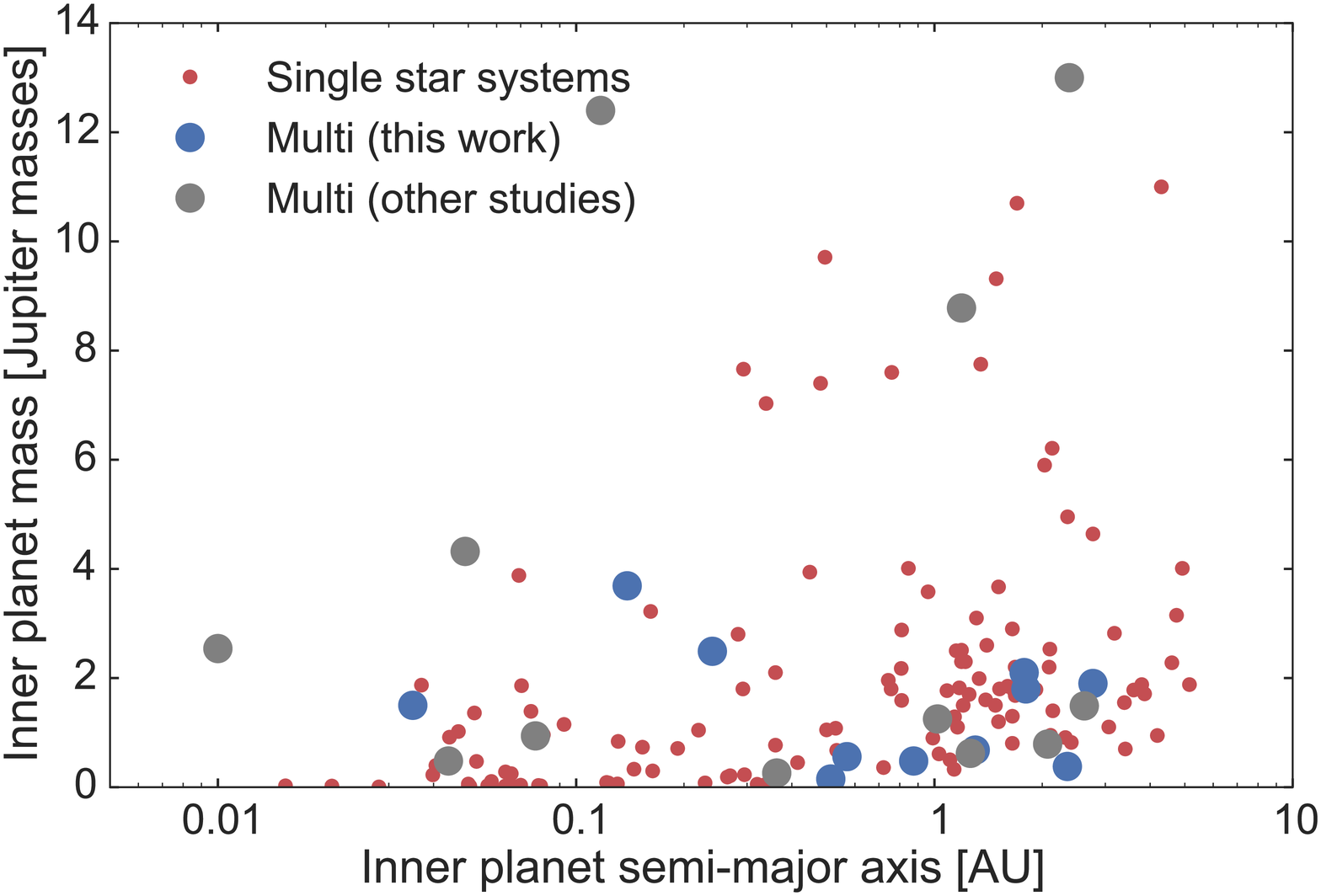}{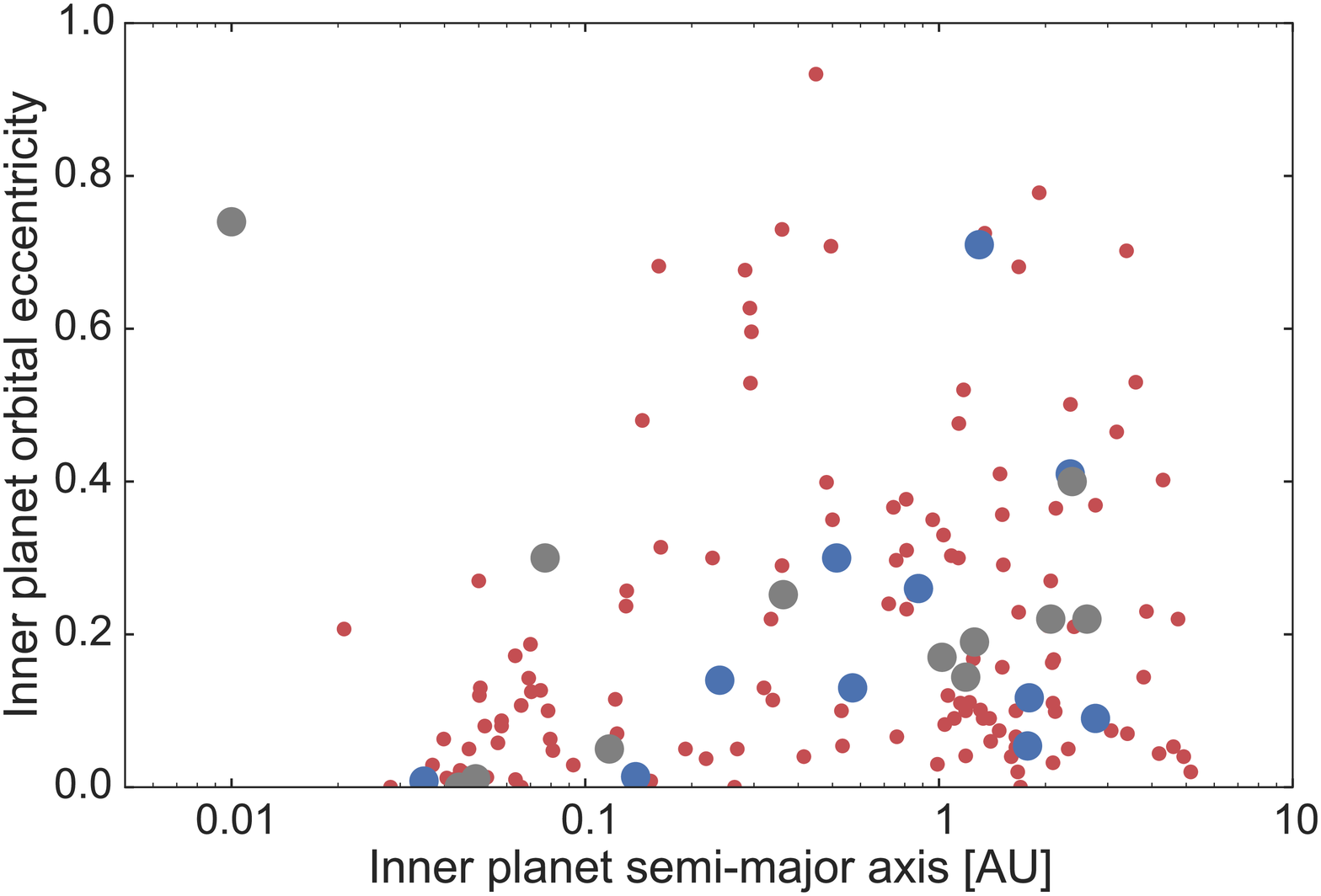}
\caption{Mass and eccentricity vs. semimajor axis of the inner-most planet for single star systems (red) and multi-stellar systems from our survey (blue) and other studies listed in Table~\ref{tab:RV_pl} (gray). 
\label{fig:rv_single_v_multi}}
\end{figure*}

\subsection{Characteristics of RV-detected planets in multi-stellar systems}
\label{sec:single_v_multi}
Although RV surveys are undoubtedly subject to different selection biases than transit surveys when it comes to multiple star systems, this bias is effectively removed when we limit ourselves to comparing different sub-samples within our RV planet survey population, assuming the selection biases affect all sub-samples in the same way. Previous studies have suggested that stellar companions are less common in systems with long-period planets than those with short-period planets~\citep[e.g][]{Kaib2013,Zuckerman2014,Wang2015c}. Here, we compare properties of planets in multi-stellar and single star systems within our survey population of RV-detected planet host stars. Figure~\ref{fig:rv_single_v_multi} and Table~\ref{tab:RV_pl} show the orbital properties of the innermost planet of each single and multi-stellar system in our survey. We wish to determine whether the distributions for innermost planet orbital eccentricity, orbital period and mass differ for single star systems from multi-star systems. Thus, we calculate whether a two-population (for single and multi-stellar systems) distribution is a better fit to the data than a one-population model. To avoid tidal circularization effects, we exclude 35 single and 5 multi-stellar systems with planets with semimajor axes less than 0.1 au in this analysis. Because we have a relatively small sample of multiple star systems in our survey, we also include similar multiple star systems (i.e., those with projected separations less than 6\arcsec; see Table~\ref{tab:RV_pl} for a complete list) from the published literature. In total, there are 98 single star systems and 17 multi-stellar systems considered in this part of the analysis.

Assuming planet eccentricities follow a beta distribution~\citep{Kipping2013}, we calculate the probability of obtaining an individual planet orbital eccentricity $e_k$ to be
\begin{equation}
\mathrm{prob}(e_k | a,b) = \frac{\Gamma(a+b)}{\Gamma(a)\,\Gamma(b)}e^{a-1}(1-e)^{b-1},
\end{equation}
where $\Gamma()$ denotes the Gamma function and $a$ and $b$ are the model parameters. We assume the planet mass and orbital period take the form of the \citet{Cumming2008} power law, so that the probability of obtaining an individual planet mass $m_k$ and orbital period $P_k$ is
\begin{equation}
\mathrm{prob}(m_k,P_k | \alpha, \beta) \propto m^\alpha P^\beta,
\end{equation}
where $\alpha$ and $\beta$ are the model parameters. We assume that the orbital eccentricity is not correlated with orbital period and planetary mass so we can determine the probability of obtaining any individual planetary system to be the product of the above probabilities. Our goal is to compute the likelihood of a model $M$ with a set of parameters $\theta = (a,b,\alpha,\beta)$ for a set of planets. For an individual system, we can write
\begin{equation}
\mathrm{prob} (e_k,m_k,P_k | \theta, M) \propto \frac{\Gamma(a+b)}{\Gamma(a)\,\Gamma(b)} e^{a-1}(1-e)^{b-1} m^\alpha P^\beta. 
\end{equation}
From Bayes' Theorem and choosing uniform priors for all model parameters, we can write the log-likelihood of a one-population model $\mathcal{L}_1$ as
\begin{equation}
\mathcal{L}_1 = \sum_k \ln \left[\mathrm{prob}(e_k,m_k,P_k | M, \theta) \right] + C,
\end{equation} 
where the sum over $k$ includes all RV planet host systems and $C$ is a constant. Similarly, we can write the log-likelihood of a two-population model $\mathcal{L}_2$ as 
\begin{eqnarray}
\mathcal{L}_2 & = & \sum_i \ln \left[\mathrm{prob}(e_i,m_i,P_i | M_s, \theta_s)\right] \nonumber\\
& + & \sum_j \ln \left[\mathrm{prob}(e_j,m_j,P_j | M_m, \theta_m)\right] + C,
\end{eqnarray}
where the sum over $i$ includes all RV planet host systems with no companion star, the sum over $j$ includes all RV planet host systems with at least one companion star, and the model parameters have subscripts $s$ and $m$ to denote separate sets of parameters for single and multi-stellar systems, respectively. $C$ is the same constant from the one-population likelihood.

\begin{figure}
\epsscale{1.0}
\plotone{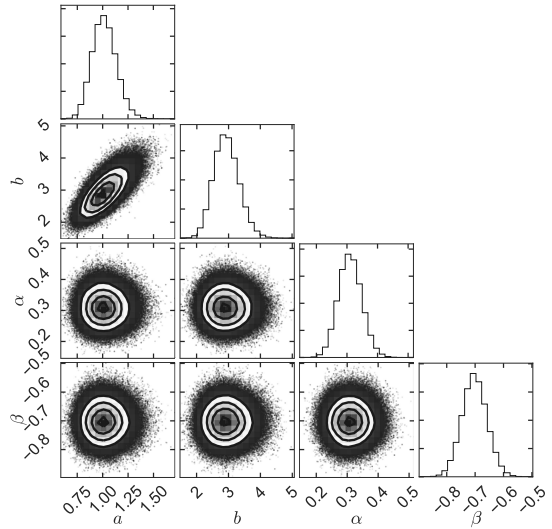}
\caption{Two-dimensional posterior probability distributions on the four model parameters describing the 98 single systems and 17 multi-stellar systems as a single population of planets. The histograms represent the one-dimensional marginalized posterior probability distribution for each parameter.
\label{fig:posterior_1pop}}
\end{figure}

\begin{figure*}
\epsscale{1.0}
\plottwo{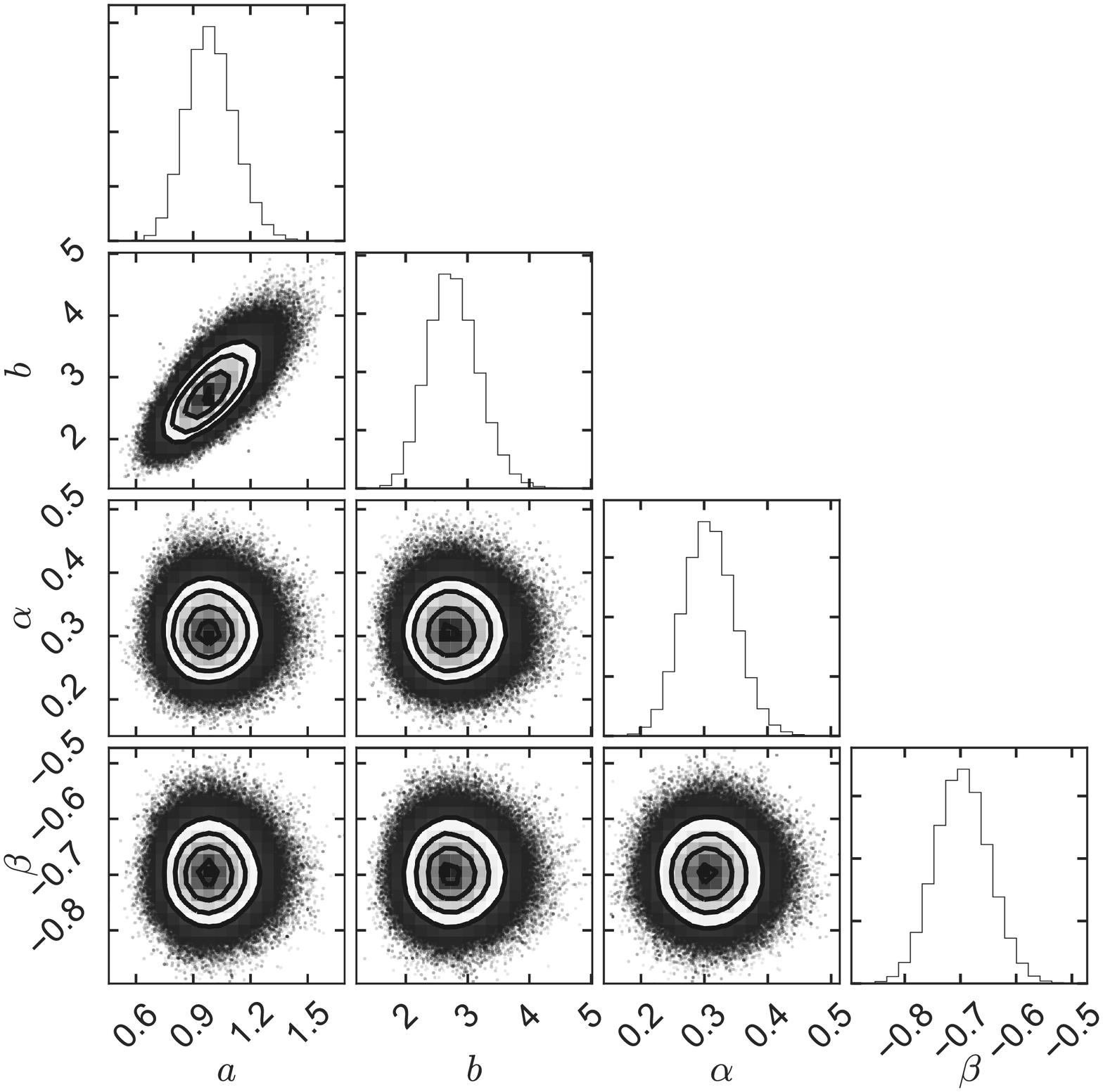}{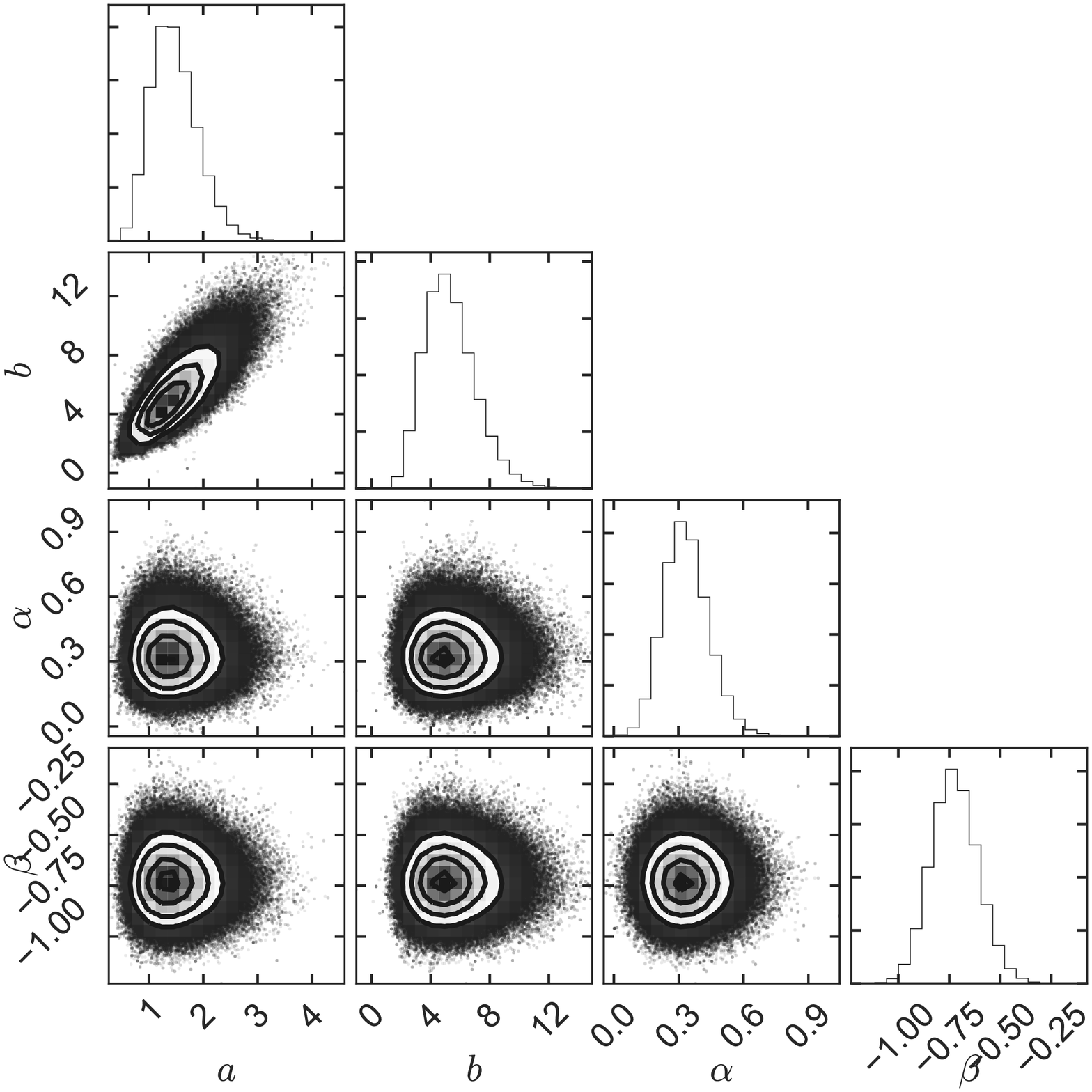}
\caption{Posterior distributions on the eight model parameters describing the 98 single systems and 17 multi-stellar systems as a two distinct populations of planets. The set of plots on the left correspond to the distribution of single systems and the set on the right corresponds to multi-stellar systems. The histograms represent the one-dimensional marginalized posterior probability distribution for each parameter.
\label{fig:posterior_2pop}}
\end{figure*}

Using the Markov Chain Monte Carlo implemented by \verb+emcee+ python package's affine-invariant sampler~\citep{ForemanMackey2013}, we compute the posterior probability distributions for the model parameters and determine the maximum likelihoods of each model, $\hat{\mathcal{L}_1}$ and $\hat{\mathcal{L}_2}$. Figure~\ref{fig:posterior_1pop} and~\ref{fig:posterior_2pop} show our calculated posteriors on each model parameter for each model. We determine whether a two-population model is justified by comparing the Bayesian Information Criteria (BIC) for these two models and by computing the Bayesian odds ratio. First, we compute the BIC as $\mathrm{BIC} = \ln(N)k - 2\hat{\mathcal{L}}$, where $N$ is the number of planets in the model fit and $k$ is the total number of parameters (i.e. four in the one-population model and eight in the two-population model). We find that the difference between the two-population model BIC and the one-population model BIC to be 17, indicating that the two-population model is very strongly disfavored~\citep{Kass1995}. Second, we compute the Bayesian odds ratio as the probability of a one-population model divided by the probability of a two-population model, assuming both models have equal prior likelihoods and uniform priors on all model parameters, i.e.
\begin{equation}
\frac{\mathrm{prob}(1\mathrm{pop})}{\mathrm{prob}(2\mathrm{pop})} = \frac{\exp{\hat{\mathcal{L}}_1}}{\exp{\hat{\mathcal{L}}_2}} \  \frac{\prod_x 2\pi\delta\theta_x / \Delta \theta_x}{\prod_{y} 2\pi\delta\theta_y / \Delta \theta_y}.
\end{equation}
This is equal to the ratio of the evidence for each model in the case where the posteriors are n-dimensional normal distribution and when the priors are uniform. The first term on the right is the Bayes factor, and the second term is the Ockham factor to account for the model parameters. The product sum over $x$ covers the four parameters of the one-population model and the product sum over $y$ covers the eight parameters of the two-model population. The $\Delta\theta$ term corresponds to range in allowable parameter values from our uniform prior and $\delta\theta$ is the region in which the parameter yields a good fit. For this calculation, we calculate $\delta\theta$ as the $\theta$ interval, centered on the median value for $\theta$, that encompasses 68\% of the posterior probability. Table~\ref{tab:bayes} shows our assumed priors and fit errors used to calculate the odds ratio. We compute the Bayes factor to be 0.64 and the Ockham factor to be 135, yielding an overall odds ratio that favors the one-population model over the two-population model 87 to 1. 

We also consider whether some of these companion stars could be more influential than others by repeating the above calculation with only the top 9 (i.e. the top half) systems in Table~\ref{tab:RV_pl}. In this case, the BIC comparison still favors the one-population model with a $\Delta$ BIC of 13. However, due to the smaller number of multi-stellar systems, the larger uncertainties on the model parameters for the multiple star component of the two-population model reduces the Ockham factor and yields an odds ratio of 2.6 to 1, indicating no strong preference for either model. This also shows that our angular separation cutoff choice does not affect our results. Finally, we also repeat the above analysis including all planets in each system instead of only the innermost planet and find no difference in our results.

Based on these calculations we conclude that there is no evidence for a difference in the eccentricity, mass and orbital period distributions of the inner giant planet between single and multi-stellar systems within 6\arcsec. Notably, in \citet{Bryan2016}, we searched for outer planetary-mass companions in these systems using long-term RV monitoring and found that the presence of such companions correlated with increased eccentricities for the inner planets in these systems, a difference significant at the $3\sigma$ level.

\subsection{Constraining additional sub-stellar companions}
For the remaining single star systems, we provide deep K-band contrast curves from our imaging campaign. The average 5-$\sigma$ \Ksfilt\ contrast at separations of 0\arcsec.25, 0\arcsec.50, and 1\arcsec.0 are 4.96, 6.64 and 8.22, respectively. These contrasts corresponds to companions with masses of 0.16, 0.09 and 0.08 solar masses, respectively, around a sun-like primary star. These upper limits on another object in a narrow field of view around these stars provide upper limits on mass and semi-major axis of any potential additional sub-stellar companions which might be detected by other techniques, such as RV~\cite[e.g.][]{Bryan2016} or astrometry. There is a strong theoretical motivation for continued RV monitoring of these systems, as \citet{Petrovich2016} presents a scenario for warm and hot Jupiter formation via planet-planet interactions and predicts additional giant planets at much wider separations that may be found via a long term RV survey. A recent study~\citep{Hamers2017b} suggest that a companion star could interact with an additional planetary mass companion to induce the hot Jupiter to migrate via planetary Kozai-Lidov oscillations. Long term RV surveys~\citep[e.g.][]{Fischer2014a,Knutson2014,Montet2014,Bryan2016} currently have RV baselines up to 25 years. \citet{Bryan2016} report 50\% completeness in their surveys for 1 Jupiter mass planets at 20\unit{au} and for 10 Jupiter mass planets at 70\unit{au}. \citet{Hamers2017b} predict Jupiter-sized planets at 40\unit{au} could cause migration. In the coming decades, RV surveys can find or rule out objects as small as a few Jupiter masses at separations up to 40\unit{au}. These published contrast curves will help constrain the masses and semimajor axes of these potential future discoveries. In the future, Gaia astrometry will reach accuracies as low as $10\mu$as (microarcseconds) for stars with $V$ magnitudes 7--12 and $25\mu$as for stars with $V=15$~\citep{Perryman2014}. This would be accurate enough to determine the mutual inclination between widely separated giant planetary companions found in transiting and RV surveys~\citep[e.g.][]{Buhler2016} and allow for constraints on planet-planet Kozai-Lidov migration.

\subsection{Astrometry of triple systems}
\label{sec:ofti}
We study six hierarchical triple systems to determine the stellar orbital architectures. In these systems, the secondary and tertiary stars (the ``inner binary'' orbit) are close enough that we can detect orbit motion over our survey's baseline, unlike the orbits of our widely separated binary systems. Because of the hierarchical architecture, when we consider the ``outer binary'' orbit, the secondary and tertiary stars behave like a single body. We fit for all the orbital parameters of both the ``inner'' (B and C components) and ``outer'' orbits (A and BC components) using the Orbits For The Impatient method (OFTI), a Bayesian rejection sampling method described in \citet{Blunt2017}. As demonstrated in \citet{deRosa2015}, \citet{Rameau2016}, \citet{Bryan2016} and \citet{Blunt2017}, OFTI calculates posterior distributions of orbital parameters that are identical to those produced by MCMC, but operates significantly faster when the input astrometry covers a short fraction of the total orbit ($<10\%$). 

OFTI generates an initial orbit with a semimajor axis $a$ of 1\unit{au}, a position angle of nodes $\Omega$ of $0^{\circ}$, and other orbital parameters drawn from appropriate priors: uniform in eccentricity $e$, argument of periastron $\omega$, epoch of periastron passage $T_0$, and uniform in $\cos(i)$ (inclination angle). System mass and distance values are drawn from Gaussian distributions with medians and standard deviations equal to the measured values and observational uncertainties, and period $P$ is calculated from Kepler's third law. OFTI then scales $a$ and rotates $\Omega$ to match a single observational epoch, with observational errors included by adding random values drawn from Gaussian distributions with FWHM equal to the observed uncertainties in projected separation and position angle. Finally, the orbit's probability is computed from $p=e^{-\chi^2 / 2}$. This value is compared with a uniform random number in (0,1). If the chi-square probability is greater than this random number, the orbit is accepted. After many iterations of this process, probability distributions are calculated by computing histograms of the accepted sets of orbital parameters.

\begin{figure*}
\epsscale{1.0}
\plottwo{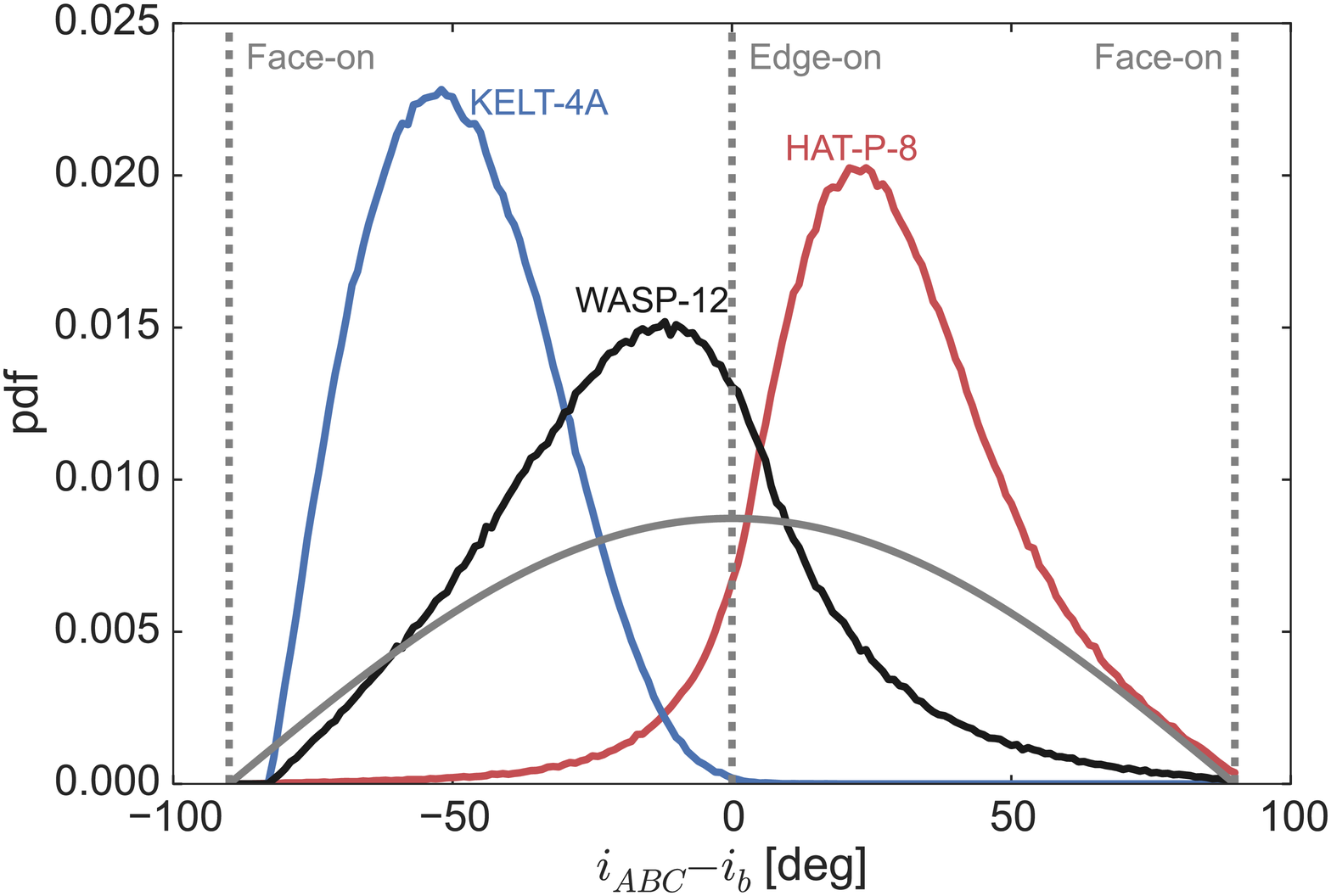}{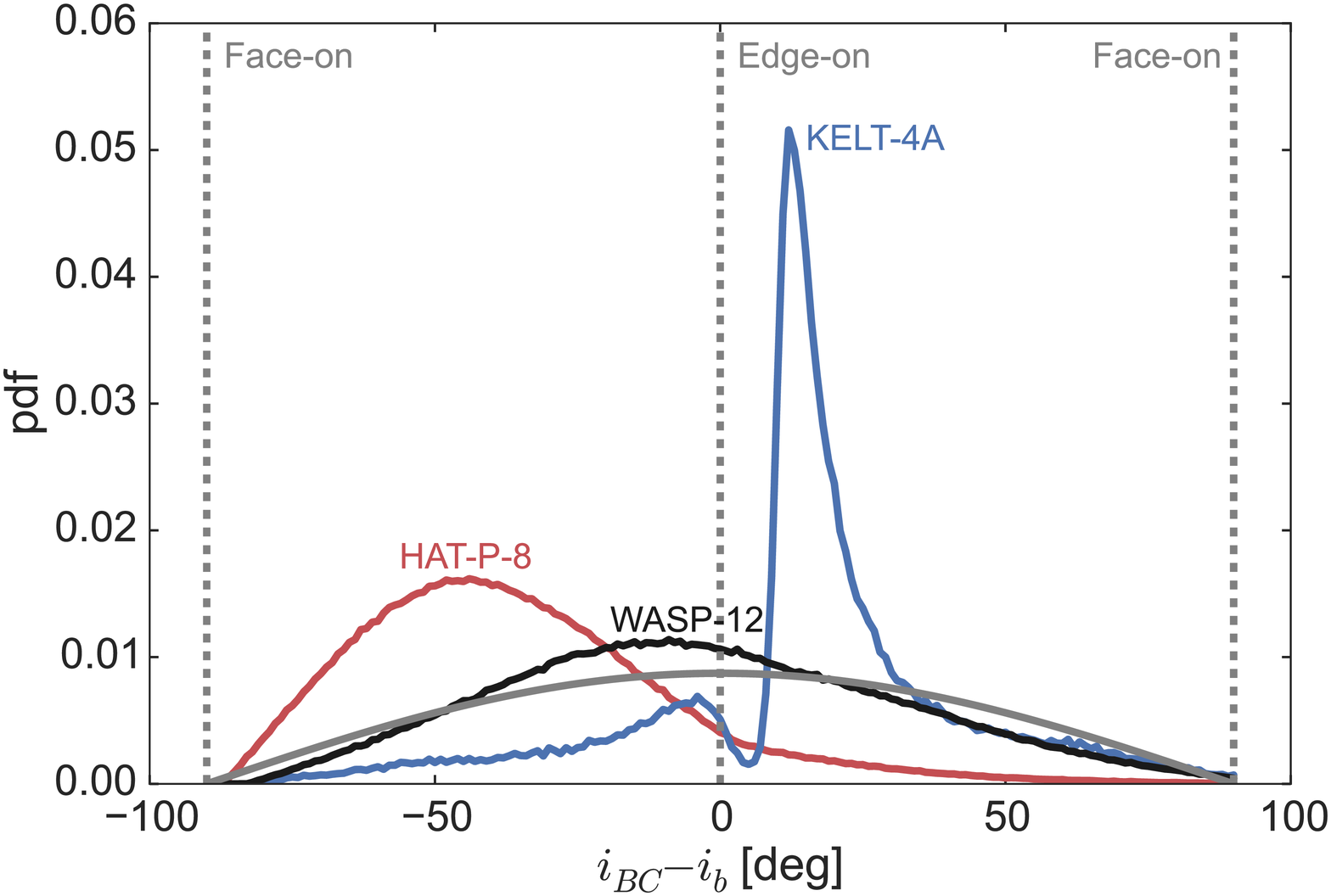}
\caption{Left: Posterior distributions on the difference between the outer binary's inclination and the transiting planet's inclination. Posteriors on the outer binary's inclination are computed from OFTI while the planet's inclination come from \citet{Latham2009}, \citet{Eastman2016} and \citet{Collins2017} for HAT-P-8b, KELT-4Ab, and WASP-12b, respectively. The solid grey pdf represents the prior on $\Delta i$ for a edge-on planet. Right: The same for the difference between the inner binary's inclination and the transiting planet's inclination.
\label{fig:delta_inc}}
\end{figure*}

Table~\ref{tab:triple_astrom} describes the posterior distributions on the orbital inclinations of the inner (BC) and outer (ABC) orbits. One system, HD 142245, has secondary and tertiary stars with a mutual orbital plane that is misaligned with the plane of their orbit around the primary star. This misalignment is significant at the 95\% confidence level. Two systems, HD 43691 and WASP-12, have well-aligned orbital planes. Out of the transiting systems where the planet's inclination is measured, KELT-4A is the only system where companion stars are misaligned with the planet at the 95\% confidence level. Note that RV fits do not provide the position angle of the nodes, so we are not sensitive to any misalignment perpendicular to our line of sight. Therefore, any offset in inclination angles in RV systems represent a minimum misalignment. Figure~\ref{fig:delta_inc} shows the posterior distribution on the difference between the binary inclination and the planetary inclination for the three triple star systems with transiting planets. We find that both the inner and outer binaries of these systems are no more or less likely to be aligned or misaligned with the transiting planet. The transiting planets have inclinations close to edge-on, so this implies that the outer and inner orbits favor neither an edge-on nor a face-on orbit. Although there are only three systems in our sample, it would be interesting to investigate the general distribution of $i_{ABC}-i_b$ and $i_{BC}-i_b$ with more transiting systems. Although we only discuss the inclination probability distribution here, we plot the probability contours for all seven orbital parameters for one sample system in Figure~\ref{fig:ofti_triangle}, we summarize the posteriors on all orbital elements for all triple systems in Tables~\ref{tab:ofti_hat8} to \ref{tab:ofti_wasp12}, and we provide posterior samples of all parameters for all six triple systems online.

\subsection{Summary of current observational constraints on the effects of companions on giant planet formation and migration}
There is a considerable body of literature focused on possible formation and migration mechanisms for hot and warm Jupiters. In this section, we review this literature and the current work to constrain these theories via surveys for planetary and stellar companions to hot and warm Jupiters. We also put the results of this work in context with these other surveys. Although there exist variations on the strict definitions of a ``hot'' vs a ``warm'' Jupiter, for brevity in this discussion, we refer to giant planets with masses greater than 0.1\Mjup\ and semimajor axes less than 0.1\unit{au} as ``hot'' Jupiters and planets with semimajor axes between 0.1-1.0\unit{au} as ``warm'' Jupiters.

Possible migration mechanisms include both interactions with the protoplanetary gas disk and with other planetary or stellar companions in the system. Formation followed by gas disk migration must occur quickly, as the gas disk only survives for 1-10 million years~\citep{Pollack1996,Haisch2001,Hernandez2009}. This formation channel is  expected to create hot Jupiters on low eccentricity orbits~\citep[e.g.][]{Goldreich1980,Lin1986,Tanaka2002}. On the other hand, interactions such as gravitational scattering with other planets~\citep[e.g.][]{Chatterjee2008,Wu2011,Beauge2012,Lithwick2014,Petrovich2015b} or stars via stellar Kozai-Lidov oscillations~\citep[e.g][]{Wu2003,Fabrycky2007,Naoz2012,Naoz2013,Storch2014} in the system could create hot Jupiters on more eccentric orbits. 

Recent studies have used characteristics of the existing giant exoplanet population to attempt to distinguish between migration mechanisms. \citet{Dawson2015} points out a lack of high-eccentricity warm Jupiters in the {\it Kepler} sample, suggesting that multi-body processes are unlikely to form hot Jupiters. However, \citet{Petrovich2014a} showed that planet-planet scattering at separations within 0.2\unit{au} would not excite high eccentricities. In this scenario, the giant planets could scatter off each other without creating high eccentricity warm Jupiters. Stellar Kozai-Lidov migration is expected to create misaligned hot Jupiters, but our recent stellar companion surveys~\citep{Ngo2015,Piskorz2015} find no correlation between the incidence of hot Jupiter misalignment and stellar multiplicity. In addition, we place an upper limit of 20\% for systems experiencing Kozai-Lidov migration~\citep{Ngo2016}. Furthermore, \citet{Schlaufman2016} argue that a planet-planet scattering scenario for hot Jupiter migration would predict that hot Jupiters would have fewer giant planet companions interior to the water-ice line as compared to warm Jupiters. They examined RV-detected hot and warm Jupiter systems and found that hot Jupiters are just as likely to host exterior giant planet companions as warm Jupiters. In addition, short-period giant planets found around young T Tauri stars, such as CI Tau b~\citep{JohnsKrull2016}, have lifetimes too short for migration via multi-body interactions. These results disfavor high-eccentricity hot Jupiter migration and would instead suggest that disk migration or {\it in situ} formation scenarios are more likely for short period giant planets. 

RV monitoring surveys have found that long period giant planet companions to transiting hot Jupiters~\citep{Knutson2014} and RV-detected giant planets~\citep{Bryan2016} are common. In \citet{Bryan2016}, we find that $52\% \pm 5\%$ of the RV giant planet systems host additional long-period planetary mass companions (5-20 AU, 1-20 Jupiter masses). In addition, the gas giant planets beyond 0.1\unit{au} have, on average, higher orbital eccentricities when they have an outer companion. This finding is consistent with work by \citet{Petrovich2016} showing that secular planet-planet interactions can account for most of the observed hot Jupiter population; however, these interactions fail to reproduce the known warm Jupiter planets. These types of interactions can also excite large mutual inclinations, resulting in misaligned planetary systems~\citep[e.g. see][]{Johansen2012,Morton2014b,Ballard2016,Becker2016,Spalding2016}. Finally, these additional planets can also interact with the inner giant planets through planet-planet Kozai-Lidov effects~\citep{Dawson2014c}.

The presence of massive planetary and/or stellar companions in these systems can also have important implications for {\it in situ} formation models. Some {\it in situ} models~\citep[e.g.][]{Boley2016} invoke a globally enhanced disk mass or a local concentration of solids in the region of interest, both of which would affect the locations and masses of other gas giant planets formed in the same disk. Alternatively, other {\it in situ} models~\citep[e.g.][]{Batygin2016} form hot Jupiters from rapid gas accretion onto super-Earth planets, which are already commonly found at short periods~\citep{Fressin2013}. \citet{Batygin2016} also predict that hot Jupiters that formed {\it in situ} should also have additional low-mass planets with orbital periods less than 100 days. RV surveys of known planetary systems find preliminary evidence that that hot Jupiters are more likely to host an additional companion than warm Jupiters~\citep{Bryan2016}. Many theoretical studies of planet formation in binary star systems predict that the presence of a second star would be detrimental for planet formation by exciting or removing planetesimals in the protoplanetary disk~\citep[e.g.][]{Mayer2005,Pichardo2005}. Stellar companions could also eject planets after formation~\citep[e.g.][]{Kaib2013,Zuckerman2014}. For close (less than 50\unit{au} separation) binaries, the current observational evidence appears to support this view. \citet{Kraus2012} found that 2/3 of young stars with stellar companions within 40 AU lose their protoplanetary disks within 1 million years, while systems with more distant companions have disk lifetimes that are comparable to single-star systems. In a followup study, \citet{Kraus2016} surveyed 386 {\it Kepler} planet host stars and showed that these stars are three times less likely to have a stellar companion within 50 AU than non-planet hosting field stars. \citet{Wang2014} also came to a similar conclusion in their survey of 56 {\it Kepler} planet host stars. 

Although current studies indicate that planet formation is suppressed in close stellar binaries, there are many examples of known planet-hosting stars in relatively wide (greater than 50\unit{au}) binaries. The two most recent directly imaged giant planet systems, 51 Eri b~\citep{Macintosh2015,Montet2015c} and HD 131399 Ab~\citep{Wagner2016}, are both part of hierarchical triple systems. \citet{Ngo2016} surveyed a sample of 77 transiting hot Jupiter host stars and found that $47\%\pm7\%$ of these systems have a directly imaged stellar companion. Other near-infrared diffraction-limited direct imaging surveys for stellar companions to transiting close-in giant planet systems have found companion fractions consistent with our result~\citep{Adams2013,Woellert2015a,Woellert2015b,Wang2015b,Evans2016}. In \citet{Ngo2016}, we found that hot Jupiter host stars have fewer close-in stellar companions (projected separations less than 50\unit{au}) than field stars; however, they are three times more likely to have a wide companion star (projected separations greater than 50\unit{au}) than field stars. These companions may play some role in enhancing planet formation.

In this work, we considered the effects of stellar companions on gas giant planets at intermediate ($0.1-5$\unit{au}) separations. We conducted a large survey for stellar companions to RV-detected warm and cool ($a<5\unit{au}$) Jupiters. We show that there is currently no evidence for a correlation between the incidence of a stellar companion and the gas giant planet's mass, orbital eccentricity or orbital period. This suggests that the presence or absence of a stellar companion do not significantly alter the formation or orbital evolution of gas giant planets at intermediate separations. Given the mass ratios and projected separations of the stellar companions in our sample, it seems unlikely that these companions could have induced Kozai-Lidov oscillations in most of the systems observed. This result is consistent with the absence of increased planet eccentricities in multi-stellar systems, and lends more weight to {\it in situ} or planet-planet scattering theories for the formation of warm Jupiters. Our results also increase the number of known RV-planet systems with companion stars; these systems can serve as case studies for models of planet formation and migration in multiple star systems.

\section{Summary}
\label{sec:summary}
We carry out an AO imaging search for stellar companions around 144 stars with RV-detected giant planets. The sample is the largest survey for stellar companions around RV planet hosts to date and includes 123 stars from our previous long-term RV monitoring study~\citep{Bryan2016}. We detect 11 comoving multi-stellar systems, corresponding to a raw companion fraction of $7.6\%\pm2.3\%$. This value is consistent with other surveys for stellar companions around RV planet systems, but is much lower than the stellar companion fraction for transiting gas giant planets because of strong biases against multi-stellar systems in sample selection for RV surveys. 

Three of the multi-stellar systems are presented for the first time in this work (HD 30856, HD 86081 and HD 207832). We confirm common proper motion for another three systems (HD 43691, HD 116029 and HD 164509). These six new confirmed multi-stellar RV systems increase the total number systems with known companions closer than 6\arcsec\ to 22. We compare the mass, orbital eccentricity and semimajor axis distribution of the innermost planet in the multi-stellar systems with those of the innermost planet in the single star systems. Our analysis indicates that these distributions are the same for both single and multi-stellar systems. This suggests the observed stellar companions do not significantly alter the properties of the giant planet in these systems. Even when limiting our comparison to the most dynamically influential (i.e., the most massive and closest in) stellar companions, we find no evidence for any difference in the distribution of planet orbital properties. These results appear to disfavor Kozai-type migration processes, and are consistent with both {\it in situ} formation and planet-planet scattering.

We also compute contrast curves for all 144 surveyed targets. These provide upper limits on remaining undetected stellar and substellar companions, and can be used to constrain the masses of any additional companions found in long term RV monitoring surveys. We note that there is great value in continued RV monitoring of these systems, as the presence or absence of more distant ($> 5-10$ au) planetary mass companions would provide invaluable insights into the likely formation and migration histories of these systems. Another potentially valuable study would be to obtain AO imaging data for a control sample of stars from the CPS survey which are not currently known to host planets.  As in \citet{Eggenberger2007}, this sample would allow us to empirically measure the selection biases against multi-stellar systems in our current planet-hosting star sample and calculate a stellar multiplicity rate for that sample that can be directly compared to that of field stars.

Finally, in our survey's hierarchical triple systems (HD 43691, HD 142245, and HD 207832), the secondary and tertiary stellar components are on very tight orbits (less than 10\unit{au}), so it is possible to measure orbital motion over the several year baseline of our survey. We fit orbital parameters for all three stars in these three triple systems as well as three additional triple systems from transiting planet surveys. We show that these orbital fits allow us to constrain the geometry of the triple system (e.g. edge-on or face-on), which has implications for the dynamical evolution of the planet orbits in these systems.



\acknowledgments

We would like to thank John A. Johnson, Christopher Spalding and Benjamin Montet for helpful discussions. We also appreciate the useful suggestions from the anonymous referee and the statistics editor. This work was supported by NASA grant NNX14AD24G. HN is grateful for funding support from the Natural Sciences and Engineering Research Council of Canada and the NASA Earth and Space Science Fellowship Program grant NNX15AR12H. HAK acknowledges support from the Sloan Foundation. ELN and SCB are supported by NASA grant NNX14AJ80G.

This work was based on observations at the W. M. Keck Observatory granted by the California Institute of Technology. We thank the observers who contributed to the measurements reported here and acknowledge the efforts of the Keck Observatory staff. We extend special thanks to those of Hawaiian ancestry on whose sacred mountain of Mauna Kea we are privileged to be guests.

Facilities: \facility{Keck:II}

Software: Astropy~\citep{AstroPyPaper}, OFTI~\citep{Blunt2017}, emcee~\citep{ForemanMackey2013}, corner.py~\citep{ForemanMackey2016}

\bibliographystyle{apj}

\clearpage
\LongTables
\begin{deluxetable*}{lclccccc}
\tablecolumns{8}
\tabletypesize{\scriptsize}
\tablecaption{Summary of NIRC2 AO Observations  \label{tab:obs}}
\tablewidth{0pt}
\tablehead{
\colhead{Target} & \colhead{$N_{cc}$} & \colhead{UT Obs. Date} & \colhead{Filter} & \colhead{Array} & \colhead{$T_{\mathrm{int}}$ (s)} & \colhead{$N_{\mathrm{fit}}$} & \colhead{$N_{\mathrm{stack}}$}
}
\startdata
\sidehead{\bf RV planet host stars}
GJ 317 & 0 & 2014 Nov 07 & \Kcfilt & 1024 & 12.0 & \nodata & 12 \\
GJ 433 & 0 & 2015 Jan 09 & \Kcfilt & 1024 & 13.6 & \nodata & 12 \\
GJ 667C & 0 & 2014 Jul 12 & \Kcfilt & 1024 & 13.6 & \nodata & 12 \\
GJ 876 & 0 & 2013 Aug 19 & \Kcfilt & 512 & 10.6 & \nodata & 12 \\
 &  & 2013 Oct 17 & \Jcfilt & 512 & 10.6 & \nodata & \nodata \\
 &  & 2013 Oct 17 & \Kcfilt & 512 & 10.6 & \nodata & 12 \\
HD 1461 & 0 & 2014 Jul 12 & \Kcfilt & 512 & 15.9 & \nodata & 12 \\
HD 1502 & 0 & 2013 Aug 19 & \Kcfilt & 1024 & 9.0 & \nodata & 12 \\
HD 3651 & 0 & 2013 Aug 19 & \Kcfilt & 256 & 9.0 & \nodata & 12 \\
HD 4203 & 0 & 2013 Aug 19 & \Kcfilt & 1024 & 9.0 & \nodata & 12 \\
HD 4208 & 0 & 2013 Nov 17 & \Kcfilt & 1024 & 10.0 & \nodata & 15 \\
HD 4313 & 0 & 2013 Oct 17 & \Kcfilt & 512 & 10.6 & \nodata & 12 \\
HD 5319 & 0 & 2014 Jul 12 & \Kcfilt & 1024 & 13.6 & \nodata & 12 \\
HD 5891 & 0 & 2013 Oct 17 & \Jcfilt & 512 & 10.6 & \nodata & \nodata \\
 &  & 2013 Oct 17 & \Kcfilt & 512 & 10.6 & \nodata & 12 \\
 &  & 2015 Dec 19 & \Ksfilt & 1024 & 15.0 & \nodata & 12 \\
HD 7924 & 0 & 2014 Oct 03 & \Kcfilt & 1024 & 13.6 & \nodata & 12 \\
HD 8574 & 0 & 2015 Jan 09 & \Kcfilt & 1024 & 13.6 & \nodata & 12 \\
HD 10697 & 0 & 2015 Dec 19 & \Kcfilt & 1024 & 15.0 & \nodata & \nodata \\
HD 11506 & 0 & 2013 Nov 17 & \Kcfilt & 1024 & 10.0 & \nodata & 15 \\
HD 11964A & 0 & 2013 Nov 17 & \Kcfilt & 512 & 10.6 & \nodata & 12 \\
HD 12661 & 0 & 2013 Aug 19 & \Kcfilt & 1024 & 9.0 & \nodata & 12 \\
HD 13931 & 0 & 2015 Jan 09 & \Kcfilt & 1024 & 13.6 & \nodata & 12 \\
HIP 14810 & 0 & 2013 Oct 17 & \Kcfilt & 1024 & 9.0 & \nodata & 10 \\
HD 16141 & 0 & 2013 Aug 19 & \Kcfilt & 512 & 10.6 & \nodata & 12 \\
HD 17156 & 0 & 2014 Oct 03 & \Kcfilt & 1024 & 15.0 & \nodata & 12 \\
HD 22049 & 0 & 2016 Sep 13 & \Kcfilt & 512 & 15.9 & \nodata & 12 \\ 
HIP 22627 & 0 & 2015 Jan 10 & \Kcfilt & 1024 & 15.0 & \nodata & 12 \\
HD 24040 & 0 & 2015 Jan 10 & \Kcfilt & 1024 & 13.6 & \nodata & 12 \\
HD 28678 & 0 & 2014 Oct 04 & \Kcfilt & 1024 & 13.6 & \nodata & 12 \\
HD 30856 & 1 & 2014 Oct 04 & \Jcfilt & 1024 & 13.6 & 12 & \nodata \\
 &  & 2014 Oct 04 & \Kcfilt & 1024 & 13.6 & 12 & 12 \\
 &  & 2014 Dec 07 & \Jcfilt & 1024 & 15.0 & 11 & \nodata \\ 
 &  & 2014 Dec 07 & \Kcfilt & 1024 & 13.6 & 12 & 12 \\
 &  & 2015 Oct 26 & \Kcfilt & 1024 & 15.0 & 12 & 12 \\
HD 32963 & 0 & 2014 Oct 04 & \Kcfilt & 1024 & 13.6 & \nodata & 12 \\
HD 33142 & 0 & 2014 Oct 04 & \Kcfilt & 1024 & 13.6 & \nodata & 12 \\
HD 33283 & 0 & 2014 Oct 04 & \Kcfilt & 1024 & 12.5 & \nodata & 12 \\
HD 34445 & 0 & 2014 Nov 07 & \Kcfilt & 1024 & 13.6 & \nodata & 12 \\
HD 37124 & 0 & 2014 Jan 12 & \Kpfilt & 1024 & 10.0 & \nodata & 10 \\
 &  & 2014 Dec 07 & \Kcfilt & 1024 & 15.0 & \nodata & 12 \\
HD 37605 & 0 & 2014 Jan 12 & \Kpfilt & 1024 & 9.0 & \nodata & 12 \\
 &  & 2014 Dec 07 & \Kcfilt & 1024 & 12.5 & \nodata & 12 \\
HD 38529 & 0 & 2015 Jan 10 & \Kcfilt & 512 & 13.2 & \nodata & 12 \\
HD 38801 & 0 & 2014 Jan 12 & \Kpfilt & 1024 & 9.0 & \nodata & 9 \\
 &  & 2014 Jan 12 & \Kcfilt & 1024 & 10.0 & \nodata & 11 \\
 &  & 2014 Dec 07 & \Kcfilt & 1024 & 12.5 & \nodata & 12 \\
HD 40979 & 0 & 2014 Nov 10 & \Kcfilt & 512 & 12.0 & \nodata & 12 \\
HD 43691 & 2 & 2013 Dec 18 & \Kcfilt & 1024 & 10.0 & 4 & 12 \\
 &  & 2014 Dec 04 & \Kcfilt & 1024 & 10.0 & 10 & 12 \\
 &  & 2014 Dec 04 & \Jcfilt & 1024 & 10.0 & 5 & \nodata \\
 &  & 2015 Oct 26 & \Kcfilt & 1024 & 25.0 & 12 & 12 \\
 &  & 2016 Sep 13 & \Kcfilt & 1024 & 45.0 & 12 & 12 \\
HD 45350 & 0 & 2013 Oct 17 & \Kcfilt & 1024 & 9.0 & \nodata & 12 \\
HD 46375 & 0 & 2013 Oct 17 & \Kcfilt & 1024 & 9.0 & \nodata & 12 \\
HD 49674 & 0 & 2013 Oct 17 & \Kcfilt & 1024 & 9.0 & \nodata & 12 \\
HD 50499 & 0 & 2014 Nov 07 & \Kcfilt & 1024 & 13.6 & \nodata & 12 \\
HD 50554 & 0 & 2013 Dec 18 & \Kcfilt & 1024 & 10.0 & \nodata & 12 \\
HD 52265 & 0 & 2013 Oct 17 & \Kcfilt & 512 & 10.6 & \nodata & 12 \\
HIP 57050 & 0 & 2014 May 21 & \Kcfilt & 1024 & 12.5 & \nodata & 12 \\
HIP 57274 & 0 & 2014 May 21 & \Kcfilt & 1024 & 12.5 & \nodata & 12 \\
 &  & 2015 Dec 20 & \Kcfilt & 1024 & 15.0 & \nodata & 12 \\
HD 66428 & 0 & 2013 Dec 18 & \Kcfilt & 1024 & 10.0 & \nodata & 12 \\
HD 68988 & 0 & 2013 Dec 18 & \Kcfilt & 1024 & 10.0 & \nodata & 12 \\
HD 69830 & 0 & 2014 Jan 12 & \Kcfilt & 1024 & 9.0 & \nodata & 8 \\
 &  & 2014 Dec 05 & \Kcfilt & 512 & 13.2 & \nodata & 12 \\
HD 72659 & 0 & 2014 Jan 12 & \Kcfilt & 1024 & 9.0 & \nodata & 15 \\
 &  & 2014 Nov 10 & \Kcfilt & 1024 & 13.6 & \nodata & 12 \\
HD 73256 & 0 & 2014 Nov 10 & \Kcfilt & 1024 & 12.5 & \nodata & 12 \\
HD 73534 & 0 & 2014 Jan 12 & \Kcfilt & 1024 & 9.0 & \nodata & 15 \\
 &  & 2014 Dec 05 & \Kcfilt & 1024 & 13.6 & \nodata & 12 \\
HD 74156 & 0 & 2014 Nov 07 & \Kcfilt & 1024 & 13.6 & \nodata & 12 \\
HIP 74995 & 0 & 2013 Jul 04 & \Kcfilt & 1024 & 9.0 & \nodata & 12 \\
HD 75732 & 0 & 2014 May 21 & \Kcfilt & 256 & 10.0 & \nodata & 12 \\
HD 75898 & 0 & 2014 May 21 & \Jcfilt & 1024 & 12.5 & \nodata & 12 \\
 &  & 2014 May 21 & \Kcfilt & 1024 & 12.5 & \nodata & 12 \\
 &  & 2015 Dec 20 & \Kcfilt & 1024 & 15.0 & \nodata & 12 \\
HIP 79431 & 0 & 2013 Jul 04 & \Kcfilt & 1024 & 9.0 & \nodata & 12 \\
HD 80606 & 0 & 2014 Jan 12 & \Kcfilt & 1024 & 10.0 & \nodata & 12 \\
 &  & 2014 Dec 07 & \Kcfilt & 1024 & 15.0 & \nodata & 12 \\
HD 82886 & 0 & 2014 Jan 12 & \Kcfilt & 1024 & 9.0 & \nodata & 15 \\
 &  & 2014 May 21 & \Kcfilt & 512 & 13.2 & \nodata & 12 \\
HD 82943 & 0 & 2015 Jan 10 & \Kcfilt & 1024 & 15.0 & \nodata & 12 \\
 &  & 2016 Jan 25 & \Kcfilt & 1024 & 15.0 & \nodata & 12 \\
HIP 83043 & 0 & 2013 Jul 04 & \Kcfilt & 1024 & 9.0 & \nodata & 12 \\
HD 83443 & 0 & 2015 Jan 09 & \Kcfilt & 1024 & 12.5 & \nodata & 12 \\
HD 86081 & 1 & 2013 Dec 18 & \Kcfilt & 1024 & 10.0 & 9 & 12 \\ 
 &  & 2014 Dec 05 & \Kcfilt & 1024 & 12.0 & 10 & 12 \\ 
HD 87883 & 0 & 2015 Jan 10 & \Kcfilt & 1024 & 13.6 & \nodata & 12 \\
HD 88133 & 0 & 2013 Dec 18 & \Kcfilt & 1024 & 10.0 & \nodata & 12 \\
HD 90043 & 0 & 2014 Jan 12 & \Kcfilt & 512 & 10.6 & \nodata & 12 \\
 &  & 2014 May 21 & \Kcfilt & 512 & 13.2 & \nodata & 12 \\
 &  & 2014 May 21 & \Jcfilt & 256 & 13.5 & \nodata & 6 \\
HD 92788 & 0 & 2014 Dec 05 & \Kcfilt & 1024 & 13.6 & \nodata & 12 \\
HD 95089 & 0 & 2014 Dec 05 & \Kcfilt & 1024 & 13.6 & \nodata & 12 \\
HD 95128 & 0 & 2014 May 21 & \Kcfilt & 256 & 15.0 & \nodata & 12 \\
HD 96063 & 0 & 2014 May 21 & \Kcfilt & 1024 & 12.5 & \nodata & 12 \\
HD 96167 & 1 & 2014 Jan 12 & \Kcfilt & 1024 & 10.0 & 4 & 12 \\
 &  & 2014 Dec 07 & \Jcfilt & 1024 & 15.0 & 12 & \nodata \\
 &  & 2014 Dec 07 & \Kcfilt & 1024 & 15.0 & 12 & 12 \\
 &  & 2015 Jan 09 & \Jcfilt & 1024 & 13.6 & 12 & \nodata \\
 &  & 2015 Jan 09 & \Kcfilt & 1024 & 15.0 & 12 & 12 \\
HD 97658 & 0 & 2013 Dec 18 & \Kcfilt & 1024 & 9.0 & \nodata & 12 \\
HD 99109 & 0 & 2014 May 21 & \Ksfilt & 1024 & 18.1 & \nodata & 12 \\
 &  & 2016 Jan 25 & \Kcfilt & 1024 & 15.0 & \nodata & 12 \\
HD 99492 & 0 & 2014 May 21 & \Kcfilt & 512 & 13.2 & \nodata & 12 \\
HD 99706 & 0 & 2014 May 21 & \Kcfilt & 512 & 13.2 & \nodata & 12 \\
HD 102195 & 0 & 2014 May 21 & \Kcfilt & 1024 & 12.5 & \nodata & 12 \\
HD 102329 & 0 & 2014 May 21 & \Kcfilt & 1024 & 13.6 & \nodata & 12 \\
HD 102956 & 0 & 2014 May 21 & \Kcfilt & 1024 & 12.5 & \nodata & 12 \\
BD-103166 & 0 & 2014 Jan 12 & \Kcfilt & 1024 & 20.0 & \nodata & 12 \\
 &  & 2015 Jan 09 & \Kcfilt & 1024 & 15.0 & \nodata & 12 \\
HD 104067 & 0 & 2014 May 21 & \Kcfilt & 1024 & 12.0 & \nodata & 12 \\
HD 106270 & 0 & 2014 May 21 & \Kcfilt & 1024 & 12.5 & \nodata & 12 \\
HD 107148 & 0 & 2014 May 21 & \Kcfilt & 1024 & 12.5 & \nodata & 12 \\
HD 108863 & 0 & 2014 Jun 09 & \Kcfilt & 1024 & 13.6 & \nodata & 12 \\
HD 108874 & 0 & 2015 Jan 09 & \Kcfilt & 1024 & 12.0 & \nodata & 12 \\
 &  & 2015 Dec 20 & \Kcfilt & 1024 & 15.0 & \nodata & 12 \\
HIP 109388 & 0 & 2014 Nov 07 & \Kcfilt & 1024 & 13.6 & \nodata & 12 \\
HD 109749 & 0 & 2014 Jun 09 & \Kcfilt & 1024 & 12.5 & \nodata & 12 \\
HD 114729 & 0 & 2015 Jan 09 & \Kcfilt & 512 & 12.0 & \nodata & 12 \\
HD 114783 & 0 & 2014 Jun 09 & \Kcfilt & 1024 & 13.6 & \nodata & 12 \\
HD 115617 & 0 & 2014 Jun 09 & \Kcfilt & 256 & 13.5 & \nodata & 13 \\
HD 116029 & 1 & 2013 Jul 04 & \Kcfilt & 512 & 10.0 & 11 & 12 \\ 
 &  & 2014 Jun 09 & \Kcfilt & 1024 & 13.6 & 12 & 12 \\
 &  & 2014 Jun 09 & \Jcfilt & 512 & 13.2 & 11 & \nodata \\
 &  & 2015 Jan 09 & \Kcfilt & 512 & 12.0 & 12 & 12 \\
 &  & 2015 Jan 09 & \Jcfilt & 512 & 12.7 & 12 & \nodata \\
HD 117176 & 0 & 2013 Jul 04 & \Kcfilt & 256 & 18.0 & \nodata & 12 \\
HD 117207 & 0 & 2014 Jun 09 & \Kcfilt & 1024 & 12.5 & \nodata & 14 \\
HD 125612 & 0 & 2014 Jun 09 & \Kcfilt & 1024 & 15.0 & \nodata & 12 \\
 &  & 2015 Jun 24 & \Kcfilt & 1024 & 15.0 & \nodata & 12 \\
HD 126614 & 1 & 2014 Jun 09 & \Ksfilt & 1024 & 13.6 & 12 & 12 \\
 &  & 2015 Jan 09 & \Kcfilt & 1024 & 12.5 & 12 & 12 \\
 &  & 2015 Jun 24 & \Kcfilt & 1024 & 12.5 & 12 & 12 \\
 &  & 2015 Jun 24 & \Jcfilt & 1024 & 12.5 & 12 & \nodata \\
HD 128311 & 0 & 2014 Jul 07 & \Kcfilt & 512 & 10.6 & \nodata & 12 \\
HD 130322 & 0 & 2014 Jul 12 & \Kcfilt & 1024 & 13.6 & \nodata & 12 \\
HD 131496 & 0 & 2013 Jul 04 & \Kcfilt & 512 & 10.0 & \nodata & 12 \\
HD 134987 & 0 & 2016 Jun 09 & \Kcfilt & 1024 & 15.0 & \nodata & 12 \\
 &  & 2016 Jun 09 & \Jcfilt & 1024 & 15.0 & \nodata & \nodata \\
HD 141399 & 0 & 2014 May 21 & \Kcfilt & 1024 & 13.6 & \nodata & 12 \\
HD 141937 & 0 & 2013 Jul 04 & \Kcfilt & 1024 & 9.0 & \nodata & 12 \\
HD 142245 & 2 & 2013 Jul 04 & \Kcfilt & 512 & 10.6 & 15 & 15 \\
 &  & 2014 Jun 09 & \Kcfilt & 512 & 13.2 & 11 & 12 \\ 
 &  & 2014 Jun 09 & \Jcfilt & 512 & 13.2 & 10 & \nodata \\ 
 &  & 2015 Jun 24 & \Kcfilt & 768 & 12.7 & 12 & 12 \\
 &  & 2016 Jun 09 & \Kcfilt & 512 & 15.9 & 11 & 12 \\ 
HD 143761 & 0 & 2014 Jun 09 & \Kcfilt & 256 & 13.5 & \nodata & 12 \\
HD 145675 & 0 & 2014 Jul 12 & \Kcfilt & 512 & 15.9 & \nodata & 12 \\
HD 145934 & 0 & 2014 May 21 & \Kcfilt & 1024 & 12.5 & \nodata & 12 \\
HD 149143 & 0 & 2013 Jul 04 & \Kcfilt & 1024 & 9.0 & \nodata & 12 \\
HD 152581 & 0 & 2013 Jul 04 & \Kcfilt & 1024 & 9.0 & \nodata & 12 \\
HD 154345 & 0 & 2014 Jul 12 & \Kcfilt & 512 & 15.9 & \nodata & 12 \\
HD 156279 & 0 & 2013 Aug 19 & \Kcfilt & 1024 & 9.0 & \nodata & 12 \\
HD 156668 & 0 & 2013 Jul 04 & \Kcfilt & 1024 & 9.0 & \nodata & 12 \\
HD 158038 & 0 & 2015 Jun 24 & \Kcfilt & 768 & 12.7 & \nodata & 12 \\
HD 163607 & 0 & 2013 Aug 19 & \Kcfilt & 1024 & 9.0 & \nodata & 12 \\
HD 164509 & 1 & 2014 Jun 09 & \Jcfilt & 1024 & 12.5 & 11 & \nodata \\ 
 &  & 2014 Jun 09 & \Kcfilt & 1024 & 12.5 & 12 & 12 \\
 &  & 2015 Jun 24 & \Kcfilt & 1024 & 12.5 & 12 & 12 \\
 &  & 2016 Jun 09 & \Jcfilt & 512 & 15.9 & 12 & \nodata \\
 &  & 2016 Jun 09 & \Kcfilt & 1024 & 15.0 & 12 & 12 \\
HD 164922 & 0 & 2015 Jun 24 & \Kcfilt & 768 & 12.7 & \nodata & 12 \\
HD 168443 & 0 & 2013 Aug 19 & \Jcfilt & 512 & 10.6 & \nodata & \nodata \\
 &  & 2013 Aug 19 & \Kcfilt & 512 & 10.0 & \nodata & 12 \\
HD 168746 & 0 & 2013 Aug 19 & \Kcfilt & 1024 & 9.0 & \nodata & 12 \\
HD 169830 & 0 & 2013 Aug 19 & \Kcfilt & 512 & 10.6 & \nodata & 12 \\
HD 170469 & 0 & 2013 Jul 04 & \Kcfilt & 1024 & 9.0 & \nodata & 12 \\
HD 175541 & 0 & 2013 Jul 04 & \Kcfilt & 1024 & 9.0 & \nodata & 12 \\
HD 177830 & 1 & 2014 Jun 09 & \Jcfilt & 512 & 13.2 & 9 & \nodata \\ 
 &  & 2014 Jun 09 & \Kcfilt & 512 & 13.2 & 12 & 12 \\
 &  & 2015 Jul 05 & \Kcfilt & 512 & 13.2 & 12 & 12 \\
 &  & 2016 Jun 09 & \Jcfilt & 512 & 15.9 & 12 & \nodata \\
 &  & 2016 Jun 09 & \Kcfilt & 512 & 15.9 & 12 & 12 \\
HD 178911B & 0 & 2013 Aug 19 & \Kcfilt & 1024 & 9.0 & \nodata & 12 \\
HD 179079 & 0 & 2013 Aug 19 & \Kcfilt & 1024 & 9.0 & \nodata & 12 \\
HD 179949 & 0 & 2013 Aug 19 & \Kcfilt & 512 & 10.6 & \nodata & 12 \\
HD 180902 & 0 & 2014 Jul 12 & \Kcfilt & 1024 & 13.6 & \nodata & 12 \\
HD 181342 & 0 & 2013 Aug 19 & \Kcfilt & 512 & 10.6 & \nodata & 12 \\
HD 183263 & 0 & 2015 Jul 05 & \Kcfilt & 1024 & 15.0 & \nodata & \nodata \\
HD 186427B & 0 & 2013 Aug 19 & \Kcfilt & 512 & 10.6 & \nodata & 12 \\
HD 187123 & 1 & 2013 Aug 19 & \Kcfilt & 1024 & 9.0 & 9 & 12 \\ 
 &  & 2015 Jun 24 & \Kcfilt & 1024 & 15.0 & 12 & 12 \\
HD 188015 & 1 & 2013 Aug 19 & \Jcfilt & 1024 & 9.0 & 6 & \nodata \\
 &  & 2013 Aug 19 & \Kcfilt & 1024 & 9.0 & 8 & 12 \\
 &  & 2014 Jul 12 & \Jcfilt & 1024 & 15.0 & 12 & \nodata \\
 &  & 2014 Jul 12 & \Kcfilt & 1024 & 15.0 & 12 & 12 \\
HD 189733 & 0 & 2013 Jun 22 & \Ksfilt & 128 & 0.4 & \nodata & 12 \\
 &  & 2013 Aug 19 & \Kcfilt & 1024 & 9.0 & \nodata & 12 \\
HD 190360 & 0 & 2014 Jul 12 & \Kcfilt & 512 & 15.9 & \nodata & 12 \\
HD 192263 & 0 & 2013 Aug 19 & \Kcfilt & 512 & 10.6 & \nodata & 12 \\
HD 192310 & 0 & 2013 Aug 19 & \Kcfilt & 256 & 9.0 & \nodata & 6 \\
HD 195019 & 1 & 2013 Aug 19 & \Kcfilt & 704 & 12.4 & 12 & 12 \\
 &  & 2013 Aug 19 & \Jcfilt & 512 & 10.6 & 6 & \nodata \\
 &  & 2014 Jun 09 & \Kcfilt & 1024 & 13.6 & 12 & 12 \\
 &  & 2014 Jul 12 & \Kcfilt & 1024 & 13.6 & 12 & 12 \\
 &  & 2014 Jul 12 & \Jcfilt & 512 & 15.9 & 12 & \nodata \\
 &  & 2015 Jun 03 & \Kcfilt & 768 & 12.7 & 12 & 12 \\
HD 200964 & 0 & 2013 Aug 19 & \Kcfilt & 512 & 10.6 & \nodata & 12 \\
HD 206610 & 0 & 2013 Aug 19 & \Kcfilt & 1024 & 9.0 & \nodata & 12 \\
HD 207832 & 2 & 2013 Aug 19 & \Kcfilt & 1024 & 12.5 & 12 & 12 \\
 &  & 2013 Aug 19 & \Ksfilt & 512 & 10.6 & 9 & 9 \\
 &  & 2013 Aug 19 & \Jcfilt & 1024 & 9.0 & 5 & \nodata \\
 &  & 2014 Jul 12 & \Kcfilt & 1024 & 15.0 & 12 & 12 \\
 &  & 2014 Jul 12 & \Jcfilt & 1024 & 15.0 & 9 & \nodata \\
 &  & 2015 Jul 05 & \Kcfilt & 1024 & 20.0 & 12 & 12 \\
 &  & 2015 Jul 05 & \Jcfilt & 1024 & 20.0 & 12 & \nodata \\
 &  & 2015 Oct 26 & \Kcfilt & 1024 & 15.0 & 12 & 12 \\ 
 &  & 2016 Sep 13 & \Kcfilt & 1024 & 12.5 & 3 & 3 \\
HD 209458 & 0 & 2013 Jun 22 & \Ksfilt & 256 & 9.0 & \nodata & 12 \\
HD 210277 & 0 & 2014 Nov 07 & \Kcfilt & 1024 & 13.6 & \nodata & 15 \\
HD 212771 & 0 & 2013 Aug 19 & \Kcfilt & 512 & 12.5 & \nodata & 12 \\
HD 217014 & 0 & 2013 Aug 19 & \Kcfilt & 256 & 9.0 & \nodata & 12 \\
HD 217107 & 0 & 2013 Nov 17 & \Kcfilt & 512 & 10.6 & \nodata & 15 \\
HD 222582 & 0 & 2013 Dec 18 & \Kcfilt & 1024 & 9.0 & \nodata & 12 \\
HD 224693 & 0 & 2013 Aug 19 & \Kcfilt & 1024 & 3.6 & \nodata & 12 \\
HD 231701 & 0 & 2013 Aug 19 & \Kcfilt & 1024 & 12.5 & \nodata & 11 \\
 &  & 2014 Jun 09 & \Kcfilt & 1024 & 12.0 & \nodata & 12 \\
 &  & 2015 Jul 05 & \Kcfilt & 1024 & 15.0 & \nodata & 15 \\
HD 285968 & 0 & 2013 Oct 17 & \Kcfilt & 512 & 10.6 & \nodata & 12\\
\sidehead{\bf Transiting planet host stars in triple systems}
HAT-P-8 & 2 & 2014 Oct 03 & \Ksfilt & 1024 & 25.0 & 12 & 11\\ 
 &  & 2015 Jul 07 & \Ksfilt & 1024 & 13.6 & 12 & 12\\ 
 &  & 2016 Sep 12 & \Ksfilt & 1024 & 12.0 & 12 & 12\\ 
KELT-4A & 2 & 2015 Dec 20 & \Ksfilt & 1024 & 15.0 & 12 & 12\\
WASP-12 & 2 & 2014 Dec 04 & \Ksfilt & 1024 & 12.0 & 13 & 13\\
 &  & 2015 Dec 26 & \Ksfilt & 1024 & 15.0 & 12 & 12\\ 
 &  & 2016 Sep 13 & \Ksfilt & 1024 & 30.0 & 4 & 4
\enddata
\tablecomments{Column $N_{cc}$ is the number of candidate companions detected. Column ``Array'' is the horizontal size, in pixel, of the NIRC2 array readout region and corresponds to subarray sizes of 1024x1024 (the full NIRC2 array), 768x760, 512x512, or 256x264. Column $T_{\mathrm{int}}$ is the total integration time, in seconds, of a single frame. Column $N_{\mathrm{fit}}$ is the number of frames used in our photometric and/or astrometric analysis, and is only given when companions are present. Column $N_{\mathrm{stack}}$ is the number of frames combined to make the contrast curve measurements. We only compute contrast curves in the \Ksfilt and \Kcfilt bandpasses so this column is not applicable for other bandpasses. In some cases, $N_{\mathrm{fit}}$ and $N_{\mathrm{stack}}$ are not equal because the companion may not be present in all frames due to the dither pattern and/or observing conditions.}
\end{deluxetable*}

\clearpage
\begin{deluxetable}{lcccccccc}
\tablecolumns{9}
\tabletypesize{\scriptsize}
\tablecaption{Stellar parameters for candidate multi-stellar systems \label{tab:system_params}}
\tablehead{
& \vspace{-0.1cm} & \colhead{$T_{\mathrm{eff}}$}  & \colhead{$M$} & \colhead{$\log g$} & \colhead{$D$} & \multicolumn{3}{c}{References for...} \\
\colhead{Target} & \colhead{Type} & \vspace{-0.2cm} & & & & & \\
& & \colhead{(K)} & \colhead{(\Msun)} & \colhead{(cgs)} & \colhead{(pc)} & \colhead{$T$} & \colhead{$M,\log g$} & \colhead{$D$}
}
\startdata
\sidehead{\bf RV planet host stars}
HD 30856  & B & $4982 \pm 44$ & $1.350 \pm 0.094$ & $3.40 \pm 0.06$ & $118^{+11}_{-9}$  & 1 & 1 & 2 \\
HD 43691  & T & $6200 \pm 40$ & $1.38 \pm 0.05$     & $4.28 \pm 0.13$ & $81^{+6}_{-5}$     & 3 & 3 & 2 \\
HD 86081  & B & $6036 \pm 23$ & $1.23 \pm 0.08$     & $4.21 \pm 0.04$ & $95^{+10}_{-8}$   & 4 & 4 & 2 \\
HD 96167  & B & $5749 \pm 25$ & $1.16 \pm 0.05$     & $4.15 \pm 0.06$ & $87^{+7}_{-6}$     & 5 & 5 & 2 \\
HD 116029 & B & $4951 \pm 44$ & $1.58 \pm 0.11$     & $3.40 \pm 0.06$ & $123^{+11}_{-9}$ & 1 & 1 & 2 \\
HD 126614 & B & $5585 \pm 44$ & $1.145 \pm 0.030$ & $4.39 \pm 0.08$ & $73 \pm 5$          & 6 & 6 & 2 \\
HD 142245 & T & $4878 \pm 44$ & $1.69 \pm 0.12$    & $3.30 \pm 0.06$  & $110^{+8}_{-7}$  & 1 & 1 & 2 \\
HD 164509 & B & $5922 \pm 44$ & $1.13 \pm 0.02$    & $4.44 \pm 0.06$  & $52 \pm 3$         & 7 & 7 & 2 \\
HD 177830 & B & $5058 \pm 35$ & $1.37 \pm 0.04$    & $3.66 \pm 0.06$  & $59 \pm 2$          & 8 & 8 & 2 \\
HD 187123 & bg & $5845 \pm 22$ & $1.037 \pm 0.025$ & $4.32 \pm 0.04$ & $48 \pm 1$        & 9 & 10 & 2 \\
HD 188015 & bg & $5746$\tablenotemark{a} & $1.056 \pm 0.09$  & $4.41^{+0.05}_{-0.04}$ & $57 \pm 3$          & 11 & 7 & 2 \\
HD 195019 & B & $5741 \pm 20$ & $1.05 \pm 0.10$    & $4.06 \pm 0.04$  & $39^{+2}_{-1}$   & 12 & 12 & 2 \\
HD 207832\tablenotemark{b} & T & $5736 \pm 27$ & $0.980 \pm 0.070$ & $4.51 \pm 0.07$ & $54 \pm 3$        & 4 & 4 & 2 \\
\sidehead{\bf Transiting planet host stars in triple systems}
HAT-P-8     & T & $6223 \pm 67$ & $1.192 \pm 0.075$ & $4.177 \pm 0.022$ & $230 \pm 15$ & 13 & 14 & 15 \\
KELT-4A    & T & $6207 \pm 75$ & $1.204 \pm 0.070$ & $4.105^{+0.029}_{-0.032}$ & $211^{+13}_{-12}$ & 16 & 16 & 16 \\
WASP-12   & T & $6118 \pm 64$ & $1.38 \pm 0.19$     & $4.159 \pm 0.024$ & $427 \pm 90$ & 13 & 17 & 15
\enddata
\tablerefs{(1) \citet{Johnson2011b}; (2) \citet{vanLeeuwen2007}; (3) \citet{daSilva2007}; (4) \citet{Santos2013}; (5) \citet{Jofre2015}; (6) \citet{Howard2010}; (7) \citet{Giguere2012}; (8) \citet{Johnson2006}; (9) \citet{Santos2004}; (10) \citet{Takeda2007}; (11) \citet{Butler2006}; (12) \citet{Ghezzi2010}; (13) \citet{Torres2012}; (14) \citet{Mancini2013}; (15) \citet{Triaud2014}; (16) \citet{Eastman2016}; (17) \citet{Southworth2012c}}
\tablecomments{The ``Type'' column indicates whether the candidate multi-stellar system is a bound binary (B), bound triple (T) or a background object (bg). The planetary parameters are listed only for the innermost planet in each system in our RV planet host star survey. All planetary parameters are listed as they appear in the cited reference.}
\tablenotetext{a}{No uncertainty available for this temperature estimate. Since the candidate companion is a background object, the temperature was not used in any further calculation.}
\end{deluxetable}

\clearpage
\begin{deluxetable*}{llcccccc}
\tablecolumns{8}
\tablecaption{Photometry of confirmed stellar companions \label{tab:comp_phot}}
\tablehead{
\colhead{Companion} & \colhead{UT Obs. Date}  & \colhead{$m_{Jc}$} & \colhead{$m_{Kc}$} & \colhead{$\Delta \Jcfilt$} & \colhead{$\Delta \Kcfilt$} & \colhead{$\Jcfilt-\Kcfilt$}
}
\startdata
\sidehead{\bf RV planet host stars}
HD 30856B & 2014 Oct 04 & $10.963 \pm 0.020$ & $10.904 \pm 0.021$ & $4.708 \pm 0.020$ & $5.247 \pm 0.021$ & $0.059 \pm 0.029$  \\ 
 & 2014 Dec 07 & $11.160 \pm 0.088$ & $10.473 \pm 0.042$ & $4.905 \pm 0.088$ & $4.816 \pm 0.042$ & $0.687 \pm 0.097$ \\
 & 2015 Oct 26 & \nodata & $10.786 \pm 0.046$ & \nodata & $5.129 \pm 0.046$ & \nodata  \\ 
HD 43691B & 2013 Dec 18 & \nodata & $13.698 \pm 0.084$ & \nodata & $6.998 \pm 0.084$ & \nodata  \\ 
 & 2014 Dec 04 & $14.38 \pm 0.38$ & $13.253 \pm 0.051$ & $7.40 \pm 0.38$ & $6.553 \pm 0.051$ & $1.12 \pm 0.38$ \\ 
 & 2015 Oct 26 & \nodata & $13.203 \pm 0.093$ & \nodata & $6.503 \pm 0.093$ & \nodata \\ 
 & 2016 Sep 13 & \nodata & $12.94 \pm 0.11$ & \nodata & $6.24 \pm 0.11$ & \nodata \\ 
HD 43691C & 2013 Dec 18 & \nodata & $12.895 \pm 0.081$ & \nodata & $6.195 \pm 0.081$ & \nodata \\ 
 & 2014 Dec 04 & $23.0 \pm 3.0$ & $12.534 \pm 0.050$ & $16.0 \pm 3.0$ & $5.834 \pm 0.050$ & $10.5 \pm 3.0$ \\ 
 & 2015 Oct 26 & \nodata & $12.483 \pm 0.044$ & \nodata & $5.783 \pm 0.044$ & \nodata \\
 & 2016 Sep 13 & \nodata & $11.847 \pm 0.084$ & \nodata & $5.147 \pm 0.084$ & \nodata \\
HD 86081B & 2013 Dec 18 & \nodata & $15.25 \pm 0.35$ & \nodata & $7.95 \pm 0.35$ & \nodata  \\
 & 2014 Dec 05 & \nodata & $14.766 \pm 0.086$ & \nodata & $7.467 \pm 0.086$ & \nodata \\ 
HD 96167B & 2014 Jan 12 & \nodata & $12.97 \pm 0.40$ & \nodata & $6.41 \pm 0.40$ & \nodata  \\ 
 & 2015 Jan 09 & $15.10 \pm 0.17$ & $12.591 \pm 0.039$ & $8.18 \pm 0.17$ & $6.037 \pm 0.039$ & $2.51 \pm 0.17$ \\ 
HD 116029B & 2013 Jul 04 & \nodata & $12.513 \pm 0.091$ & \nodata & $6.979 \pm 0.091$ & \nodata \\
 & 2014 Jun 09 & $13.63 \pm 0.16$ & $12.414 \pm 0.066$ & $7.49 \pm 0.16$ & $6.880 \pm 0.066$ & $1.22 \pm 0.17$ \\ 
 & 2015 Jan 09 & $13.85 \pm 0.11$ & $12.55 \pm 0.10$ & $7.71 \pm 0.11$ & $7.01 \pm 0.10$ & $1.30 \pm 0.15$ \\ 
HD 126614B & 2015 Jan 09 & \nodata & $11.132 \pm 0.018$ & \nodata & $4.072 \pm 0.018$ & \nodata \\
 & 2015 Jun 24 & $11.65 \pm 0.12$ & $10.9740 \pm 0.0100$ & $4.18 \pm 0.12$ & $3.9140 \pm 0.0100$ & $0.68 \pm 0.12$ \\ 
HD 142245B & 2013 Jul 04 & \nodata & $11.159 \pm 0.034$ & \nodata & $6.049 \pm 0.034$ & \nodata \\ 
 & 2014 Jun 09 & $11.241 \pm 0.080$ & $10.632 \pm 0.085$ & $5.552 \pm 0.080$ & $5.522 \pm 0.085$ & $0.61 \pm 0.12$ \\ 
 & 2015 Jun 24 & \nodata & $10.896 \pm 0.037$ & \nodata & $5.786 \pm 0.037$ & \nodata \\ 
 & 2016 Jun 09 & \nodata & $10.988 \pm 0.031$ & \nodata & $5.878 \pm 0.031$ & \nodata \\ 
HD 142245C & 2013 Jul 04 & \nodata & $10.794 \pm 0.022$ & \nodata & $5.684 \pm 0.022$ & \nodata \\ 
 & 2014 Jun 09 & $11.164 \pm 0.060$ & $10.635 \pm 0.024$ & $5.475 \pm 0.060$ & $5.525 \pm 0.024$ & $0.529 \pm 0.064$ \\ 
 & 2015 Jun 24 & \nodata & $10.690 \pm 0.013$ & \nodata & $5.580 \pm 0.013$ & \nodata \\ 
 & 2016 Jun 09 & \nodata & $10.645 \pm 0.025$ & \nodata & $5.535 \pm 0.025$ & \nodata \\ 
HD 164509B & 2014 Jun 09 & $11.66 \pm 0.27$ & $10.172 \pm 0.024$ & $4.72 \pm 0.27$ & $3.586 \pm 0.024$ & $1.49 \pm 0.27$ \\
 & 2015 Jun 24 & \nodata & $10.022 \pm 0.053$ & \nodata & $3.436 \pm 0.053$ & \nodata \\ 
 & 2016 Jun 09 & $10.654 \pm 0.025$ & $10.0151 \pm 0.0056$ & $3.716 \pm 0.025$ & $3.4291 \pm 0.0056$ & $0.639 \pm 0.026$ \\ 
HD 177830B & 2014 Jun 09 & $12.81 \pm 0.19$ & $11.861 \pm 0.081$ & $7.45 \pm 0.19$ & $7.052 \pm 0.081$ & $0.95 \pm 0.21$ \\ 
 & 2015 Jul 05 & \nodata & $11.897 \pm 0.089$ & \nodata & $7.088 \pm 0.089$ & \nodata  \\
 & 2016 Jun 09 & $12.339 \pm 0.045$ & $11.539 \pm 0.061$ & $6.972 \pm 0.045$ & $6.730 \pm 0.061$ & $0.800 \pm 0.076$ \\ 
HD 195019B & 2013 Aug 19 & $8.511 \pm 0.018$ & $7.9276 \pm 0.0084$ & $2.911 \pm 0.018$ & $2.6676 \pm 0.0084$ & $0.583 \pm 0.020$ \\ 
 & 2014 Jun 09 & \nodata & $7.970 \pm 0.059$ & \nodata & $2.710 \pm 0.059$ & \nodata \\ 
 & 2014 Jul 12 & $8.492 \pm 0.014$ & $7.9395 \pm 0.0069$ & $2.892 \pm 0.014$ & $2.6795 \pm 0.0069$ & $0.553 \pm 0.016$ \\ 
 & 2015 Jun 03 & \nodata & $7.878 \pm 0.040$ & \nodata & $2.618 \pm 0.040$ & \nodata \\ 
HD 207832B & 2013 Aug 19 & $14.58 \pm 0.18$ & $13.148 \pm 0.055$ & $7.00 \pm 0.18$ & $5.941 \pm 0.055$ & $1.44 \pm 0.19$ \\ 
 & 2014 Jul 12 & $14.336 \pm 0.094$ & $13.143 \pm 0.042$ & $6.749 \pm 0.094$ & $5.936 \pm 0.042$ & $1.19 \pm 0.10$ \\  
 & 2015 Jul 05 & \nodata & $12.870 \pm 0.086$ & \nodata & $5.663 \pm 0.086$ & \nodata \\ 
 & 2015 Oct 26 & \nodata & $13.005 \pm 0.077$ & \nodata & $5.798 \pm 0.077$ & \nodata \\
 & 2016 Sep 13 & \nodata & $13.47 \pm 0.12$ & \nodata & $6.26 \pm 0.12$ & \nodata \\ 
 & 2013 Aug 19 & $14.58 \pm 0.18$ & $13.166 \pm 0.033$ & $7.00 \pm 0.18$ & $5.959 \pm 0.033$ & $1.42 \pm 0.19$ \\
HD 207832C & 2013 Aug 19 & $14.077 \pm 0.096$ & $12.813 \pm 0.048$ & $6.490 \pm 0.096$ & $5.606 \pm 0.048$ & $1.26 \pm 0.11$ \\ 
 & 2014 Jul 12 & $14.06 \pm 0.14$ & $12.826 \pm 0.070$ & $6.47 \pm 0.14$ & $5.619 \pm 0.070$ & $1.23 \pm 0.15$ \\ 
 & 2015 Jul 05 & \nodata & $12.909 \pm 0.056$ & \nodata & $5.702 \pm 0.056$ & \nodata \\ 
 & 2015 Oct 26 & \nodata & $12.921 \pm 0.044$ & \nodata & $5.714 \pm 0.044$ & \nodata \\
 & 2016 Sep 13 & \nodata & $12.62 \pm 0.13$ & \nodata & $5.41 \pm 0.13$ & \nodata \\ 
 & 2013 Aug 19 & $14.077 \pm 0.096$ & $12.770 \pm 0.054$ & $6.490 \pm 0.096$ & $5.563 \pm 0.054$ & $1.31 \pm 0.11$ \\ 
\sidehead{\bf Transiting planet host stars in triple systems}
HAT-P-8B & 2014 Oct 03 & \nodata & $14.883 \pm 0.082$ & \nodata & $5.930 \pm 0.082$ & \nodata \\ 
 & 2015 Jul 07 & \nodata & $14.811 \pm 0.065$ & \nodata & $5.858 \pm 0.065$ & \nodata \\ 
 & 2016 Sep 12 & \nodata & $14.803 \pm 0.058$ & \nodata & $5.850 \pm 0.058$ & \nodata  \\ 
HAT-P-8C & 2014 Oct 03 & \nodata & $15.204 \pm 0.089$ & \nodata & $6.251 \pm 0.089$ & \nodata \\
& 2015 Jul 07 & \nodata & $15.469 \pm 0.025$ & \nodata & $6.516 \pm 0.025$ & \nodata \\ 
& 2016 Sep 12 & \nodata & $15.462 \pm 0.021$ & \nodata & $6.509 \pm 0.021$ & \nodata \\
KELT-4A B & 2015 Dec 20 & \nodata & $10.70 \pm 0.36$ & \nodata & $2.02 \pm 0.36$ & \nodata \\
KELT-4A C & 2015 Dec 20 & \nodata & $11.79 \pm 0.15$ & \nodata & $3.10 \pm 0.15$ & \nodata \\
WASP-12B & 2014 Dec 04 & \nodata & $13.251 \pm 0.077$ & \nodata & $3.063 \pm 0.077$ & \nodata \\ 
 & 2015 Dec 26 & \nodata & $13.342 \pm 0.026$ & \nodata & $3.154 \pm 0.026$ & \nodata \\
 & 2016 Sep 13 & \nodata & $13.294 \pm 0.035$ & \nodata & $3.106 \pm 0.035$ & \nodata \\
WASP-12C & 2014 Dec 04 & \nodata & $13.70 \pm 0.10$ & \nodata & $3.51 \pm 0.10$ & \nodata \\
 & 2015 Dec 26 & \nodata & $13.411 \pm 0.023$ & \nodata & $3.223 \pm 0.023$ & \nodata \\
 & 2016 Sep 13 & \nodata & $13.300 \pm 0.041$ & \nodata & $3.112 \pm 0.041$ & \nodata
\enddata
\end{deluxetable*}

\clearpage
\begin{deluxetable}{llcccc}
\tablecolumns{5}
\tablecaption{Astrometric measurements of all candidate companions\label{tab:comp_astr}}
\tablehead{
& & & \colhead{$\rho$} & \colhead{PA} \vspace{-0.1cm}\\
\colhead{Candidate\tablenotemark{a}} & \colhead{UT Obs. Date} & \colhead{Band} &\vspace{-0.2cm}  & & \colhead{Reference}\\
& & & \colhead{(mas)} & \colhead{(\deg)}
}
\startdata
\sidehead{\bf RV planet host stars}
HD 30856 B & 2014 Oct 04 & \Kcfilt & $ 789.4 \pm    1.5$ & $108.6 \pm   0.1$ & this work\\
 & 2014 Dec 07 & \Kcfilt & $ 788.7 \pm    1.5$ & $108.7 \pm   0.1$ & this work\\
 & 2015 Oct 26 & \Kcfilt & $ 786.3 \pm    1.5$ & $108.8 \pm   0.1$ & this work\\
HD 43691 B & 2013 Dec 18 & \Kcfilt & $4550.3 \pm    1.8$ & $40.50 \pm  0.02$ & this work\\
 & 2014 Dec 04 & \Kcfilt & $4546.6 \pm    2.2$ & $40.40 \pm  0.02$ & this work\\
 & 2015 Oct 26 & \Kcfilt & $4540.4 \pm    2.5$ & $40.27 \pm  0.03$ & this work\\
 & 2016 Sep 13 & \Kcfilt & $4536.5 \pm    2.9$ & $39.91 \pm  0.04$ & this work\\
HD 43691 C & 2013 Dec 18 & \Kcfilt & $4452.6 \pm    1.8$ & $41.08 \pm  0.02$ & this work\\
 & 2014 Dec 04 & \Kcfilt & $4456.8 \pm    1.9$ & $41.08 \pm  0.02$ & this work\\
 & 2015 Oct 26 & \Kcfilt  & $4462.2 \pm    2.5$ & $41.08 \pm  0.03$ & this work\\
 & 2016 Sep 13 & \Kcfilt & $4463.5 \pm    2.5$ & $40.90 \pm  0.03$ & this work\\
HD-43691 BC & 2015 Mar 10 & \ipfilt & $4435 \pm   16$\tablenotemark{b} & $ 40.8 \pm   0.2$\tablenotemark{b} & \citet{Ginski2016}\\
 & 2013 Dec 18 & \Kcfilt & $4513.5 \pm    1.9$ & $40.72 \pm  0.02$ & this work\\
 & 2014 Dec 04 & \Kcfilt & $4512.7 \pm    2.3$ & $40.65 \pm  0.02$ & this work\\
 & 2015 Oct 26 & \Kcfilt  & $4510.9 \pm    2.7$ & $40.57 \pm  0.03$ & this work\\
 & 2016 Sep 13 & \Kcfilt & $4508.9 \pm    3.0$ & $40.28 \pm  0.05$ & this work\\
HD 86081 B & 2013 Dec 18 & \Kcfilt & $2904.8 \pm    2.3$ & $89.29 \pm  0.06$ & this work\\
& 2014 Dec 05 & \Kcfilt & $2901.3 \pm    2.9$ & $89.35 \pm  0.06$ & this work\\
HD 96167 B & 2013 Jan 24 & \Ksfilt & $5873.0 \pm    1.8$ & $297.1 \pm   0.1$ & \citet{Mugrauer2015}\\
 & 2014 Jan 12 & \Kcfilt & $5889.7 \pm    3.4$ & $297.18 \pm  0.03$ & this work\\
 & 2015 Jan 09 & \Kcfilt & $5884.1 \pm    2.1$ & $297.11 \pm  0.02$ & this work\\
 HD 116029 B & 2013 Jun 30 & \ipfilt & $1387.1 \pm    5.8$ & $209.1 \pm   0.3$ & \citet{Ginski2016}\\
 & 2013 Jul 04 & \Kcfilt & $1391.6 \pm    1.7$ & $209.32 \pm  0.07$ & this work\\
 & 2014 Jun 09 & \Kcfilt & $1393.0 \pm    1.6$ & $209.37 \pm  0.06$ & this work\\
 & 2015 Jan 09 & \Kcfilt & $1391.4 \pm    1.6$ & $209.37 \pm  0.07$ & this work\\
HD 126614 B & 2009 Apr 13 & \Ksfilt & $ 489.0 \pm    1.9$ & $ 56.1 \pm   0.3$ & \citet{Howard2010}\\
 & 2011 Jan 14 &\ipfilt & $ 499 \pm   67$ & $ 60.7 \pm   5.6$ & \citet{Ginski2012}\\
 & 2015 Jan 09 & \Kcfilt & $ 486.1 \pm    1.5$ & $ 69.1 \pm   0.2$ & this work\\
 & 2015 Jun 24 & \Kcfilt & $ 485.3 \pm    1.5$ & $ 70.4 \pm   0.2$ & this work\\
HD 142245 B & 2013 Jul 04 & \Kcfilt & $2484.7 \pm    1.6$ & $168.79 \pm  0.04$ & this work\\
 & 2014 Jun 09 & \Kcfilt & $2499.5 \pm    1.9$ & $168.21 \pm  0.04$ & this work\\
 & 2015 Jun 24 & \Kcfilt & $2501.3 \pm    1.8$ & $168.23 \pm  0.04$ & this work\\
 & 2016 Jun 09 & \Kcfilt & $2507.5 \pm    1.8$ & $168.07 \pm  0.04$ & this work\\
HD 142245 C & 2013 Jul 04 & \Kcfilt & $2524.7 \pm    1.6$ & $169.54 \pm  0.03$ & this work\\
 & 2014 Jun 09 & \Kcfilt & $2516.7 \pm    1.7$ & $169.66 \pm  0.04$ & this work\\
 & 2015 Jun 24 & \Kcfilt & $2513.2 \pm    1.8$ & $169.71 \pm  0.04$ & this work\\
 & 2016 Jun 09 & \Kcfilt & $2505.9 \pm    1.9$ & $169.74 \pm  0.04$ & this work\\
HD-142245 BC & 2012 Aug 31 & \Ksfilt & $2498 \pm    6$\tablenotemark{b} & $169.2 \pm   0.2$\tablenotemark{b} & \citet{Mugrauer2015}\\
 & 2013 Jul 24 & \Ksfilt & $2494 \pm    6$\tablenotemark{b} & $169.1 \pm   0.1$\tablenotemark{b} & \citet{Mugrauer2015}\\
 & 2013 Jul 04 & \Kcfilt & $2503.3 \pm    1.6$ & $169.14 \pm  0.04$ & this work\\
 & 2014 Jun 09 & \Kcfilt & $2507.4 \pm    2.0$ & $168.89 \pm  0.04$ & this work\\
 & 2015 Jun 24 & \Kcfilt & $2506.7 \pm    1.8$ & $168.92 \pm  0.04$ & this work\\
 & 2016 Jun 09 & \Kcfilt & $2506.5 \pm    1.9$ & $168.85 \pm  0.04$ & this work\\
HD 164509 B & 2014 Jun 09 & \Kcfilt & $ 698.8 \pm    1.5$ & $202.4 \pm   0.1$ & this work\\
 & 2014 Jul 22 & 880\unit{nm} & $ 697 \pm    2$ & $202.6 \pm   0.2$ & \citet{Wittrock2016}\\ 
 & 2015 Jun 24 & \Kcfilt & $ 703.3 \pm    1.5$ & $202.9 \pm   0.1$ & this work\\
 & 2016 Jun 09 & \Kcfilt & $ 707.7 \pm    1.5$ & $203.3 \pm   0.1$ & this work\\
HD 177830 B & 2002 Jul 19 & \Ifilt & $1620 \pm   10$ & $ 84.1 \pm   1.0$ & \citet{Roberts2011}\\
 & 2005 May 08 & \Kfilt & $1640 \pm   10$ & $ 84.6 \pm   0.4$ & \citet{Eggenberger2007}\\
 & 2012 May 9 & \Ksfilt & $1670 \pm   10$ & $ 84.3 \pm   0.2$ & \citet{Roberts2015}\\
 & 2012 Jun 12 & $YJH$ & $1680 \pm    2$ & $ 86.0 \pm   0.1$ & \citet{Roberts2015}\\
 & 2014 May 14 & \Ksfilt & $1670 \pm   10$ & $ 85.3 \pm   0.2$ & \citet{Roberts2015}\\
 & 2014 Jun 09 & \Kcfilt & $1665.9 \pm    1.7$ & $84.27 \pm  0.06$ & this work\\
 & 2015 Jul 05 & \Kcfilt & $1664.6 \pm    1.7$ & $84.19 \pm  0.06$ & this work\\
 & 2016 Jun 09 & \Kcfilt & $1665.9 \pm    1.7$ & $84.12 \pm  0.06$ & this work\\
 HD 187123 bg & 2008 Jul 11 & \ipfilt & $2926 \pm   11$ & $ 48.1 \pm   0.3$ & \citet{Ginski2012}\\
 & 2009 Sep 07 & \ipfilt & $2917 \pm   13$ & $ 43.9 \pm   0.3$ & \citet{Ginski2012}\\
 & 2013 Aug 19 & \Kcfilt &$2947.4 \pm    7.4$ & $ 29.5 \pm   0.2$ & this work\\
 & 2015 Jun 24 & \Kcfilt & $3006.8 \pm    4.1$ & $22.73 \pm  0.05$ & this work\\
HD 188015 bg & 2013 Jun 30 & \ipfilt & $4063 \pm   13$ & $113.7 \pm   0.2$ & \citet{Ginski2016}\\
 & 2013 Aug 19 & \Kcfilt & $4079.2 \pm    1.8$ & $113.45 \pm  0.02$ & this work\\
 & 2014 Jul 12 & \Kcfilt & $3992.2 \pm    1.7$ & $112.84 \pm  0.03$ & this work\\
 & 2014 Aug 20 & \ipfilt & $4006 \pm   67$ & $112.5 \pm   0.8$ & \citet{Ginski2016}\\
HD 195019 B & 2013 Aug 19 & \Kcfilt & $3416.3 \pm    1.7$ & $333.80 \pm  0.03$ & this work\\
 & 2014 Jun 09 & \Kcfilt & $3406.9 \pm    1.8$ & $333.87 \pm  0.03$ & this work\\
 & 2014 Jul 12 & \Kcfilt & $3403.9 \pm    1.7$ & $333.89 \pm  0.03$ & this work\\
 & 2015 Jun 03 & \Kcfilt & $3391.1 \pm    2.0$ & $333.95 \pm  0.03$ & this work\\
HD 207832 B & 2013 Aug 19 & \Kcfilt & $2044.3 \pm    1.6$ & $218.24 \pm  0.04$ & this work\\
 & 2013 Aug 19 & \Ksfilt & $2042.8 \pm    1.5$ & $218.23 \pm  0.05$ & this work\\
 & 2014 Jul 12 & \Kcfilt & $2051.7 \pm    1.6$ & $218.25 \pm  0.04$ & this work\\
 & 2015 Jul 05 & \Kcfilt & $2059.0 \pm    1.9$ & $218.26 \pm  0.05$ & this work\\
 & 2015 Oct 26 & \Kcfilt & $2059.9 \pm    1.8$ & $218.21 \pm  0.05$ & this work\\
 & 2016 Sep 13 & \Kcfilt & $2037.8 \pm    6.4$ & $217.3 \pm   0.2$ & this work\\
HD 207832 C & 2013 Aug 19 & \Kcfilt & $2065.0 \pm    1.6$ & $216.11 \pm  0.04$ & this work\\
 & 2013 Aug 19 & \Ksfilt & $2065.3 \pm    1.5$ & $216.08 \pm  0.04$ & this work\\
 & 2014 Jul 12 & \Kcfilt & $2075.3 \pm    1.6$ & $216.37 \pm  0.04$ & this work\\
 & 2015 Jul 05 & \Kcfilt & $2087.5 \pm    1.8$ & $216.73 \pm  0.05$ & this work\\
 & 2015 Oct 26 & \Kcfilt & $2093.6 \pm    1.8$ & $216.82 \pm  0.05$ & this work\\
 & 2016 Sep 13 & \Kcfilt & $2101.2 \pm    5.0$ & $217.79 \pm  0.08$ & this work\\
HD 207832 BC & 2013 Aug 19 & \Kcfilt & $2054.4 \pm    1.7$ & $217.17 \pm  0.04$ & this work\\
 & 2013 Aug 19 & \Ksfilt & $2053.8 \pm    1.5$ & $217.14 \pm  0.05$ & this work\\
 & 2014 Jul 12 & \Kcfilt & $2063.3 \pm    1.7$ & $217.30 \pm  0.04$ & this work\\
 & 2015 Jul 05 & \Kcfilt & $2073.2 \pm    2.0$ & $217.48 \pm  0.06$ & this work\\
 & 2015 Oct 26 & \Kcfilt & $2076.7 \pm    1.9$ & $217.51 \pm  0.05$ & this work\\
 & 2016 Sep 13 & \Kcfilt & $2069.7 \pm    7.9$ & $217.5 \pm   0.2$ & this work\\
\sidehead{\bf Transiting planet host stars in triple systems}
HAT-P-8 B & 2012 Jul 27 & \Kpfilt & $1037.9 \pm    1.5$ & $137.6 \pm   0.1$ & \citet{Ngo2015}\\
 & 2013 Aug 19 & \Ksfilt & $1041.0 \pm    1.5$ & $137.8 \pm   0.1$ & \citet{Ngo2015}\\
 & 2014 Oct 03 & \Ksfilt & $1043.9 \pm    1.6$ & $137.83 \pm  0.09$ & this work\\
 & 2015 Jul 05 & \Ksfilt & $1044.2 \pm    1.5$ & $137.86 \pm  0.08$ & this work\\
 & 2016 Sep 12 & \Ksfilt & $1045.4 \pm    1.6$ & $137.42 \pm  0.09$ & this work\\
HAT-P-8 C & 2012 Jul 27 & \Kpfilt & $1047.8 \pm    1.6$ & $140.9 \pm   0.1$ & \citet{Ngo2015}\\
 & 2013 Aug 19 & \Ksfilt & $1044.7 \pm    1.5$ & $141.1 \pm   0.1$ & \citet{Ngo2015}\\
 & 2014 Oct 03 & \Ksfilt & $1040.3 \pm    1.9$ & $141.17 \pm  0.09$ & this work\\
 & 2015 Jul 05 & \Ksfilt & $1037.5 \pm    1.7$ & $141.20 \pm  0.09$ & this work\\
 & 2016 Sep 12 & \Ksfilt & $1036.7 \pm    1.7$ & $140.8 \pm  0.1$ & this work\\
HAT-P-8 BC & 2012 Jul 27 & \Kpfilt & $1043.1 \pm    1.6$ & $139.5 \pm   0.1$ & \citet{Ngo2015}\\
 & 2013 Aug 19 & \Ksfilt & $1042.7 \pm    1.6$ & $139.6 \pm   0.1$ & \citet{Ngo2015}\\
 & 2014 Oct 03 & \Ksfilt & $1041.4 \pm    2.0$ & $139.7 \pm  0.1$ & this work\\
 & 2015 Jul 05 & \Ksfilt & $1039.9 \pm    1.7$ & $139.76 \pm  0.09$ & this work\\
 & 2016 Sep 12 & \Ksfilt & $1040.0 \pm    1.7$ & $139.3 \pm  0.1$ & this work\\
KELT-4A B & 2012 May 07 & \Kfilt & $1562.5 \pm    8.3$ & $ 29.7 \pm   0.2$ & \citet{Eastman2016}\\
 & 2015 Dec 20 & \Ksfilt & $1569.2 \pm    1.8$ & $29.80 \pm  0.07$ & this work\\
KELT-4A C & 2012 May 07 & \Kfilt & $1584.3 \pm    8.4$ & $ 28.1 \pm   0.1$ & \citet{Eastman2016}\\
 & 2015 Dec 20 & \Ksfilt & $1561.6 \pm    2.0$ & $28.18 \pm  0.07$ & this work\\
KELT-4A BC & 2012 May 07 & \Kfilt & $1573.2 \pm 1.8$ & $28.89 \pm 0.07$ & \citet{Eastman2016}\\
& 2015 Dec 20 & \Ksfilt & $1564.7 \pm 2.1$ & $28.88 \pm 0.08$ & this work\\
WASP-12 B & 2012 Feb 02 & \Kpfilt & $1058.8 \pm    1.5$ & $251.2 \pm   0.1$ & \citet{Ngo2015}\\
 & 2013 Mar 02 & \Ksfilt & $1058.6 \pm    1.5$ & $251.4 \pm   0.1$ & \citet{Ngo2015}\\
 & 2014 Dec 04 & \Ksfilt & $1059.5 \pm    1.9$ & $251.46 \pm  0.08$ & this work\\
 & 2015 Dec 26 & \Ksfilt & $1058.3 \pm    1.5$ & $251.50 \pm  0.08$ & this work\\
 & 2016 Sep 13 & \Ksfilt & $1057.7 \pm    1.6$ & $251.51 \pm  0.09$ & this work\\
WASP-12 C & 2012 Feb 02 & \Kpfilt & $1067.1 \pm    1.5$ & $246.8 \pm   0.1$ & \citet{Ngo2015}\\
 & 2013 Mar 02 & \Ksfilt & $1068.4 \pm    1.5$ & $246.9 \pm   0.1$ & \citet{Ngo2015}\\
 & 2014 Dec 04 & \Ksfilt & $1070.1 \pm    1.6$ & $246.89 \pm  0.09$ & this work\\
 & 2015 Dec 26 & \Ksfilt & $1070.2 \pm    1.5$ & $246.94 \pm  0.08$ & this work\\
 & 2016 Sep 13 & \Ksfilt & $1069.1 \pm    1.6$ & $246.90 \pm  0.09$ & this work\\
WASP-12 BC & 2012 Feb 02 & \Kpfilt & $1062.1 \pm    1.5$ & $248.97 \pm   0.09$ & \citet{Ngo2015}\\
 & 2013 Mar 02 & \Ksfilt & $1062.7 \pm    1.5$ & $246.16 \pm   0.08$ & \citet{Ngo2015}\\
 & 2014 Dec 04 & \Ksfilt & $1064.0 \pm    1.9$ & $249.14 \pm  0.09$ & this work\\
 & 2015 Dec 26 & \Ksfilt & $1063.4 \pm    1.5$ & $249.19 \pm  0.08$ & this work\\
 & 2016 Sep 13 & \Ksfilt & $1062.6 \pm    1.7$ & $249.17 \pm  0.09$ & this work
\enddata
\tablecomments{Separations ($\rho$) and position angle (PA) measurements of candidate companions in this work and other studies with published uncertainties. For triple systems, astrometry for both individual components and their center of mass (``BC'') are included. These values are also plotted in Figures~\ref{fig:astrom_bound1} and \ref{fig:astrom_bound2}.}
\tablenotetext{a}{Candidate companions are labelled with uppercase letters (B, C) when our analysis determine that they are comoving stellar companions and as ``bg'' when they are distant background objects.}
\tablenotetext{b}{This astrometric epoch from the literature did not resolve the individual companions, so the center of light, rather than the center of mass, are reported.}
\end{deluxetable}

\clearpage
\begin{deluxetable*}{llcccc}
\tablecolumns{6}
\tabletypesize{\scriptsize}
\tablecaption{Derived stellar parameters of confirmed stellar companions\label{tab:interp_sec}}
\tablehead{
& &  \colhead{$T_{\mathrm{eff}}$} & \colhead{$M$} & \colhead{$\log g$} & \colhead{$D$} \vspace{-0.1cm}\\
\colhead{Companion} & \colhead{UT Obs. Date} & \vspace{-0.2cm} & & & \\
& & \colhead{(K)} & \colhead{(\Msun)} & \colhead{(cgs)} & \colhead{(AU)} 
}
\startdata
\sidehead{\bf RV planet host stars}
HD 30856B & 2014 Oct 04 & $3940 \pm 200$ & $0.595 \pm 0.070$ & $4.693 \pm 0.060$ & $93.2^{+8.7}_{-7.1}$ \\
 & 2014 Dec 07 & $3945 \pm 70$ & $0.614 \pm 0.022$ & $4.688 \pm 0.018$ & $93.1^{+8.7}_{-7.1}$ \\
 & 2015 Oct 26* & $3731 \pm 29$ & $0.537 \pm 0.013$ & $4.759 \pm 0.013$ & $92.8^{+8.7}_{-7.1}$ \\
HD 43691B & 2013 Dec 18 & $2750 \pm 150$ & $0.0964 \pm 0.0091$ & $5.265 \pm 0.036$ & $366^{+26}_{-23}$ \\
 & 2014 Dec 04 & $2900 \pm 130$ & $0.108 \pm 0.013$ & $5.225 \pm 0.039$ & $366^{+26}_{-23}$ \\
 & 2015 Oct 26 & $2940 \pm 120$ & $0.113 \pm 0.015$ & $5.211 \pm 0.041$ & $365^{+26}_{-23}$ \\
 & 2016 Sep 13* & $3020 \pm 110$ & $0.126 \pm 0.019$ & $5.180 \pm 0.042$ & $365^{+26}_{-23}$ \\
HD 43691C & 2013 Dec 18 & $3040 \pm 110$ & $0.128 \pm 0.019$ & $5.174 \pm 0.041$ & $358^{+25}_{-22}$ \\
 & 2014 Dec 04 & $4100 \pm 1200$ & $0.18 \pm 0.18$ & $4.35 \pm 0.44$ & $358^{+25}_{-22}$ \\
 & 2015 Oct 26 & $3162 \pm 95$ & $0.153 \pm 0.025$ & $5.128 \pm 0.039$ & $359^{+25}_{-22}$ \\
 & 2016 Sep 13* & $3308 \pm 71$ & $0.209 \pm 0.037$ & $5.056 \pm 0.042$ & $359^{+25}_{-22}$ \\
HD 86081B & 2013 Dec 18 & $2400 \pm 1700$ & $0.08 \pm 0.54$ & $5.34 \pm 0.63$ & $277^{+29}_{-24}$ \\
 & 2014 Dec 05* & $2562 \pm 61$ & $0.0876 \pm 0.0019$ & $5.305 \pm 0.010$ & $276^{+29}_{-24}$ \\
HD 96167B & 2014 Jan 12* & $3080 \pm 140$ & $0.135 \pm 0.028$ & $5.159 \pm 0.053$ & $512^{+43}_{-37}$ \\
 & 2015 Jan 09 & $2820 \pm 240$ & $0.102 \pm 0.025$ & $5.224 \pm 0.081$ & $512^{+43}_{-37}$ \\
HD 116029B & 2013 Jul 04* & $3387 \pm 18$ & $0.259 \pm 0.014$ & $5.004 \pm 0.014$ & $171^{+15}_{-13}$ \\
 & 2014 Jun 09 & $3378 \pm 21$ & $0.252 \pm 0.015$ & $5.010 \pm 0.015$ & $171^{+15}_{-13}$ \\
 & 2015 Jan 09 & $3343 \pm 29$ & $0.229 \pm 0.017$ & $5.033 \pm 0.019$ & $171^{+15}_{-13}$ \\
HD 126614B & 2015 Jan 09* & $3382 \pm 23$ & $0.255 \pm 0.017$ & $5.008 \pm 0.016$ & $35.3^{+2.6}_{-2.3}$ \\
 & 2015 Jun 24 & $3434 \pm 28$ & $0.300 \pm 0.028$ & $4.966 \pm 0.023$ & $35.2^{+2.6}_{-2.3}$ \\
HD 142245B & 2013 Jul 04* & $3589 \pm 14$ & $0.455 \pm 0.011$ & $4.842 \pm 0.011$ & $273^{+20}_{-17}$ \\
 & 2014 Jun 09 & $3824 \pm 63$ & $0.573 \pm 0.025$ & $4.723 \pm 0.022$ & $275^{+20}_{-17}$ \\
 & 2015 Jun 24 & $3651 \pm 20$ & $0.501 \pm 0.011$ & $4.796 \pm 0.011$ & $275^{+20}_{-18}$ \\
 & 2016 Jun 09 & $3628 \pm 14$ & $0.484 \pm 0.011$ & $4.812 \pm 0.010$ & $276^{+20}_{-18}$ \\
HD 142245C & 2013 Jul 04* & $3687 \pm 20$ & $0.5172 \pm 0.0091$ & $4.7791 \pm 0.0094$ & $278^{+20}_{-18}$ \\
 & 2014 Jun 09 & $3847 \pm 80$ & $0.580 \pm 0.030$ & $4.717 \pm 0.027$ & $277^{+20}_{-18}$ \\
 & 2015 Jun 24 & $3725 \pm 20$ & $0.5343 \pm 0.0090$ & $4.7619 \pm 0.0088$ & $276^{+20}_{-18}$ \\
 & 2016 Jun 09 & $3741 \pm 21$ & $0.5417 \pm 0.0097$ & $4.7547 \pm 0.0094$ & $276^{+20}_{-18}$ \\
HD 164509B & 2014 Jun 09 & $3403 \pm 40$ & $0.268 \pm 0.035$ & $4.990 \pm 0.034$ & $36.6^{+2.0}_{-1.7}$ \\
 & 2015 Jun 24* & $3480 \pm 14$ & $0.355 \pm 0.015$ & $4.924 \pm 0.013$ & $36.9^{+2.0}_{-1.8}$ \\
 & 2016 Jun 09 & $3515 \pm 29$ & $0.388 \pm 0.029$ & $4.895 \pm 0.023$ & $37.1^{+2.0}_{-1.8}$ \\
HD 177830B & 2014 Jun 09 & $3159 \pm 33$ & $0.1524 \pm 0.0087$ & $5.129 \pm 0.014$ & $98.3^{+3.8}_{-3.5}$ \\\
 & 2015 Jul 05 & $3139 \pm 42$ & $0.1475 \pm 0.0099$ & $5.137 \pm 0.016$ & $98.2^{+3.8}_{-3.5}$ \\
 & 2016 Jun 09* & $3260 \pm 33$ & $0.185 \pm 0.014$ & $5.083 \pm 0.017$ & $98.3^{+3.8}_{-3.5}$ \\
HD 195019B & 2013 Aug 19 & $4000 \pm 120$ & $0.631 \pm 0.036$ & $4.673 \pm 0.031$ & $131.5^{+5.1}_{-4.8}$ \\
 & 2014 Jun 09* & $3890 \pm 100$ & $0.599 \pm 0.034$ & $4.702 \pm 0.030$ & $131.2^{+5.1}_{-4.8}$ \\
 & 2014 Jul 12 & $4000 \pm 130$ & $0.632 \pm 0.038$ & $4.672 \pm 0.032$ & $131.1^{+5.1}_{-4.8}$ \\
 & 2015 Jun 03 & $3940 \pm 110$ & $0.614 \pm 0.033$ & $4.689 \pm 0.029$ & $130.6^{+5.1}_{-4.7}$ \\
HD 207832B & 2013 Aug 19 & $2680 \pm 140$ & $0.0931 \pm 0.0070$ & $5.278 \pm 0.030$ & $111.2^{+5.9}_{-5.3}$ \\
 & 2014 Jul 12 & $2716 \pm 90$ & $0.0944 \pm 0.0051$ & $5.273 \pm 0.021$ & $111.6^{+6.0}_{-5.3}$ \\
 & 2015 Jul 05* & $2870 \pm 81$ & $0.1053 \pm 0.0076$ & $5.234 \pm 0.024$ & $112.0^{+6.0}_{-5.4}$ \\
 & 2015 Oct 26 & $2819 \pm 86$ & $0.1007 \pm 0.0065$ & $5.249 \pm 0.023$ & $112.1^{+6.0}_{-5.4}$ \\
 & 2016 Sep 13 & $2623 \pm 100$ & $0.0894 \pm 0.0040$ & $5.295 \pm 0.020$ & $110.9^{+5.9}_{-5.3}$ \\
HD 207832C & 2013 Aug 19 & $2864 \pm 98$ & $0.1049 \pm 0.0094$ & $5.234 \pm 0.030$ & $112.3^{+6.0}_{-5.4}$ \\
 & 2014 Jul 12 & $2838 \pm 85$ & $0.1023 \pm 0.0071$ & $5.243 \pm 0.024$ & $112.9^{+6.0}_{-5.4}$ \\
 & 2015 Jul 05* & $2855 \pm 80$ & $0.1040 \pm 0.0069$ & $5.238 \pm 0.023$ & $113.6^{+6.1}_{-5.4}$ \\
 & 2015 Oct 26 & $2850 \pm 80$ & $0.1035 \pm 0.0067$ & $5.240 \pm 0.022$ & $113.9^{+6.1}_{-5.4}$ \\
 & 2016 Sep 13 & $2958 \pm 75$ & $0.116 \pm 0.010$ & $5.205 \pm 0.027$ & $114.3^{+6.1}_{-5.5}$ \\
\sidehead{\bf Transiting planet host stars in triple systems}
HAT-P-8B & 2014 Oct 03 & $3186 \pm 26$ & $0.1604 \pm 0.0076$ & $5.118 \pm 0.011$ & $238 \pm 13$  \\
 & 2015 Jul 07 & $3205 \pm 23$ & $0.1659 \pm 0.0068$ & $5.1099 \pm 0.0097$ & $238 \pm 13$ \\
 & 2016 Sep 12* & $3206 \pm 22$ & $0.1664 \pm 0.0065$ & $5.1092 \pm 0.0093$ & $238 \pm 13$ \\
HAT-P-8C & 2014 Oct 03 & $3100 \pm 31$ & $0.1393 \pm 0.0065$ & $5.152 \pm 0.012$ & $237 \pm 13$  \\
 & 2015 Jul 07 & $3022 \pm 21$ & $0.1250 \pm 0.0031$ & $5.1810 \pm 0.0076$ & $236 \pm 13$ \\
 & 2016 Sep 12* & $3024 \pm 21$ & $0.1254 \pm 0.0031$ & $5.1801 \pm 0.0076$ & $236 \pm 13$ \\
KELT-4A B & 2015 Dec 20* & $4560 \pm 320$ & $0.776 \pm 0.081$ & $4.565 \pm 0.066$ & $331^{+20}_{-19}$ \\
KELT-4A C & 2015 Dec 20* & $3846 \pm 83$ & $0.584 \pm 0.029$ & $4.715 \pm 0.026$ & $330^{+20}_{-19}$ \\
WASP-12B & 2014 Dec 04 & $3861 \pm 66$ & $0.589 \pm 0.022$ & $4.711 \pm 0.019$ & $462 \pm 40$ \\
 & 2015 Dec 26* & $3820 \pm 51$ & $0.575 \pm 0.019$ & $4.723 \pm 0.017$ & $462 \pm 39$ \\
 & 2016 Sep 13 & $3842 \pm 55$ & $0.583 \pm 0.019$ & $4.716 \pm 0.017$ & $461 \pm 39$ \\
WASP-12C & 2014 Dec 04 & $3690 \pm 53$ & $0.518 \pm 0.024$ & $4.778 \pm 0.024$ & $467 \pm 40$ \\
 & 2015 Dec 26* & $3791 \pm 47$ & $0.564 \pm 0.019$ & $4.733 \pm 0.018$ & $467 \pm 40$ \\
 & 2016 Sep 13 & $3839 \pm 56$ & $0.582 \pm 0.020$ & $4.717 \pm 0.018$ & $466 \pm 40$ 
\enddata
\tablecomments{For observations on each date, we report error weighted averages of all measurements on $T_{\mathrm{eff}}$, $M$, $\log g$, and $D$. Our uncertainties account for measurement error and the published primary star's stellar parameters but do not include uncertainties introduced from use of stellar models and our assumptions on stellar composition. Therefore, the true uncertainties are larger than the values presented in this table. For each target, we choose the epoch with the lowest measurement error as the representative value for further dynamical analysis. This epoch is marked with an asterisk.}
\end{deluxetable*}

\clearpage
\begin{deluxetable*}{lccccc}
\tablecolumns{6}
\tablecaption{Stellar companion fraction of stars in RV-planet surveys \label{tab:RVsurveys}}
\tablehead{
\colhead{Survey} & \colhead{Multi-stellar systems}  & \colhead{Companion fraction} & \colhead{Observatory / Instrument} & \colhead{Range\tablenotemark{a} (\arcsec)} & Overlap
}
\startdata
\sidehead{\bf RV-detected planet host stars}
This work & 11 out of 144 & $7.6\% \pm 2.3\%$ & Keck / NIRC2 & 0.3--6.0 & --\\
\citet{Eggenberger2007} & 6 out of 56\tablenotemark{b} & $10.7\% \pm 4.4\%$\tablenotemark{b} & VLT / NACO & 0.2--13.0 & 20/56\\
\citet{Ginski2012} & 3 out of 70 & $4.3\% \pm 2.5\%$ & Calar Alto / AstraLux & 0.5--12.0 & 29/70\\
\cite{Ginski2016} & 4 out of 51 & $7.8\% \pm 3.9\%$ & Calar Alto / AstraLux & 1.2--12.0 & 22/51 \\
& 6 out of 51\tablenotemark{c} & $11.8\% \pm 4.4\%$\tablenotemark{c} & & & \\
\citet{Mugrauer2015} & 2 out of 32\tablenotemark{d} & $6.3\% \pm 4.32\%$\tablenotemark{d} & VLT / NACO & 0.3--13.0 & 5/32 \\
\sidehead{\bf Control group (RV planet survey stars without a planet)}
\citet{Eggenberger2007} & 13 out of 74\tablenotemark{b} & $17.6\% \pm 4.9\%$\tablenotemark{b} & VLT / NACO & 0.2--13.0 & 0/74
 \enddata
\tablecomments{Companion fractions were computed from the raw number of confirmed multi-stellar systems reported by each survey, assuming Poisson uncertainties. These companion fractions are not corrected for survey completeness. Only stars that have RV-detected planet hosts are counted. The ``overlap'' column indicates the number of targets from the cited survey that are also in our survey. 73 out of our 144 targets have also been imaged in the other surveys; however, only 26 out of 144 targets have been previously imaged at observatories comparable to Keck.}
\tablenotetext{a}{The inner limit of the range corresponds to a separation where the survey is sensitive to a contrast of $\Delta K \sim 5$ ($\Delta I \sim 7$ for the AstraLux surveys). We determine the outer limit of our survey to be the separation where we have 90\% directional completeness, however our full chip size is 10\arcsec. For other surveys, we report their instrument's full chip size as the outer limit.}
\tablenotetext{b}{One target in \citet{Eggenberger2007}, HD 33636, was originally in our survey list until we learned the detection of HD 33636 was retracted. Therefore, we count this system as part of \citet{Eggenberger2007}'s control group instead of their planet hosting sample.}
\tablenotetext{c}{\citet{Ginski2016} reported 4 confirmed multi-stellar system and several candidate multi-stellar systems. Two of these ambiguous cases (HD 43691 and HD 116029) were also in our survey and we confirmed them to be co-moving systems. In this row, we report an updated companion fraction rate with these two additional systems.}
\tablenotetext{d}{We were unable to confirm the planetary status for the HD 9578 system surveyed by \citet{Mugrauer2015} in the peer-reviewed literature. Thus, we exclude it from this count.}
\end{deluxetable*}

\clearpage
\begin{deluxetable*}{lcccccccccc}
\tablecolumns{11}
\tabletypesize{\scriptsize}
\tablecaption{Stellar and planetary parameters of RV-planet host systems with stellar companions within 6\arcsec \label{tab:RV_pl}}
\tablehead{
& \colhead{$M_{\mathrm{host}}$} & \colhead{$a_{\mathrm{pl}}$} & & \colhead{$m_{\mathrm{pl}}$} & & \colhead{$M_{\mathrm{comp}}$} & \colhead{$a_{\mathrm{comp}}$} & \colhead{$\rho_{\mathrm{comp}}$} & \vspace{-0.1cm}\\
\colhead{Planet host} & & & \colhead{$e_{\mathrm{pl}}$} & & \colhead{Ref.} & & & & \colhead{Ref.} & \colhead{$\nu/n$} \vspace{-0.2cm} \\
& \colhead{(\Msun)} & \colhead{(au)} & & \colhead{(\Mjup)} & & \colhead{(\Msun)} & \colhead{(au)} & \colhead{(\arcsec)} & & 
}
\startdata
{\bf HD 164509} & 1.13 (2) & 0.875 (8)   & 0.3 (1)     & 0.48 (9)  & Gig12 & {\bf 0.36 (2)} & {\bf 37 (2)} & {\bf 0.703 (2)} & {\bf this work} & 4\ee{-6}\\
{\bf HD 30856}  & 1.35 (9)  & 1.8 (2)       & 0.117 ()   & 1.8 (2)    & J11 & {\bf 0.54 (1)} & {\bf 93 (8)} & {\bf 0.786 (2)} & {\bf this work} & 3\ee{-6}\\
HD 197037 & 1.06 (2) & 2.07 (5) & 0.22 (7) & 0.79 (5) 	& S15,R12 & 0.34 (5) & 119 (2) & 3.688 (9) & G16 & 2\ee{-6}\\ 
HD 217786 & 1.02 (3) & 2.38 (4) & 0.40 (5) & 13.0 (8) 	& Mo11	 & 0.162 (7) & 157 (7) & 2.856 (7) & G16 & 6\ee{-7}\\
HD 142245\tablenotemark{(3*)} & 1.7 (1)   & 2.77 (9)     & 0.09 ()     & 1.90 (2)  & J11 & 0.97 (1) & 276 (14) & 2.505 (2) & this work & 6\ee{-7}\\
HD 126614 & 1.15 (3) & 2.35 (2)     & 0.41 (1)   & 0.38 (4)  & Ho10 & 0.26 (2) & 35 (3) & 0.486 (1) & this work & 5\ee{-7}\\
HD 142\tablenotemark{a}       & 1.2 (2) & 1.02 (3) & 0.17 (6) & 1.3 (0.2)	& W12 & 0.59 (2) & 105 (2) & 4.08 (2) & E07 & 5\ee{-7}\\
HD 132563B\tablenotemark{(3*)} & 1.01 (1) & 2.62 (4) & 0.22 (9) & 1.49 (9) & D11 & 1.64 (2) & 400 (50) & 4.11 (2) & D11 &  5\ee{-7}\\ 
{\bf HD 116029} & 1.6 (1)   & 1.78 (5)     & 0.054 ()   & 2.1 (2)    & J11 & {\bf 0.26 (1)} & {\bf 171 (15)} & {\bf 1.392 (1)} & {\bf this work} & 2\ee{-7}\\
HD 87646 & 1.12 (9) & 0.117 (3) & 0.05 (2) & 12.4 (7) & M16 & 0.6\tablenotemark{(est)} & 15.7 (2.1) & 0.213 () & M16, HIP & 2\ee{-7}\\
HD 89484 & 1.2 (2) & 1.19 (2) & 0.14 (5) & 9 (1) & Ha10 & 0.9\tablenotemark{(est)} & 178 (6) & 4.629 () & Ma06 & 2\ee{-7}\\
{\bf HD 207832\tablenotemark{(3*,b)}} & 0.98 (7) & 0.570 (2)   & $0.13^{+0.18}_{-0.05}$ & $0.56^{+0.06}_{-0.03}$ & S13,H12 & {\bf 0.21 (1)} & {\bf 113 (4)} &{\bf  2.073 (2)} & {\bf this work} & 3\ee{-8}\\
HD 177830\tablenotemark{c} & 1.37 (4) & 0.5137 (3) & 0.3 (1)     & 0.15 (2)  & J06,Me11 & 0.19 (1) & 98 (4) & 1.666 (2) & this work & 2\ee{-8}\\
HD 2638     & 0.9 (1) & 0.044\tablenotemark{x}  () & 0 () & 0.48 ()  & G10,M05 & 0.43 (1) & 26 (2) & 0.520 (4) & G16 & 2\ee{-9}\\ 
HD 96167  & 1.16 (5)  & 1.30 (7)     & 0.71 (4)   & 0.7 (2)    & J15,P09 & 0.14 (3) & 512 (43) & 5.890 (3) & this work & 2\ee{-9}\\
HD 16141   & 1.1 (1) & 0.36 (2) & 0.25 (5) & 0.26 (3) 	& G10,B06 & 0.31 (2) & 223 (11) & 6.22 (3) & E07 & 1\ee{-9}\\ 
HD 195019 & 1.1 (1)   & 0.139 (8)   & 0.014 (4) & 3.7 (3)    & G10,B06 & 0.60 (3) & 131 (5) & 3.407 (2) & this work & 7\ee{-10}\\
{\bf HD 86081}  & 1.23 (8)  & 0.035\tablenotemark{x} ()       & 0.008 (4) & 1.5 ()      & S13,J06 & {\bf 0.088 (2)} & {\bf 276 (29)} & {\bf 2.901 (2)} & {\bf this work} & 1\ee{-10}\\
{\bf HD 43691\tablenotemark{(3*)}}  & 1.38 (5)  & 0.24 ()       & 0.14 (2)   & 2.49 ()    & ds07 & {\bf 0.33 (4)} & {\bf 362 (18)} & {\bf 4.500 (2)} & {\bf this work} & 7\ee{-11}\\
HD 41004A & 0.7\tablenotemark{(est)} & 0.006\tablenotemark{x} (2) & 0.7 (2) & 2.5 (7) & S02,Z04 & 0.4\tablenotemark{(est)} & 22.5 () & 0.5 () & S02,HIP & 1\ee{-11}\\
HD 185269 & 1.3 (1) & 0.077\tablenotemark{x} () & 0.23 (3) & 1.03 (3) 	& Mo06	  & 0.23 (1) & 215 (8) & 4.53 (1) & G16 & 8\ee{-12}\\ 
tau Boo      & 1.34 (5) & 0.049\tablenotemark{x}  (3) & 0.011 (6) & 4.32 (4) 	& B12,B15 & 0.4 () & 225 (1) & 2.18 (1) & Gin12 & 3\ee{-12}
\enddata
\tablerefs{(B06) \citet{Butler2006}; (B12) \citet{Brogi2012}; (B15) \citet{Borsa2015}; (D11) \citet{Desidera2011}; (E07) \citet{Eggenberger2007}; (G10) \citet{Ghezzi2010}; (Gig12) \citet{Giguere2012}; (Gin12) \citet{Ginski2012}; (G16) \citet{Ginski2016}; (Ha10) \citet{Han2010}; (Ho10) \citet{Howard2010}; (H12) \citet{Haghighipour2012}; (HIP) Hipparcos Catalogue \citet{Perryman1997}; (J06) \citet{Johnson2006}; (J11) \citet{Johnson2011b}; (J15) \citet{Jofre2015}; (M05) \citet{Moutou2005}; (Ma06) \citet{Mason2006}; (Mo06) \citet{Moutou2006}; (Me11) \citet{Meschiari2011}; (Mo11) \citet{Moutou2011}; (M16) \citet{Ma2016}; (P09) \citet{Peek2009}; (S02) \citet{Santos2002}; (dS07) \citet{daSilva2007}; (S13) \citet{Santos2013}; (S15) \citet{Sousa2015}; (W12) \citet{Wittenmyer2012}; (W13) \citet{Wittenmyer2013}; (Z04) \citet{Zucker2004}}
\tablecomments{All RV-planet hosting systems with stellar companions detected within 6\arcsec. As this is a soft limit (see text), we included HD 16141 with a companion at 6\arcsec$\!$.22. For brevity, the number(s) in parentheses are uncertainties on the last digit(s) (i.e. 2.35 (2) is $2.35\pm0.02$). Missing values indicate no uncertainty provided by the source. References for the system and companion parameters follow each set of columns. We rank this list by a metric, $\nu/n = \frac{M_{\mathrm{comp}}}{M_{\mathrm{host}}}\left(\frac{a_{\mathrm{pl}}}{a_{\mathrm{comp}}}\right)^3$, to represent strength of the precession induced on the planet by the companion star. Systems with new or newly confirmed stellar companions have boldface names. and companion properties}
\tablenotetext{(3*)}{This is a triple system. The reported $M_{\mathrm{comp}}$ is the combined mass of both stars and $a_{\mathrm{comp}}$ is the average projected separation of each component.}
\tablenotetext{a}{The innermost planet, HD 142b, is reported here. This system also hosts a second planet, HD 142c, at 6.8\unit{au} and 5.3\unit{\Mjup}.}
\tablenotetext{b}{The innermost planet, HD 207832b, is reported here. This system also hosts a second planet, HD 207832c, at 2.112\unit{au} and 0.73\unit{\Mjup}.}
\tablenotetext{c}{The innermost planet, HD 177830c, is reported here. This system also hosts a second planet, HD 177830b, at 1.2218\unit{au} and 1.49\unit{\Mjup}.}
\tablenotetext{(est)}{These stellar masses were not measured. Instead, they were estimated from the star's spectral type.}
\tablenotetext{x}{This system is not included in the comparison between single and multi stellar systems described in Section~\ref{sec:single_v_multi} because the planet's semimajor axis is less than 0.1\unit{au}.}
\end{deluxetable*}

\clearpage
\begin{deluxetable}{lccccccc}
\tablecolumns{4}
\tablecaption{Bayesian model comparison \label{tab:bayes}}
\tablehead{
 & & \multicolumn{2}{c}{Two-population model} \vspace{-0.1cm} \\
 & \colhead{One-population model} & \vspace{-0.2cm}\\
& & \colhead{single systems} & \colhead{multi-stellar systems}
}
\startdata
\sidehead{Uniform prior ranges and $\Delta \theta$}
$a$ & (0,10), 10 & (0,10), 10 & (0,10), 10 \\
$b$ & (0,10), 10 & (0,10), 10 & (0,15), 15 \\
$\alpha$ & (-2,2), 4 & (-2,2), 4 & (-2,2), 4 \\
$\beta$ & (-2,2), 4 & (-2,2), 4 & (-2,2), 4 \\
\sidehead{Fit results and uncertainties $\delta \theta$ for all single and multi-stellar systems}
$\hat{\mathcal{L}}$ & -1331 & -1136 & -195 \\
N parameters & 4 & 4 & 4 \\
N systems & 115 & 98 & 17 \\
$\delta a$ & 0.119 & 0.126 & 0.431 \\
$\delta b$ & 0.397 & 0.409 & 1.745 \\
$\delta \alpha$ & 0.038 & 0.041 & 0.104 \\
$\delta \beta$ & 0.044 & 0.048 & 0.114 \\
\sidehead{Fit results and uncertainties $\delta \theta$ for all single and the top 9 multi-stellar systems}
$\hat{\mathcal{L}}$ & -1236 & -1136 & -97 \\
N parameters & 4 & 4 & 4 \\
N systems & 107 & 98 & 9 \\
$\delta a$ & 0.125 & 0.125 & 0.852 \\
$\delta b$ & 0.411 & 0.407 & 2.948 \\
$\delta \alpha$ & 0.039 & 0.041 & 0.148 \\
$\delta \beta$ & 0.047 & 0.048 & 0.224
\enddata
\tablecomments{Maximum likelihood, prior ranges and fit uncertainties for Bayesian model comparison of a one-population vs. two-population model of giant planets discussed in Section~\ref{sec:single_v_multi}.}
\end{deluxetable}

\begin{deluxetable*}{lccccccc}
\tablecolumns{8}
\tablecaption{Orbital inclination fits of hierarchical triple systems \label{tab:triple_astrom}}
\tablehead{
 & Planet inclination & \multicolumn{3}{c}{Outer orbit (ABC) inclination (deg)}  & \multicolumn{3}{c}{Inner orbit (BC) inclination (deg)}  \vspace{-0.1cm}\\
\colhead{System} & & & & & & & \vspace{-0.2cm}\\
& (deg) & \colhead{$i_{\mathrm{P,max}}$} & \colhead{68\% C.I.} & \colhead{95\% C.I.} & \colhead{$i_{\mathrm{P,max}}$} & \colhead{68\% C.I.} & \colhead{95\% C.I.} 
}
\startdata
HAT-P-8      & $87.5^{+1.9}_{-0.9}$      & 118.8 & (94.8,136.3) & (69.6,161.2)      & 44.7 & (24.9,75.6) & (9.4,118.6)\\ 
HD 43691   & unknown                        & 121.4 & (131.9,163.4) & (116.4,173.9)  & 127.0 & (131.0,162.8) & (115.2,173.6)\\ 
HD 142245 & unknown                         & 116.2 & (107.3,139.7) & (97.7,162.8)   & 52.7 & (21.8,53.9) & (8.3,64.3) \\ 
HD 207832 & unknown                         & 43.8 & (18.5,53.9) & (6.8,65.0)              & 69.5 & (54.2,75.7) & (37.7,82.6) \\ 
KELT-4A     & $83.16^{+0.22}_{-0.21}$ & 41.2 & (16.6,50.6) & (6.2,66.8)              & 111.0 & (74.9,119.0) & (28.3,155.9) \\ 
WASP-12   & $83.37^{+0.72}_{-0.64}$ & 76.3 & (40.1,94.8) & (15.9,133.6)          & 84.7 & (44.2,117.0) & (17.5,153.0) 
\enddata
\tablecomments{The most likely value ($i_{\mathrm{P,max}}$) and 68\% and 95\% confidence intervals (C.I.) for orbital inclinations of the triple star systems are computed from the orbital fits described in Section~\ref{sec:ofti}. HAT-P-8, HD 142245, and KELT-4A have the orientation of the binary pair misaligned with the entire triple system. HD 43691 and WASP-12 are well aligned. HD 207832 is only marginally aligned (within the 68\% confidence intervals). All of the transiting planet systems have binary companions on an orbital plane consistent with the planet's inclination at the 68\% confidence interval (KELT-4A and WASP-12) or the 95\% confidence interval (HAT-P-8).}
\end{deluxetable*}

\clearpage
\appendix

\begin{deluxetable*}{lccccc}
\tablecolumns{6}
\tablecaption{Contrast curves for all targets \label{tab:contrast}}
\tablehead{
 & & & \colhead{Separation}  & \colhead{$5\sigma$ contrast} & \colhead{Directional} \vspace{-0.1cm}\\
 \colhead{Star} & \colhead{Bandpass} & \colhead{Date} & & \vspace{-0.2cm}\\
& & & \colhead{(arcsec)} & \colhead{(magnitudes)} & \colhead{completeness}
}
\startdata
GJ 317 & Kc & 2014-11-07 & 0.154 & 2.87 & 100.0 \\ 
 & & & 0.252 & 4.52 & 100.0 \\ 
 & & & 0.350 & 5.53 & 100.0 \\ 
 & & & 0.449 & 6.33 & 100.0 \\ 
 & & & \nodata & \nodata  & \nodata\\
GJ-433 & Kc & 2015-01-09 & 0.154 & 3.07 & 100.0 \\ 
 & & & 0.251 & 4.90 & 100.0 \\ 
 & & & 0.350 & 5.77 & 100.0 \\ 
 & & & 0.449 & 6.52 & 100.0 \\ 
 & & & \nodata & \nodata  & \nodata\\
\nodata & \nodata & \nodata & \nodata & \nodata  & \nodata
 \enddata
\tablecomments{Best $5\sigma$ contrast curve for each target in the indicated bandpass. The full table with all 144 targets will be available in comma separated value (CSV) format.}
\end{deluxetable*}

\clearpage
\begin{figure*}
\epsscale{1.0}
\plotone{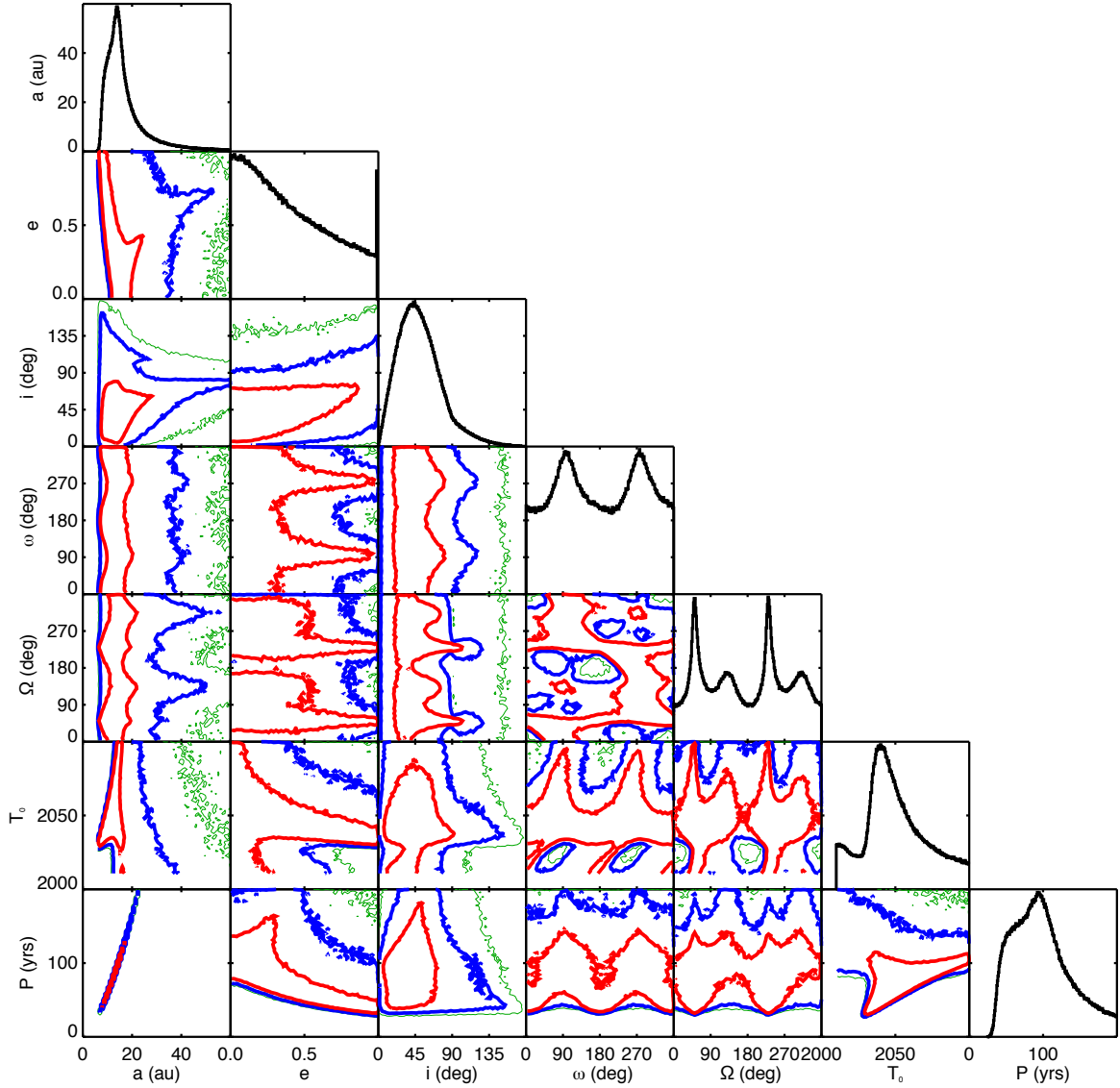}
\caption{Two-dimensional probability contours for each pair of the seven OFTI-fitted orbital parameters. The red, blue and green contours represent regions containing 68\%, 95\% and 99.7\% of the marginalized probabilities. The black histograms show the one-dimensional marginalized probability distributions for each orbital parameter. This representative set of panels are for the inner orbit of the system with the best constraints, HAT-P-8. Tables~\ref{tab:ofti_hat8} through \ref{tab:ofti_wasp12} summarize the fit results for this system and all triple systems.
\label{fig:ofti_triangle}}
\end{figure*}

\begin{deluxetable*}{lccccccc}
\tablecolumns{7}
\tablecaption{OFTI fit summary for triple system HAT-P-8 \label{tab:ofti_hat8}}
\tablehead{
\colhead{$x$} & \colhead{Median} & \colhead{$x_{\chi^2,\mathrm{min}}$} & \colhead{$x_{\mathrm{P,max}}$} &  \colhead{68\% C.I.} &  \colhead{95\% C.I.} &  \colhead{99.7\% C.I.}
}
\startdata
\sidehead{{\bf Outer orbit} (min $\chi^2$ = 25.701, $\chi^2$ at highest probability = 25.854)}
$a$ (au)       & 241.9 & 178.7 & 111.5 & (160.2,491.4) & (128.3,1713.3) & (112.9,8324.6) \\
$e$            & 0.582 & 0.413 & 0.832 & (0.210,0.868) & (0.031,0.980) & (0.002,0.990) \\
$i$ (deg)      & 113.4 & 115.3 & 118.8 & (94.8,136.3) & (69.6,161.2) & (25.6,175.2) \\
$\omega$ (deg) & 91.3 & 162.6 & 22.4 & (36.4,144.3) & (5.5,174.6) & (0.3,179.7) \\
$\Omega$ (deg) & 128.5 & 134.4 & 327.8 & (50.6,148.8) & (6.8,173.7) & (0.4,179.6) \\
$T_0$ (yr)     & 3312 & 2892 & 2556 & (2687,6377) & (2393,30869) & (2035,325732) \\
$P$ (yr)      & 3091 & 1916 & 942 & (1665,8956) & (1191,58222) & (978,625040)\\
\sidehead{{\bf Inner orbit} (min $\chi^2$ = 0.320, $\chi^2$ at highest probability = 0.787)}
$a$ (au)       & 14.9 & 15.8 & 9.3 & (10.5,25.3) & (7.9,73.4) & (6.8,366.7) \\
$e$            & 0.347 & 0.193 & 0.457 & (0.098,0.728) & (0.014,0.955) & (0.001,0.990) \\
$i$ (deg)      & 47.7 & 3.5 & 44.7 & (24.9,75.6) & (9.4,118.6) & (2.3,160.5) \\
$\omega$ (deg) & 94.2 & 242.9 & 288.4 & (34.9,147.6) & (4.9,175.1) & (0.3,179.7) \\
$\Omega$ (deg) & 80.3 & 326.6 & 93.4 & (39.2,143.0) & (6.7,173.4) & (0.4,179.6) \\
$T_0$ (yr)     & 2056 & 2010 & 2034 & (2035,2135) & (2014,2603) & (2010,8439) \\
$P$ (yr)      & 105 & 116 & 52 & (63,234) & (41,1159) & (32,12930)
\enddata
\tablecomments{Summary of the posteriors from OFTI orbit fitting (see Section~\ref{sec:ofti}) for this triple system's outer and inner orbits. For each parameter, we show the median value, the value with the lowest $\chi^2$ ($x_{\chi^2,\mathrm{min}}$), the most likely value ($x_{\mathrm{P,max}}$) and three confidence intervals (C.I.). Samples of the posteriors for all fit parameters are available online as FITS tables.}
\end{deluxetable*}

\begin{deluxetable*}{lccccccc}
\tablecolumns{7}
\tablecaption{OFTI fit summary for triple system HD 43691 \label{tab:ofti_hd43691}}
\tablehead{
\colhead{$x$} & \colhead{Median} & \colhead{$x_{\chi^2,\mathrm{min}}$} & \colhead{$x_{\mathrm{P,max}}$} &  \colhead{68\% C.I.} &  \colhead{95\% C.I.} &  \colhead{99.7\% C.I.}
}
\startdata
\sidehead{{\bf Outer orbit} (min $\chi^2$ = 7.089, $\chi^2$ at highest probability = 7.444)}
$a$ (au)       & 452.1 & 401.5 & 162.2 & (341.5,709.3) & (254.5,1547.9) & (190.4,4882.8) \\
$e$            & 0.229 & 0.169 & 0.312 & (0.068,0.474) & (0.010,0.730) & (0.001,0.909) \\
$i$ (deg)      & 148.3 & 165.0 & 121.4 & (131.9,163.4) & (116.4,173.9) & (103.1,178.5) \\
$\omega$ (deg) & 92.0 & 143.5 & 223.4 & (28.4,152.4) & (4.0,176.0) & (0.2,179.8) \\
$\Omega$ (deg) & 98.3 & 210.6 & 49.3 & (28.9,153.5) & (4.1,175.9) & (0.2,179.7) \\
$T_0$ (yr)     & 5071 & 8249 & 3059 & (2930,11372) & (2130,34638) & (2017,187612) \\
$P$ (yr)    & 7518 & 6569 & 1649 & (4935,14775) & (3172,47655) & (2050,266357) \\
\sidehead{{\bf Inner orbit} (min $\chi^2$ = 2.256, $\chi^2$ at highest probability = 2.974)}
$a$ (au)       & 11.2 & 16.6 & 7.2 & (8.5,17.9) & (6.5,39.8) & (5.0,127.7) \\
$e$            & 0.214 & 0.333 & 0.310 & (0.064,0.456) & (0.009,0.727) & (0.001,0.910) \\
$i$ (deg)      & 147.4 & 119.8 & 127.0 & (131.0,162.8) & (115.2,173.6) & (102.3,178.5) \\
$\omega$ (deg) & 79.1 & 148.6 & 256.2 & (25.6,148.7) & (3.9,176.0) & (0.2,179.8) \\
$\Omega$ (deg) & 79.7 & 278.9 & 272.8 & (33.1,128.8) & (5.9,173.4) & (0.4,179.6) \\
$T_0$ (yr)     & 2029 & 2027 & 2032 & (2018,2056) & (2012,2172) & (2010,2952) \\
$P$ (yr)    & 74 & 138 & 39 & (49,151) & (33,499) & (22,2863)
\enddata
\tablecomments{Summary of the posteriors from OFTI orbit fitting (see Section~\ref{sec:ofti}) for this triple system's outer and inner orbits. For each parameter, we show the median value, the value with the lowest $\chi^2$ ($x_{\chi^2,\mathrm{min}}$), the most likely value ($x_{\mathrm{P,max}}$) and three confidence intervals (C.I.). Samples of the posteriors for all fit parameters are available online as FITS tables.}
\end{deluxetable*}

\begin{deluxetable*}{lccccccc}
\tablecolumns{7}
\tablecaption{OFTI fit summary for triple system HD 142245 \label{tab:ofti_hd142245}}
\tablehead{
\colhead{$x$} & \colhead{Median} & \colhead{$x_{\chi^2,\mathrm{min}}$} & \colhead{$x_{\mathrm{P,max}}$} &  \colhead{68\% C.I.} &  \colhead{95\% C.I.} &  \colhead{99.7\% C.I.}
}
\startdata
\sidehead{{\bf Outer orbit} (min $\chi^2$ = 21.154, $\chi^2$ at highest probability = 21.337)}
$a$ (au)       & 273.2 & 309.5 & 111.9 & (182.0,556.9) & (140.6,1694.7) & (115.1,6011.4) \\
$e$            & 0.543 & 0.568 & 0.711 & (0.211,0.785) & (0.032,0.918) & (0.002,0.982) \\
$i$ (deg)      & 120.4 & 118.3 & 116.2 & (107.3,139.7) & (97.7,162.8) & (91.6,175.7) \\
$\omega$ (deg) & 85.8 & 268.6 & 161.2 & (38.4,139.8) & (6.0,173.9) & (0.4,179.6) \\
$\Omega$ (deg) & 91.3 & 207.6 & 163.9 & (24.2,168.6) & (2.5,177.8) & (0.1,179.9) \\
$T_0$ (yr)     & 3060 & 4428 & 2312 & (2573,5746) & (2343,23231) & (2034,167483) \\
$P$ (yr)      & 2747 & 3351 & 712 & (1493,8005) & (1012,42425) & (749,285552) \\
\sidehead{{\bf Inner orbit} (min $\chi^2$ = 6.811, $\chi^2$ at highest probability = 8.294)}
$a$ (au)       & 6.4 & 44.1 & 5.1 & (5.1,8.7) & (4.4,15.1) & (3.9,43.0) \\
$e$            & 0.440 & 0.924 & 0.705 & (0.210,0.619) & (0.033,0.774) & (0.002,0.897) \\
$i$ (deg)      & 39.5 & 64.5 & 52.7 & (21.8,53.9) & (8.3,64.3) & (1.9,74.6) \\
$\omega$ (deg) & 97.2 & 133.8 & 90.1 & (38.9,145.9) & (5.5,174.5) & (0.4,179.6) \\
$\Omega$ (deg) & 75.5 & 33.4 & 15.0 & (37.3,115.6) & (7.7,170.7) & (0.5,179.5) \\
$T_0$ (yr)     & 2012 & 2013 & 2012 & (2011,2021) & (2010,2030) & (2010,2060) \\
$P$ (yr)      & 16 & 289 & 11 & (11,25) & (9,58) & (8,279) 
\enddata
\tablecomments{Summary of the posteriors from OFTI orbit fitting (see Section~\ref{sec:ofti}) for this triple system's outer and inner orbits. For each parameter, we show the median value, the value with the lowest $\chi^2$ ($x_{\chi^2,\mathrm{min}}$), the most likely value ($x_{\mathrm{P,max}}$) and three confidence intervals (C.I.). Samples of the posteriors for all fit parameters are available online as FITS tables.}
\end{deluxetable*}

\begin{deluxetable*}{lccccccc}
\tablecolumns{7}
\tablecaption{OFTI fit summary for triple system HD 207832 \label{tab:ofti_hd207832}}
\tablehead{
\colhead{$x$} & \colhead{Median} & \colhead{$x_{\chi^2,\mathrm{min}}$} & \colhead{$x_{\mathrm{P,max}}$} &  \colhead{68\% C.I.} &  \colhead{95\% C.I.} &  \colhead{99.7\% C.I.}
}
\startdata
\sidehead{{\bf Outer orbit} (min $\chi^2$ = 3.760, $\chi^2$ at highest probability = 4.301)}
$a$ (au)       & 171.5 & 158.0 & 60.8 & (116.4,299.0) & (86.9,684.2) & (68.8,2131.0) \\
$e$            & 0.713 & 0.701 & 0.680 & (0.567,0.809) & (0.359,0.900) & (0.122,0.966) \\
$i$ (deg)      & 36.2 & 30.6 & 43.8 & (18.5,53.9) & (6.8,65.0) & (1.6,72.1) \\
$\omega$ (deg) & 64.0 & 232.3 & 20.4 & (23.1,133.7) & (4.0,175.8) & (0.3,179.8) \\
$\Omega$ (deg) & 67.9 & 229.8 & 42.1 & (41.3,104.3) & (9.9,168.0) & (0.7,179.4) \\
$T_0$ (yr)     & 3925 & 3765 & 2357 & (3007,6644) & (2604,18356) & (2057,92862) \\
$P$ (yr)    & 2083 & 1900 & 456 & (1159,4821) & (743,16659) & (522,91976) \\
\sidehead{{\bf Inner orbit} (min $\chi^2$ = 7.067, $\chi^2$ at highest probability = 7.376)}
$a$ (au)       & 6.7 & 21.9 & 5.0 & (4.5,12.1) & (3.4,26.3) & (2.8,81.1) \\
$e$            & 0.305 & 0.538 & 0.597 & (0.093,0.613) & (0.013,0.847) & (0.001,0.958) \\
$i$ (deg)      & 65.9 & 75.5 & 69.5 & (54.2,75.7) & (37.7,82.6) & (15.2,87.5) \\
$\omega$ (deg) & 61.2 & 204.7 & 63.1 & (28.1,141.0) & (4.8,175.1) & (0.3,179.7) \\
$\Omega$ (deg) & 105.2 & 285.3 & 275.9 & (89.9,120.8) & (50.6,135.9) & (7.8,170.0) \\
$T_0$ (yr)     & 2026 & 2236 & 2024 & (2019,2051) & (2011,2160) & (2010,2852) \\
$P$ (yr)    & 39 & 231 & 25 & (22,95) & (14,302) & (10,1639) 
\enddata
\tablecomments{Summary of the posteriors from OFTI orbit fitting (see Section~\ref{sec:ofti}) for this triple system's outer and inner orbits. For each parameter, we show the median value, the value with the lowest $\chi^2$ ($x_{\chi^2,\mathrm{min}}$), the most likely value ($x_{\mathrm{P,max}}$) and three confidence intervals (C.I.). Samples of the posteriors for all fit parameters are available online as FITS tables.}
\end{deluxetable*}

\begin{deluxetable*}{lccccccc}
\tablecolumns{7}
\tablecaption{OFTI fit summary for triple system KELT-4A \label{tab:ofti_kelt4a}}
\tablehead{
\colhead{$x$} & \colhead{Median} & \colhead{$x_{\chi^2,\mathrm{min}}$} & \colhead{$x_{\mathrm{P,max}}$} &  \colhead{68\% C.I.} &  \colhead{95\% C.I.} &  \colhead{99.7\% C.I.}
}
\startdata
\sidehead{{\bf Outer orbit} (min $\chi^2$ = 0.039, $\chi^2$ at highest probability = 0.900)}
$a$ (au)       & 425.1 & 478.9 & 369.1 & (312.4,725.5) & (239.2,1657.8) & (193.5,5227.0) \\
$e$            & 0.406 & 0.561 & 0.229 & (0.145,0.660) & (0.021,0.846) & (0.001,0.951) \\
$i$ (deg)      & 32.7 & 5.8 & 41.2 & (16.6,50.6) & (6.2,66.8) & (1.4,80.3) \\
$\omega$ (deg) & 89.0 & 63.3 & 289.7 & (29.0,150.8) & (4.2,175.8) & (0.3,179.7) \\
$\Omega$ (deg) & 114.7 & 41.8 & 169.8 & (18.9,163.6) & (2.5,177.6) & (0.1,179.9) \\
$T_0$ (yr)     & 2683 & 2408 & 2501 & (2421,3736) & (2122,12556) & (2018,70154) \\
$P$ (yr)    & 5467 & 6377 & 4279 & (3443,12198) & (2304,42183) & (1673,236009) \\
\sidehead{{\bf Inner orbit} (min $\chi^2$ = 0.001, $\chi^2$ at highest probability = 0.180)}
$a$ (au)       & 7.4 & 7.9 & 4.6 & (5.5,12.5) & (4.7,27.8) & (4.2,85.2) \\
$e$            & 0.865 & 0.460 & 0.894 & (0.568,0.969) & (0.335,0.990) & (0.135,0.990) \\
$i$ (deg)      & 97.8 & 96.1 & 111.0 & (74.9,119.0) & (28.3,155.9) & (6.9,174.1) \\
$\omega$ (deg) & 48.8 & 16.8 & 1.8 & (13.7,163.9) & (2.0,178.0) & (0.1,179.9) \\
$\Omega$ (deg) & 109.5 & 106.5 & 113.1 & (89.5,127.8) & (44.7,151.1) & (2.5,177.6) \\
$T_0$ (yr)     & 2012 & 2010 & 2012 & (2011,2013) & (2010,2053) & (2010,2318) \\
$P$ (yr)    & 19 & 21 & 9 & (12,42) & (9,138) & (8,739)
\enddata
\tablecomments{Summary of the posteriors from OFTI orbit fitting (see Section~\ref{sec:ofti}) for this triple system's outer and inner orbits. For each parameter, we show the median value, the value with the lowest $\chi^2$ ($x_{\chi^2,\mathrm{min}}$), the most likely value ($x_{\mathrm{P,max}}$) and three confidence intervals (C.I.). Samples of the posteriors for all fit parameters are available online as FITS tables.}
\end{deluxetable*}

\begin{deluxetable*}{lccccccc}
\tablecolumns{7}
\tablecaption{OFTI fit summary for triple system WASP-12 \label{tab:ofti_wasp12}}
\tablehead{
\colhead{$x$} & \colhead{Median} & \colhead{$x_{\chi^2,\mathrm{min}}$} & \colhead{$x_{\mathrm{P,max}}$} &  \colhead{68\% C.I.} &  \colhead{95\% C.I.} &  \colhead{99.7\% C.I.}
}
\startdata
\sidehead{{\bf Outer orbit} (min $\chi^2$ = 19.894, $\chi^2$ at highest probability = 20.116)}
$a$ (au)       & 512.6 & 432.2 & 122.0 & (322.7,990.3) & (222.8,2992.2) & (155.2,13293.9) \\
$e$            & 0.531 & 0.202 & 0.858 & (0.173,0.856) & (0.025,0.980) & (0.001,0.990) \\
$i$ (deg)      & 67.9 & 74.6 & 76.3 & (40.1,94.8) & (15.9,133.6) & (3.9,167.0) \\
$\omega$ (deg) & 92.4 & 141.6 & 344.8 & (32.5,148.9) & (4.7,175.4) & (0.3,179.7) \\
$\Omega$ (deg) & 73.8 & 69.4 & 72.8 & (46.9,126.8) & (8.3,171.7) & (0.5,179.5) \\
$T_0$ (yr)     & 5230 & 7474 & 2409 & (3412,12532) & (2609,57434) & (2051,530896) \\
$P$ (yr)    & 7364 & 5874 & 811 & (3675,19784) & (2102,103755) & (1219,967232)\\
\sidehead{{\bf Inner orbit} (min $\chi^2$ = 0.106, $\chi^2$ at highest probability = 0.600)}
$a$ (au)       & 40.7 & 30.5 & 11.2 & (25.8,79.3) & (18.0,251.4) & (12.9,1166.0) \\
$e$            & 0.521 & 0.736 & 0.976 & (0.169,0.853) & (0.024,0.979) & (0.001,0.990) \\
$i$ (deg)      & 77.9 & 64.7 & 84.7 & (44.2,117.0) & (17.5,153.0) & (4.4,173.2) \\
$\omega$ (deg) & 99.1 & 299.5 & 194.0 & (35.9,148.4) & (5.1,175.0) & (0.3,179.7) \\
$\Omega$ (deg) & 128.6 & 19.2 & 164.8 & (26.5,167.9) & (2.8,177.5) & (0.2,179.8) \\
$T_0$ (yr)     & 2131 & 2086 & 2034 & (2068,2374) & (2032,4005) & (2011,22015) \\
$P$ (yr)    & 245 & 158 & 35 & (124,666) & (72,3764) & (43,37519)
\enddata
\tablecomments{Summary of the posteriors from OFTI orbit fitting (see Section~\ref{sec:ofti}) for this triple system's outer and inner orbits. For each parameter, we show the median value, the value with the lowest $\chi^2$ ($x_{\chi^2,\mathrm{min}}$), the most likely value ($x_{\mathrm{P,max}}$) and three confidence intervals (C.I.). Samples of the posteriors for all fit parameters are available online as FITS tables.}
\end{deluxetable*}

\end{document}